%% file: main.tex
\documentclass[acmsmall,screen,nonacm]{acmart}
\AtBeginDocument{%
  }

\setcopyright{acmlicensed}
\copyrightyear{2018}
\acmYear{2018}
\acmDOI{XXXXXXX.XXXXXXX}

\acmJournal{JACM}
\acmVolume{37}
\acmNumber{4}
\acmArticle{111}
\acmMonth{8}

\usepackage{xspace}         %
\usepackage{mathtools}      %
\usepackage{amssymb}        %
\usepackage{mathpartir}     %
\usepackage{listings}       %
\usepackage{colortbl}       %
\usepackage[nointegrals]{wasysym} %
\usepackage{subcaption}     %
\usepackage{cleveref}       %
\usepackage[inline]{enumitem}
\input{stele-macros}         %

\newcommand{\EquiVM}{\textsc{EquiVM}\xspace}
\newcommand{\lang}{\textsf{Stele}\xspace} %

\newcommand{\nrealworld}{thirty-five\xspace}
\newcommand{\EVM}{\textsc{EVM}\xspace}
\newcommand{\Lean}{\textsc{Lean}\xspace}
\newcommand{\Xexec}{\ensuremath{\Xi}}                 %
\newcommand{\code}[1]{\lstinline[basicstyle=\ttfamily]{#1}}
\newif\ifshowcomments
\showcommentsfalse
\ifshowcomments
\newcommand{\todo}[1]{\textcolor{red}{[TODO: #1]}}
\newcommand{\zoe}[1]{\textcolor{purple}{[Zoe: #1]}}
\newcommand{\lefteris}[1]{\textcolor{teal}{[Lefteris: #1]}}
\else
\newcommand{\todo}[1]{}
\newcommand{\zoe}[1]{}
\newcommand{\lefteris}[1]{}
\fi
\newcommand{\ie}{\textit{i.e.}\xspace}
\newcommand{\eg}{\textit{e.g.}\xspace}

\lstdefinestyle{lean}{
  basicstyle=\ttfamily\small,
  keywordstyle=\bfseries,
  columns=fullflexible,
  showstringspaces=false,
  breaklines=true,
  extendedchars=true,
  literate={⟨}{{\ensuremath{\langle}}}1 {⟩}{{\ensuremath{\rangle}}}1,
}
\begin{document}

\title{Foundational Refinement Proofs for EVM Bytecode, at the Price of Tokens}

\author{Eleftherios Lazaropoulos}
\affiliation{%
  \institution{National Technical University of Athens}
  \city{Athens}
  \country{Greece}}

\author{Zoe Paraskevopoulou}
\affiliation{%
  \institution{National Technical University of Athens}
  \city{Athens}
  \country{Greece}}
\email{zoe.paraskevopoulou@gmail.com}

\makeatletter
\def\@ACM@checkaffil{%
  \if@ACM@instpresent\else
    \ClassWarningNoLine{\@classname}{No institution present for an affiliation}%
  \fi
  \if@ACM@countrypresent\else
    \ClassWarningNoLine{\@classname}{No country present for an affiliation}%
  \fi}
\makeatother

\begin{abstract}
Relating executable code to a high-level model of its behavior has been a
central concern of programming-language research for decades.
Formally verified compilers, translation validators and certifying compilers
address this problem through different trade-offs between proof effort,
automation, generality, and the size of the trusted base.
Recently, large language models (LLMs) have begun to change the economics of
formal verification. Agentic proof development is now capable of producing
machine-checked proofs, sometimes even unsupervised, at a scale and speed that
were previously out of reach.
We propose a verification architecture that takes advantage
of this shift.
An executable obtains a machine-checked proof that it behaves according to a
high-level model, independently of the compiler that produced it.
The proof is written by an LLM agent.
We evaluate the feasibility of this architecture for the Ethereum Virtual
Machine (\EVM), a stack-based virtual machine that executes smart contracts on
the Ethereum blockchain.
We build \EquiVM, a foundational framework in \Lean comprising an executable
\EVM semantics and a formal specification language, \lang, that gives a high-level
authoritative model of the bytecode's behavior, but commits to no source
language or compilation toolchain. 
We state what it means for \EVM bytecode to refine a \lang spec and we build a
proof library to facilitate proof construction.
We evaluate the capabilities of frontier LLMs to construct such proofs, as well
as the cost of proof production and repair.
Our results show that agents prove \nrealworld real production contracts end to
end, without modifying their source or bytecode, among them core contracts of
Safe, Morpho, Compound and Uniswap protocols, with bytecode of size close to the
maximum the EVM allows.
Several proofs need only the initial prompt, the rest a handful of human inputs,
and existing proofs are repaired in a fraction of the original time when the
source or the compiler configuration changes.
\end{abstract}

\maketitle

\section{Introduction}
\label{sec:intro}
Reasoning about programs, whether formally or informally, is most effective at a
high level of abstraction, where a program's data and control flow are
structured and the details of its machine-level implementation are hidden.
What runs on the machine, though, is not the high-level program but the
low-level code a compiler produces from it. 
A theorem about the source carries over to the running system only insofar as
compilation preserves the semantics of the model that was used in the proof.
Closing the gap between a verified high-level program and the low-level code
that actually executes has been a major focus of programming-languages research
for decades.

A \emph{verified compiler} proves once and for all that every program it
compiles is refined by the compiled code, amortizing the cost of proof across
all programs. This is the approach of
CompCert~\cite{Compcert:Leroy09:formalverification} and
CakeML~\cite{CakeML:Kumar:cakeml}.
Such efforts are labor-intensive, but provide a very strong guarantee, albeit
tied to a single source language and toolchain.
\emph{Translation validation}~\cite{TransVal:Pnueli98:translationvalidation}
checks each compiler run after the fact with an automated decision procedure,
often using an SMT solver.
Some translation validators require no compiler cooperation or knowledge of the
compiler internals, and can thus validate translation from different evolving
toolchains (\eg, \code{gcc} and \code{clang})~\cite{TransVal:Rose24:dynamo}.
A third alternative is a \emph{certifying
compiler}~\cite{PCC:Necula98:certifyingcompiler, PCC:OConnor16:cogent,
PCC:Kang18:crellvm, Cakeml:Myreen12:proofproducing} that can generate, along with the
executable, a proof that it preserves the semantics of the source.
This requires knowledge of the compiler's internals and, thus, it is tied 
to one toolchain.
Each design trades off generality, proof labor, automation, and the size of the
trusted base.

Recently, large language models started drastically changing the economics of
formal verification. LLM agents can now produce formal proofs at a fraction of the
time that human experts required~\cite{LLM:Paraskevopoulou26:mgproofs,
LLM:Ioannidis26:proofspromptly, sergey2026movelean}.
In this paper we propose a different design, that leverages the ability of
LLM agents to produce formal proofs.
Each executable gets its own machine-checked proof that it behaves according to a
high-level model, generated post hoc by an LLM agent with no knowledge of or
assumptions about the compiler that produced it.

We develop the architecture of exactly this \emph{post hoc, per-executable
certification} discipline for the Ethereum Virtual Machine.
On the \EVM, the code that executes is bytecode, and its provenance is genuinely
heterogeneous: different compilers and versions, and even
hand-written assembly. 
No end-to-end verified compiler exists for the \EVM, let alone compiler
toolchains that support interaction with arbitrary bytecode, which is common in
\EVM.
At the same time, Solidity's optimizing compiler has a public record of
miscompilation bugs~\cite{SC:Ma24:solbugs, SC:Solidity:buglist}. 
Scrutiny of deployed code happens mostly at the bytecode level, with heuristic
decompilers~\cite{SC:Grech19:gigahorse, SC:Grech22:elipmoc} and vulnerability
analyzers~\cite{SC:Luu16:oyente, SC:Tsankov18:securify}, and SMT-based bytecode
verifiers~\cite{SC:Dxo24:hevm, a16z:halmos, rv:kontrol, SC:Grossman24:certora}.
A refinement proof can transfer this scrutiny, soundly, to a high-level program.
Auditors read the specification instead of decompiler output, and verification
and analysis tools can operate at a high level of abstraction.
The \EVM also makes such a certificate worthwhile. Once deployed, bytecode is
immutable so one proof holds for the lifetime of the contract.
Additionally, the financial value a contract controls justifies per-contract
effort. 
Contracts are small programs, capped by the protocol at 24KB of deployed code,
and run in a formalized execution environment. This makes per-executable proofs
feasible.

\EquiVM is a \Lean framework that realizes the architecture as follows. 
We introduce \lang, an imperative specification language with formal semantics,
inspired by Solidity but tied to no source language (\cref{sec:stele}).%
It is parametric in the contract's storage layout, so a single specification can
characterize bytecode emitted by different compilers. 
To model \EVM, we use an extension of Nethermind's \EVM mechanization in \Lean
(\cref{sec:evm}).
\EquiVM defines a refinement relation between the deployed bytecode and its
specification (\cref{sec:equiv}).
Refinement proves that the specification is a sound abstraction of the bytecode,
for the behaviors the relation observes, so scrutiny of the specification
soundly covers the deployed bytecode.
Contracts interact, by design, with bytecode of arbitrary provenance. Refinement
must build that interaction in.
To do so, we use multi-language
semantics~\cite{General:Matthews07:operationalsemantics,
Compcorr:Perconti14:verifyinganopencompiler, Compcorr:Ahmed15:multilangworld,
Compcorr:Patterson22:seminterop, Compcorr:Sammler23:dimsum}.
\lang interprets an external call using the \EVM's own call semantics, making
interoperation with arbitrary bytecode part of the semantics and the refinement
statement.

In summary, our contributions are the following:
\begin{itemize}%
\item \EquiVM, a foundational framework for proving that \EVM bytecode refines a
high-level specification of its behavior. 
Its refinement relation assumes nothing about where the bytecode came from and
nothing about the code it calls. 
The framework provides a comprehensive reasoning library for constructing such
refinement proofs.
\item An evaluation of LLM agents producing refinement proofs in this
architecture (\cref{sec:casestudies}).
We prove \nrealworld unmodified real-world contracts (11,146 compiled lines),
across several \code{solc} versions and configurations, and \code{Vyper}.
We report cost in tokens, time, and human input for proof construction,
specification, and proof repair after source or compiler changes.
To our knowledge, this is the first use of LLM agents to produce such
proofs.
\end{itemize}
We begin with an overview of the \EVM and \EquiVM (\cref{sec:overview}) and end
with related work (\cref{sec:related}) and conclusion (\cref{sec:conclusion}). 

\section{Overview}
\label{sec:overview}
\input{Overview}

\section{The EVM, Formally}
\label{sec:evm}
\input{EVM}

\section{The \lang Surface Language}
\label{sec:stele}
\input{Stele}

\section{Refinement}\label{sec:equiv}
\input{Refinement}
\subsection{Reasoning Library}
\label{sec:reasoning}
\input{Reasoning}

\section{Evaluation}
\label{sec:casestudies}
\input{Evaluation}

\section{Related Work}
\label{sec:related}
\input{related.tex}

\section{Conclusion and Future Work}
\label{sec:conclusion}
We suggest a new approach of connecting low-level bytecode to a high-level
account of its behavior, building on recent advances in the ability of LLMs to
conduct formal proofs.
We build \EquiVM, a \Lean framework for stating and proving that \EVM bytecode
of any provenance refines a high-level \lang specification. It comes with a
reasoning library and scripts that draft specifications and proof scaffolds.
We then evaluate the ability of agents running frontier models to conduct such
proofs for real production \EVM contracts, and the cost
of repairing a proof when the source or the compiler configuration changes.
The technique requires no cooperation from the compiler and we did not modify the
source code or the generated bytecode.
Our results show that agents can construct such proofs for large contracts, up
to 24 kilobytes of bytecode, just under the \EVM's limit on deployed code,
comprising a challenging set of constructs (48 loops, six of them nested),
sometimes from the initial prompt alone.
Writing and proving the \lang specification surfaces subtle compiler behavior,
such as the checks \code{solc} inserts and permissive calldata decoding, and in two
cases even a known compiler bug that random testing alone did not catch.
We demonstrate that after establishing refinement, properties proved on the
high-level \lang specification can be transferred all the way down to the bytecode that runs.

In the future, we plan to investigate methods for lowering the cost of such
proofs, in tokens, time and proof size, and to build verification infrastructure
on top of \lang. 

\begin{acks}
This research has been funded by Argot Collective.
\end{acks}

\bibliographystyle{ACM-Reference-Format}
\bibliography{bib/cakeML,bib/closure,bib/compcert,bib/compcorr,bib/compilers,bib/coq,bib/cps,bib/evm,bib/GC,bib/general,bib/lisp,bib/llm,bib/logrel,bib/ML,bib/online,bib/pcc,bib/proofs,bib/resource,bib/scheme,bib/sepcomp,bib/space,bib/synth,bib/theses,bib/transval,bib/types}
\appendix
\input{Appendix}

\end{document}

%% file: stele-macros.tex
\newcommand{\cfg}{\ensuremath{\chi}\xspace}          %
\newcommand{\evmst}{\ensuremath{\Sigma}}             %
\newcommand{\frm}{\ensuremath{F}}                    %
\newcommand{\locals}{\ensuremath{\rho}}              %
\newcommand{\imms}{\ensuremath{\iota}}                %
\newcommand{\anexp}{\ensuremath{e}}
\newcommand{\astmt}{\ensuremath{s}}
\newcommand{\stmts}{\ensuremath{\overline{s}}}
\newcommand{\aval}{\ensuremath{v}}
\newcommand{\avals}{\ensuremath{\overline{v}}}
\newcommand{\aref}{\ensuremath{\ell}}                %

\newcommand{\CallEVM}{\ensuremath{\Theta}}           %
\newcommand{\CreateEVM}{\ensuremath{\Lambda}}        %
\newcommand{\TxEVM}{\ensuremath{\Upsilon}}           %
\newcommand{\XEVM}{\ensuremath{\mathsf{X}}}          %
\newcommand{\stepEVM}{\ensuremath{\mathsf{O}}}       %
\newcommand{\evmyullean}{\textsc{EVMYulLean}\xspace}

\newcommand{\rok}[1]{\ensuremath{\mathsf{ok}\,#1}}
\newcommand{\rrev}{\ensuremath{\mathsf{revert}}\xspace}
\newcommand{\rerr}{\ensuremath{\mathsf{error}}\xspace}

\newcommand{\oOk}[2]{\ensuremath{\mathsf{ok}\langle #1,\, #2\rangle}}
\newcommand{\oRet}[3]{\ensuremath{\mathsf{ret}\langle #1,\, #2,\, #3\rangle}}
\newcommand{\oBrk}[2]{\ensuremath{\mathsf{brk}\langle #1,\, #2\rangle}}
\newcommand{\oCont}[2]{\ensuremath{\mathsf{cont}\langle #1,\, #2\rangle}}
\newcommand{\oRev}{\ensuremath{\mathsf{rev}}\xspace}
\newcommand{\oStatic}{\ensuremath{\mathsf{static}}\xspace}   %
\newcommand{\oNoMatch}{\ensuremath{\mathsf{nomatch}}\xspace}  %
\newcommand{\oNoDecode}{\ensuremath{\mathsf{nodecode}}\xspace} %
\newcommand{\outc}{\ensuremath{R}}                   %

\newcommand{\evalE}[3]{\ensuremath{[\![#3]\!]^{#1}_{#2}}}
\newcommand{\execS}[4]{\ensuremath{\langle #1,\, #2\rangle \vdash #3 \Downarrow #4}}
\newcommand{\execB}[4]{\ensuremath{\langle #1,\, #2\rangle \vdash #3 \Downarrow^{*} #4}}
\newcommand{\execF}[4]{\ensuremath{\langle #1,\, #2\rangle \vdash #3 \Downarrow_{\mathsf{body}} #4}}
\newcommand{\steleexec}[4]{\ensuremath{#1;\, #2 \vdash (#3) \Downarrow_{\mathsf{msg}} #4}}

\newcommand{\kw}[1]{\ensuremath{\mathsf{#1}}}        %
\newcommand{\rulename}[1]{\textsc{#1}}               %

\newcommand{\layout}{\ensuremath{\mathcal{L}}\xspace}      %
\newcommand{\backend}{\ensuremath{\mathcal{B}}\xspace}     %
\newcommand{\readSt}[3]{\ensuremath{\mathsf{read}_{#1}(#2,\, #3)}}
\newcommand{\writeSt}[4]{\ensuremath{\mathsf{write}_{#1}(#2,\, #3,\, #4)}}

\newcommand{\eqdef}{\mathrel{\overset{\mathrm{def}}{=}}}
\newcommand{\judgbox}[1]{\fbox{$\,#1\,$}}   %

\newcommand{\reachJ}[4]{\ensuremath{#1 \rightsquigarrow^{#2,\,#3} #4}}
\newcommand{\reachRet}[3]{\ensuremath{#1 \Downarrow_{\mathsf{ret}} (#2,\, #3)}}
\newcommand{\reachRev}[1]{\ensuremath{#1 \Downarrow_{\mathsf{rev}}}}
\newcommand{\cursor}[6]{\ensuremath{\langle #1 \mid #2 \mid #3,\,#4 \mid #5 \mid #6\rangle}}
\newcommand{\oog}{\ensuremath{\mathsf{oog}}}

\newcommand{\bnfdef}{\ensuremath{\;{: :=}\;}}
\newcommand{\bnfalt}{\ensuremath{\;\mid\;}}
\newcommand{\gcat}[1]{\textit{#1}}                   %

\lstdefinelanguage{Stele}{
  morekeywords={contract,constructor,function,external,public,payable,view,
    returns,receive,fallback,event,indexed,require,return,break,continue,
    delete,is,if,else,while,for,try,catch,new,var,storage,memory,constant,
    immutable,transient},
  keywordstyle=\bfseries,
  morekeywords=[2]{address,bool,uint256,uint160,uint8,bytes32,bytes4,bytes,
    string,mapping},
  keywordstyle=[2]\color[gray]{0.30},
  sensitive=true,
  morecomment=[l]{//},
  morecomment=[s]{/*}{*/},
  commentstyle=\itshape\color{gray},
  morestring=[b]",
  columns=fullflexible,
  keepspaces=true,
  basicstyle=\ttfamily\small,
  showstringspaces=false,
  breaklines=true,
  mathescape=true,
}
\lstdefinestyle{stele}{language=Stele}

\lstdefinelanguage{Solidity}{
  morekeywords={contract,mapping,public,external,returns,function,constructor,
    require,return,address,uint256,bool,event,emit,indexed,pragma,solidity},
  sensitive=true,
  morecomment=[l]{//},
  morecomment=[s]{/*}{*/},
  commentstyle=\itshape\color[gray]{0.35},
  morestring=[b]",
  columns=fullflexible,
  keepspaces=true,
  basicstyle=\ttfamily\small,
  keywordstyle=\bfseries,
  showstringspaces=false,
  breaklines=true,
}

%% file: Overview.tex
Ethereum executes programs, called smart contracts, on the Ethereum Virtual
Machine (\EVM).
The machine's persistent state is a map from \emph{addresses} to
\emph{accounts}.
A user account holds a balance of \emph{Ether}, the platform's currency. A smart
contract account holds a balance, together with its \emph{runtime bytecode} and
its own \emph{storage}.
Only the contract's own code can write its own storage, so a contract
encapsulates its state much as an object in an object-oriented language does.
A contract's account is created by \emph{deployment}: the \EVM runs the
contract's \emph{initcode} once, which runs its constructor and installs the
account's runtime bytecode.
Execution is triggered by transactions. A user submits a transaction naming a
deployed contract's address, together with an input that selects the method to
run and supplies its arguments. The machine then runs that contract's runtime
code.
The bytecode is typically obtained by compiling a smart contract written in a
high-level language.
The dominant language of the ecosystem is Solidity~\cite{SC:Solidity:docs}, but
other languages exist, \eg, Vyper~\cite{SC:Vyper:docs} and
Fe~\cite{SC:Fe:docs}.

\subsection{Execution by Example}\label{sec:overview-contract}
\Cref{fig:overview-token} shows \code{ERC20}, a token contract in Solidity. 
It implements a ledger: the storage variable \code{balanceOf} is a
\code{mapping}, a key-value store that records how many tokens each address
holds. 
The constructor credits the deployer with the initial supply. The public
function \code{transfer} moves an amount from the caller to a recipient,
if there is sufficient balance.

\input{figs/overview-example}

\textbf{Executing a Call.}\label{sec:overview-exec}
A user calls a contract by sending a transaction to the contract's address.
The transaction specifies an input byte string called \emph{calldata}, an amount
of Ether, and a \emph{gas} budget. 
Once the transaction's own bookkeeping is done (signature and nonce checks, fee
accounting), the execution begins with a \emph{message call} to the destination
address.
A message call is a request to run the code held at an address. It specifies the
caller (\code{msg.sender} in Solidity), the Ether transferred, the calldata, and
the gas the execution may consume.
The code reads and writes the called account's storage, and produces a success
flag and return data.

How callers and contracts agree on the meaning of calldata is the subject of the
application binary interface (ABI).
The EVM has a stable compiler-independent ABI that specifies the calling
convention and the representation of data across calls so that contracts
compiled by different toolchains call each other freely.
Calldata begins with a four-byte \emph{selector}, the first four bytes of the
Keccak-256 hash of the function's signature followed by the ABI encoding of the
arguments. 
A statically sized argument is encoded in place, in as many $32$-byte words as
it needs. 
A dynamic one, such as a dynamic array, puts a single offset word there instead,
pointing into a trailing region that carries its length and contents.
Return values are encoded similarly.
Execution can complete and return a byte string as its result or revert (\eg, a
 failed \code{require}). 
A revert abandons the executing context and rolls back every state change that
has been made.
Every instruction consumes gas. If gas runs out, execution aborts and its
changes are rolled back as well. Gas is the platform's payment for computation,
and it rules out non-termination.

Storage variables are stored in the contract's account storage, a flat
word-to-word map.
How each variable is laid out is a choice of the compiler.
In the \code{ERC20} example, \code{balanceOf}, declared first, gets slot $0$. A
mapping stores the entry for key $k$ at the slot obtained by hashing $k$
together with the mapping's slot number. So the balance of address $a$ lives at
slot $\mathsf{keccak}(a, 0)$. Other compilers or versions of \code{solc} may lay
storage out differently. 

\subsection{\EquiVM}
\label{sec:overview-equivm}

\EquiVM consists of the following ingredients.

\textbf{The EVM semantics.} We use a formal model of the EVM that extends Nethermind's
\evmyullean~\cite{SC:Nethermind25:evmyullean}, a \Lean~4 executable
formalization of the \EVM and Yul at the Cancun fork.
Our extension ports the development to a current \Lean toolchain, restructures
the EVM state and the signatures of the semantic functions to more closely follow
the Yellow Paper, and replaces various FFI implementations of computations
with pure \Lean definitions.
The EVM mechanization passes all Cancun conformance tests of the Ethereum
Tests suite, release v17.2~\cite{SC:Ethereum25:tests}.

\textbf{The \lang specification language.} We define \lang, an imperative
language with a formal semantics, to serve as an authoritative high-level
model of a contract's behavior.
In principle any language with a formal semantics expressive enough would do.
\lang deliberately stays close to Solidity so that specifications are easy to
write and to read.
Pinning its syntax and semantics to Solidity, however, would commit us to a
particular compilation toolchain and version.
For example, before version~0.8, Solidity's arithmetic wrapped silently and
contracts guarded it with libraries such as SafeMath. 
From version~0.8 on every operation is checked and reverts on overflow, and
\code{unchecked} blocks are how a contract opts back into wrapping.
The same source text therefore denotes different behavior under different
compiler versions. Other languages draw the line elsewhere (\eg Vyper spells
wrapping arithmetic \code{unsafe_add} and \code{unsafe_sub}).
Instead, \lang is \emph{language and toolchain agnostic}.
Its integers are mathematical rather than $256$-bit words, and overflow or wrap
around has to be explicit in the specification (\eg, the expression
$\anexp\;\kw{is}\;\kw{uint}n$ reverts unless the value of $\anexp$ lies in
$[0,2^{n})$, and is the identity otherwise).
In the \lang version of the ERC20 contract (\cref{fig:overview-spec}) the only
difference from Solidity is the range guard \code{is uint256} on the recipient's
new balance. The operation above needs no guard, because the \code{require}
prevents any reverts. Integers written to storage are truncated to the slot's
width.
For the same reason \lang is parametrized by each contract's storage layout.

\lang covers most of what deployed contracts do.
We explicitly list what is not supported.
\lang leaves out the ABI's \kw{function} type, and the fixed-point decimal types
that the ABI defines but Solidity does not implement.
\lang does not model events and does not carry revert payloads (adding them
is a matter of engineering and enriching the outcomes of \lang execution).
\lang has neither inheritance nor modifiers. These are not expressivity
limitations: inheritance is flattened into the contract and a modifier's body
appears inline at each function that carries it.
This is in line with \lang's aim to be language agnostic, distinguishing the
behavior from the way its source happens to be organized.
Gas is also not modeled in \lang and giving a sound gas model to EVM is outside
the scope of this paper. 
Because \lang does not track gas, the gas forwarded to external calls and the
value returned by \code{gasleft()} are picked non-deterministically. The
semantics admits any choice, and a refinement proof instantiates them with the
values the bytecode uses. Thus, a specification may admit more executions than
the bytecode.

\textbf{The refinement relation.} The refinement relation connects executions of
the bytecode with those of the \lang specification (\cref{sec:equiv}).
Refinement implies that for any input of a message call starting from any \EVM
state the bytecode's execution either runs out of gas or is matched by a
specification run on the same call from the same state.
A successful bytecode execution is matched by a successful \lang
execution with identical final accounts and return data (up to ABI encoding).
A bytecode revert is matched by a \lang revert.
A bytecode halt at an invalid instruction, which \code{solc} before version 0.8
emits for division by zero and failed assertions, must also be matched by a
specification revert.
Exceptional halts such as a stack underflow or an illegal jump cannot be
matched, so they never occur in bytecode that refines a specification.
Gas, revert data and the substate, which records events and refunds, are not
compared in the relation, because \lang does not model them.
Executions that run out of gas are unconstrained.
Because refinement must hold for every gas budget, a call escapes the relation
only if the bytecode runs out of gas on it under every budget.
Such an execution has no observable effect on the EVM state beyond the gas it
consumes.

The relation therefore ensures that the specification is a sound abstraction of the
bytecode for the behaviors it observes.
A mistake in a \lang specification can make a proof fail, but it cannot make a
false statement about the bytecode provable.
To demonstrate the adequacy of the refinement relation, we show that it can be 
used to transfer properties proved about the high-level \lang spec to 
the bytecode that refines it. 
For an ERC20 token, we prove on the spec alone that the constructor establishes
the supply invariant $\mathit{totalSupply} = \sum_a \mathit{balanceOf}[a]$ and
that every call preserves it, assuming only that the storage layout maps
distinct storage references to distinct slots.
We then use the proved refinement to transfer this to the bytecode: every
successful deployment establishes the invariant and every successful EVM
execution of the runtime code preserves the invariant.
The bytecode theorem does not mention the high-level spec or the refinement 
relation.

\textbf{A reasoning library.}
We provide a library of lemmas and tactics to support construction of 
refinement proofs.
The library comprises $49$k lines of \Lean .
Its core abstraction is a relation stating that the whole execution from a start
state runs out of gas, or that after a number of steps and an amount of gas it
reaches a state with a specified program counter, stack, memory, return data,
and account map.
One lemma per opcode advances the relation by one instruction, and chaining
these lemmas performs forward symbolic execution of the bytecode without
unfolding the \EVM's execution function.
The library further provides lemmas about arithmetic, memory, storage encodings,
ABI encoding and decoding, and contract deployment.
Sequences that various versions of \code{solc} emit in every contract, such as
the dispatcher and checked arithmetic are proved once and reused in contracts'
proofs.

\textbf{Automation.}
We provide scripts that automate routine parts of specification and proof
construction.
For Solidity contracts, a translator produces a draft \lang specification
from the Solidity abstract syntax tree emitted by \code{solc} in JSON
format.
It leaves holes, most notably inline assembly whose behavior must be
expressed using \lang's high-level operations.
Differential testing runs the specification and bytecode on the same inputs and
compares their results, to help catch specification mistakes before proof
construction.
The scripts also generate a proof scaffold consisting of the contract-specific
definitions, theorem statements for each function and the constructor, with
their proofs left as holes for the agent to fill, and the top-level refinement
theorem, with a proof that combines these results.
A block-summary generator composes the reasoning library's opcode lemmas
into proofs describing the effect of straight-line bytecode sequences.
These summaries let the agent reason about a sequence of bytecode
instructions as a single step.

With these at hand, the \EquiVM workflow is the following.

\textbf{Phase 1: Specification and proof scaffolding.}
Given the source of a contract and the compiler configuration that produces
its bytecode, the first step is to construct the \lang specification.
We give an agent these inputs together with a fixed prompt explaining how to
use the scripts, complete the generated specification, and check it against the
bytecode.
The agent runs differential tests, corrects specification mistakes, and then
audits each function and the constructor against the bytecode.
The output is a completed specification and a proof scaffold that
type-checks in \Lean, with the remaining proofs marked as holes.

\textbf{Phase 2: Formal proof.}
An agent receives the specification and proof scaffold, together with a
fixed proof-development prompt, and is asked to complete all remaining
proof holes.
The prompt, common for all contracts, explains how to organize the proof, use the
reasoning library and block summaries, and draw on existing proofs as examples.
Completion requires a successful \Lean build with no remaining proof holes and
an inspection of the theorem's axiom dependencies against the accepted trusted
base.

\textbf{Evaluation.} 
We evaluate \EquiVM on 35 real contracts (\cref{sec:casestudies}), totalling
11,146 lines of compiled source code, written for \code{solc} versions from
0.5.16 to 0.8.37 and compiled through both compilation pipelines (legacy and
\code{--via-ir}) under different optimizer settings, as well as one contract
compiled with Vyper.
We modify neither the source code nor the bytecode that the compilation
pipelines produce.
For nine benchmark contracts, totalling 8,229 compiled lines, we report detailed
telemetry on specification and proof construction, measuring
token consumption, proof time and human input. 
These are the implementation contract of the Safe smart-account wallet, the core
contracts of the Morpho Blue and Compound~III lending protocols, a MetaMorpho
vault, the Uniswap v3 exchange pool, the pool manager Uniswap v4 contract, the
registry of the Ethereum Attestation Service and an example contract built on
it, and Curve's LlamaLend V2 vault, written in Vyper.
Agents proved all nine, four of them from the initial prompt alone and the rest
with a handful of human inputs, in six to 35 hours of active proof time per
contract.
Specification took one to five hours per contract, and its differential tests
and audit caught specification errors before proving started.
In two cases, specification surfaced a known compiler
bug~\cite{SC:Solidity25:sol20251}.
Repairing a proof after a source change took under two hours from a single
prompt, against eighteen for the original proof.
Another 26 smaller contracts, totalling 2,917 compiled lines, were proved with
earlier versions of the framework and proofs were ported incrementally.
We submit the code of \EquiVM and proved contracts, the scripts, the general
prompts, and an additional appendix as supplementary material. 
The sessions from which we extract the telemetry with scripts are several GB, 
and exceed the supplementary material size limit.

\textbf{Trusted base.} 
The \EVM semantics are trusted to faithfully capture the EVM execution.
Proof checking relies on the \Lean kernel and three standard logical
axioms: propositional extensionality (\code{propext}), classical choice
(\code{Classical.choice}), and quotient soundness (\code{Quot.sound}).
For computations inside proofs we often need to use \code{native_decide},
which compiles and runs a computation outside of the \Lean kernel and then
accepts a successful result as an axiom.
We need \code{native_decide} for computations over large byte arrays, which in
the kernel are represented as lists, and evaluating them exhausts its stack on
real contracts.
This adds \Lean's compiler and native evaluator to the trusted base.
As in compiler verification, one must also trust that the refinement relation
expresses the intended correctness guarantee.
The LLM agents, the scripts, and the compiler that produced the bytecode are
outside this trusted base.

%% file: figs/overview-example.tex
\begin{figure}[t]
\begin{minipage}[b]{0.48\textwidth}
\begin{lstlisting}[language=Solidity,basicstyle=\ttfamily\footnotesize,
    aboveskip=2pt,belowskip=0pt]
contract ERC20 {
  mapping(address => uint256) public balanceOf;
  uint256 public totalSupply;
  constructor(uint256 initialSupply) {
    balanceOf[msg.sender] = initialSupply;
    totalSupply = initialSupply;
    ...
  }
  function transfer(address to, uint256 value)
      external returns (bool) {
    require(balanceOf[msg.sender] >= value);
    balanceOf[msg.sender] -= value;
    balanceOf[to] += value;
    return true;
  }
  ...
}
\end{lstlisting}
\caption{\code{ERC20} in Solidity (excerpts).}
\label{fig:overview-token}
\end{minipage}%
\hfill
\begin{minipage}[b]{0.50\textwidth}
\begin{lstlisting}[style=stele,basicstyle=\ttfamily\footnotesize,
    keywordstyle={[2]\bfseries},aboveskip=2pt,belowskip=0pt]
contract ERC20 {
  mapping(address => uint256) balanceOf;
  uint256 totalSupply;
  constructor(uint256 initialSupply) {
    balanceOf[msg.sender] = initialSupply;
    totalSupply = initialSupply;
    ...
  }
  function transfer(address to, uint256 value)
      external returns (bool) {
    require(balanceOf[msg.sender] >= value);
    balanceOf[msg.sender] -= value;
    balanceOf[to] =
        (balanceOf[to] + value) is uint256;
    return true;
  }
  ...
}
\end{lstlisting}
\caption{\code{ERC20} in \lang (excerpts).}
\label{fig:overview-spec}
\end{minipage}
\end{figure}

%% file: EVM.tex
In this section, we give an overview of the formal semantics of the EVM. The model
follows the EVM Yellow Paper's~\cite{SC:Wood14:yellowpaper} structure and
notation closely enough that they can be read side by side.

\textbf{Machine and World State.}
\Cref{fig:evm-state} gives the state. A single record $\evmst$ bundles the
account map $\sigma$, its transaction-start checkpoint $\sigma_{0}$, the accrued
substate $A$ (bookkeeping for refunds, access records, and logs), the execution
environment $I$ of the current call, and the machine state $\mu$, together with
the block context that environment opcodes read.
\input{figs/evm-state}

\textbf{Semantics.}
Execution is organized, as in the Yellow Paper, into a hierarchy of
semantic functions (\cref{fig:evm-functions}). 
They are total, deterministic and executable.
The transaction function $\TxEVM$ performs the administrative layer: signature
and nonce checks, fee deduction, and final settlement.
Deployed bytecode is run only through the message-call function $\CallEVM$.
Every transaction that names a recipient and every \code{CALL}, \code{CALLCODE},
\code{DELEGATECALL}, and \code{STATICCALL} instruction enters through it. 
It receives the world state and the parameters of the call (sender, recipient,
value, calldata, gas, depth, and the code to run), transfers the value, builds
the callee's execution environment $I$, and runs the code (through $\Xexec$).
If the recipient is one of the ten precompiled contracts (built-in contracts at
10 fixed addresses providing hash functions and elliptic-curve operations), it
runs the corresponding built-in function.
It returns a world state, a substate, the remaining gas, the output, and a
success flag. 
On success, the world state and substate are the callee's final
ones. 
On revert or exceptional halt, the initial world state and substate are
returned, undoing all state changes.
The creation function $\CreateEVM$ derives the new account's address, enforces a
size cap on initcode, and installs the code that the initcode returns. 
Bytecode is run through the code-execution function
\[
  \Xexec(\sigma, \sigma_{0}, g, A, I)
  \;\in\;
  \bigl\{\; \mathsf{success}\,(\sigma', g', A')\; o
  \;\mid\; \mathsf{revert}\; g'\, o
  \;\mid\; \mathsf{exception}\; \bigr\}.
\]
$\Xexec$ assembles a fresh machine state over the given world state, computes
the valid jump destinations of the code with a static scan $D_{J}$, and iterates
single instructions with the iterator $\XEVM$.
Before each instruction, the check $Z$ decides exceptional halts, among them
unknown opcodes, stack under- and overflow, insufficient gas, and invalid jump
targets. 
Then the $\stepEVM$ function performs the instruction and charges its gas.

\input{figs/evm-functions}

%% file: figs/evm-state.tex
\begin{figure}[t]
\footnotesize
\renewcommand{\arraystretch}{1.15}
\begin{tabular}{@{}l l l@{}}
\toprule
\textbf{Record} & \textbf{Field (glyph)} & \textbf{Contents} \\
\midrule
state $\evmst$
  & $\sigma$, $\sigma_{0}$ & account map; checkpoint taken at transaction start \\
  & $A$, $I$, $\mu$ & accrued substate, execution environment, machine state \\
\midrule
account $\sigma[a]$
  & $\sigma[a]_{n}$, $\sigma[a]_{b}$, $\sigma[a]_{s}$ & nonce, balance, persistent storage (a finite word-to-word map) \\
  & ---, $\sigma[a]_{c}$ & transient storage (EIP-1153), deployed code (the bytes themselves) \\
\midrule
environment $I$
  & $I_{a}$, $I_{s}$, $I_{o}$, $I_{v}$, $I_{d}$ & executing account, caller, transaction origin, call value, calldata \\
  & $I_{b}$, $I_{p}$, $I_{H}$, $I_{e}$, $I_{w}$ & executing code, gas price, block header, call depth ($< 1025$), write permission \\
\midrule
machine $\mu$
  & $\mathit{pc}$, $\mu_{s}$, $\mu_{g}$ & program counter, word stack ($\le 1024$), remaining gas (saturating) \\
  & $\mu_{m}$, $\mu_{i}$, $\mu_{o}$ & byte-addressed memory, active-word count, return data of the last sub-call \\
\midrule
substate $A$
  & $A_{s}$, $A_{t}$, $A_{r}$ & self-destruct set, touched accounts, refund balance \\
  & $A_{a}$, $A_{K}$, $A_{l}$, --- & accessed accounts and storage keys (EIP-2929), log series, created accounts (EIP-6780) \\
\bottomrule
\end{tabular}
\caption{The \EVM state $\evmst$, with the Yellow Paper's glyphs. An account
carries its deployed code as bytes, where real clients store a hash into a
global database.}
\label{fig:evm-state}
\end{figure}

%% file: figs/evm-functions.tex
\begin{figure}[t]
\footnotesize
\renewcommand{\arraystretch}{1.2}
\begin{tabular}{@{}l l p{\dimexpr\textwidth-13em\relax}@{}}
\toprule
\textbf{Function} & \textbf{Level} & \textbf{Role} \\
\midrule
$\TxEVM$ & transaction &
  signature and nonce checks, intrinsic gas, EIP-1559/4844 fee deduction,
  access lists, checkpoint $\sigma_{0}$; delegates to $\CallEVM$ or
  $\CreateEVM$ and settles refunds and fees \\
$\CallEVM$ & message call &
  value transfer, callee environment construction, precompiled contracts
  ($1$--$10$); runs the callee's code via $\Xexec$; yields
  $(\sigma', g', A', z, o)$ with success flag $z$ and output $o$ \\
$\CreateEVM$ & creation &
  \code{CREATE}/\code{CREATE2} address derivation by Keccak-256, collision
  and init-code-size checks (EIP-7610/3860); runs init code via $\Xexec$ and
  deploys the returned bytes \\
$\Xexec$ & code execution &
  runs one code body: iterates $\XEVM$ with fuel $g+1$ and the jump-destination
  set $D_{J}(I_{b})$; returns $\mathsf{success}(\sigma', g', A')\,o$,
  $\mathsf{revert}\;g'\,o$, or an exceptional halt \\
$\XEVM$ & instruction loop &
  $\mathsf{decode}$ the opcode at $\mathit{pc}$ (with \code{PUSH}
  immediates); check $Z$; apply $\stepEVM$; detect halting on \code{RETURN},
  \code{REVERT}, \code{STOP}, and \code{SELFDESTRUCT} \\
$Z$ & exceptional halt &
  invalid opcode, stack under-/overflow, out-of-gas for memory and
  instruction charges, bad jump destination, \code{RETURNDATACOPY} bounds,
  static-mode violation, init-code cap \\
$\stepEVM$ & one opcode &
  the per-instruction state transformer, one case per opcode of the Cancun
  instruction set, including full gas accounting \\
\bottomrule
\end{tabular}
\caption{The semantic functions of the \EVM model (total \Lean
functions). The mutual nest
$\stepEVM$/$\XEVM$/$\Xexec$/$\CallEVM$/$\CreateEVM$ terminates by a
lexicographic measure: the remaining call depth, bounded by $1024$,
decreases at every nested call, and within one call frame the fuel of
$\XEVM$ decreases at every instruction.}
\label{fig:evm-functions}
\end{figure}

%% file: Stele.tex
As we explain in \cref{sec:overview-equivm}, \lang is designed to be language
and compiler agnostic so that it can be used in the specification of smart
contracts compiled from different source languages, or different versions of the
same source language.
Its syntax is designed to be close to Solidity's conventions, the most widely
used EVM language, so the specification can be easily generated and read.
We define a surface DSL in \Lean so that specifications can be written in concrete
syntax.
We explain \lang's syntax (\cref{fig:stele-syntax}) and its semantics (\cref{fig:stele-judgments}).

\input{figs/stele-syntax}

\subsection{Syntax}
\textbf{Types.} 
Elementary types $\tau$ are integers, booleans, addresses and fixed-size byte
strings ($\kw{bytes}n$).
ABI types $\alpha$ add static and dynamic arrays, dynamic byte strings
($\kw{bytes}$), strings and tuples ($(\overline{\alpha})$).
Storage types $T$ add mappings ($\kw{mapping}(\tau\;\texttt{=>}\;T)$), structs
(declared at the beginning of a contract and denoted by their name $S$), and
contract types ($C$). 
Mapping keys are elementary.
Solidity also admits \kw{bytes}, \kw{string}, contract, enum and user-defined
value types as mapping keys.
\lang omits two ABI types \kw{function}, a 24-byte pair of an address and a
selector, and fixed-point decimals, which Solidity accepts in declarations
but cannot assign to or from.
Enums and user-defined value types are absent as well. 
The ABI represents an enum as \kw{uint8} and a user-defined value type as its
underlying type, and both can be written in a \lang specification in the same way.

\textbf{Paths.}
A path $\aref$ is a variable $x$ possibly followed by field selectors of a
struct ($\aref.f$) or element selectors of mappings and arrays
($\aref[\anexp]$).
If the root is a local variable, a constant or an immutable then the path
denotes the value of the variable or the component it selects.
If the root is a storage reference, persistent or transient, then it is a
reference to a storage slot. 
Paths can be read and are the targets of assignment, \kw{push}, \kw{pop} and
\kw{delete}.

\textbf{Expressions.}
In \lang, expressions have no side effects, similar to CompCert's
Clight~\cite{CompCert:Blazy06:frontend}. 
The basic expressions are literals, paths and environment variables. 
The environment variables give access to message call information (its sender,
value, signature and data), the transaction (its origin and gas price), the
executing contract (its address and balance) and the block (its number,
timestamp, chain id, coinbase, gas limit, prevrandao and base fee).
Compound expressions include operators on integers, booleans and fixed-size byte
strings, and conditionals. 
Bitwise operators on integers are annotated with the type of their operands
($\oplus[\tau]$), which fixes the width and signedness they operate at.
The range guard $\anexp\;\kw{is}\;\tau$ checks that an integer expression is
within the range of a type, and the cast $T(\anexp)$ converts between types.
Compound values are built with array, tuple and struct literals, and $\kw{new}$
for in-memory arrays and byte strings holding default values.
The remaining expressions are the keccak function (\kw{keccak256}), ABI encoding
and decoding (exposed as in Solidity), and queries on accounts and blocks
(\kw{balance}, \kw{code.length}, \kw{codehash}, a code prefix and
\kw{blockhash}).

\textbf{Statements.}
Local variables are declared with $\alpha\;x = \anexp$ or
$T\;\kw{storage}\;x = \aref$.
The former holds an in-memory value and the latter a reference to storage.
Storage types do not in general have value forms since mappings cannot be copied.
An assignment $\aref = \anexp$ updates a local variable, persistent or transient
storage, or, in the constructor, an immutable.
Control mirrors Solidity and includes \kw{if}, \kw{while}, \kw{for}, \kw{break},
\kw{continue} and \kw{return}. 
The guard $\kw{require}(\anexp)$ reverts when the condition is not true.
Dynamic storage arrays are modified with \kw{push} and \kw{pop}, and a storage
path is cleared with \kw{delete}.
Calls bind their result to a variable. 
Call forms include internal calls ($T\;\kw{storage}^{?}\;x = f(\overline{\anexp})$),
external calls ($\alpha\;x = \anexp.f\{\kw{value}{:}\,\anexp\}(\overline{\anexp})$),
and external view calls ($\alpha\;x = \anexp.f\{\kw{view}\}(\overline{\anexp})$),
which run the callee without write permission.
The low-level \kw{call}, \kw{staticcall} and \kw{delegatecall} take raw calldata
and bind a success flag and the raw return data, as in Solidity.
Contract creation ($\kw{address}\;x =
\kw{new}\;C\{\kw{value}{:}\,\anexp,\,\kw{salt}{:}\,\anexp\}(\overline{\anexp})$)
creates a new contract  $C$ and binds its address.

\textbf{Top-level declarations.}
A contract declaration groups together its persistent and transient storage
fields with their types ($\overline{T\;x}$), its constants and immutables, a
constructor with its parameters and body, and the contract's transitions and
functions.
A constant is bound to a compile-time constant expression. An immutable has an
elementary type and is assigned once in the constructor body.
External and internal functions are both declared with \kw{function} and have
the same shape: a name, a list of typed parameters, a return type (possibly
multi-valued or empty), and a statement body. 
External functions are declared \kw{external} and have ABI types.
Internal functions are declared \kw{internal} and have storage types.
\kw{receive} is a payable function without parameters and results that runs on a
call with empty calldata.
\kw{fallback} runs on a call matching no external transition, or on a call with empty
calldata when no \kw{receive} is defined.
It may have \kw{bytes}-typed parameters and results.

\subsection{Semantics}
\lang has a relational and an executable semantics connected by an equivalence theorem.
The former is used in proofs and the latter in specification execution, which 
is especially useful in differential testing against the bytecode.
Relational semantics treat gas non-deterministically while executable semantics
treat it as an oracle.
\Cref{fig:stele-judgments} shows the main semantics definitions of \lang.

\input{figs/stele-judgments}\label{fig:semantics}

\textbf{Values and frames.}
Integer values in \lang are mathematical integers and overflow and wrapping 
must be written explicitly.
A storage reference $\kw{ref}\,(\hat{\aref} : T)$ is an evaluated path, with $T$
the type of its slot.
A frame $\frm$ holds the contract declaration, the local store $\locals$ and
the immutables store $\imms$.

\textbf{State and configuration.} 
The state $\Sigma$ of the \lang semantics is the \EVM state
of \cref{sec:evm} itself: the account map $\sigma$ with its checkpoint
$\sigma_0$, the substate $A$, the execution environment $I$ and the machine
state $\mu$.
\lang reaches the \EVM storage through a storage interface of six operations,
each over a storage reference $\hat{\aref}$, the declared type $T$ of the cell it
reaches, and the state $\evmst$.
$\mathsf{read}(\hat{\aref}, T, \evmst)$ yields the data stored there as a
\lang value, a revert or an error.
$\mathsf{write}(\hat{\aref}, T, \aval, \evmst)$ takes the value to store and
yields the new state, a revert or an error.
The remaining four, clear, length, push and pop, are used for \kw{delete},
\kw{length} and the dynamic-array statements.
This interface depends on the specific layout, $\layout$, which maps references
to storage slots, which has to be instantiated per contract.

A configuration carries
\begin{enumerate*}
\item the storage interfaces for the persistent and transient storage,
\item the ABI operations,
\item $\mathsf{initcode}$, a map from contract names and constructor arguments
  to their init code, so that they can be created with $\kw{new}$, and
\item $\mathsf{deploy}$, the compiler's rule for forming the init code of the
  contract under verification from its creation bytecode and its constructor
  arguments.
\end{enumerate*}
ABI encoders and decoders in general need not change across contracts. 
However, the ABI specification does not fix what a decoder does with malformed
calldata, such as a non-canonical value word or a wrong length, and compilers
treat this differently (recent \code{solc} rejects it, \code{solc} before 0.8
cleans the value instead, and Vyper checks sizes differently).
In practice, our ABI operations are parameterized by a mode which selects the
decoder's validation.

\textbf{Semantic definitions.} 
Expression evaluation $\evalE{\frm}{\evmst}{\anexp}$ is a total function returning 
a result.
Resolving a path evaluates its indices and yields the evaluated path with the
declared type of its cell. 
Reads and writes then go through the storage interface, or update the frame for
a path rooted at a local.
Statement execution is big-step and extends to block execution that threads
normal completion and stops at any other outcome.
External calls and creation bridge to the \EVM's $\CallEVM$ and $\CreateEVM$ on
the live account map, with gas and input substate existentially quantified.
Message-level execution dispatches on the selector, decodes the calldata and
runs the transition's body. It returns \oNoMatch when no entry point accepts
the call or \oNoDecode when a selector matches but the calldata does not decode
at the transition's parameter types.

Most semantic definitions are standard, and we only include some excerpts in the
appendix provided as supplementary material.
Here, we only show the more interesting external call rules
(\cref{fig:stele-extcall}).
They are defined in the style of multi-language
semantics~\cite{General:Matthews07:operationalsemantics,
Compcorr:Perconti14:verifyinganopencompiler}, making the interaction with
arbitrary bytecode part of the semantics.
A call encodes its arguments with the callee's ABI, invokes the \EVM's
message-call function $\CallEVM$ on the live account map, and decodes what
comes back.
The callee is whatever code the account map holds at the target address, run by
the \EVM semantics itself.
The call gas $g$ and the input substate $A_{\mathit{in}}$ are existentially
quantified, since \lang tracks neither.
The refinement proof instantiates them with the gas and the substate that the
bytecode's \code{CALL} actually passes.
This allows the refinement relation to talk about a single contract and still
cover any interaction it may have with arbitrary bytecode.

\label{sec:stele-semantics}
\input{figs/stele-extcall}

%% file: figs/stele-syntax.tex
\begin{figure}[t]
\footnotesize
\[
\begin{array}{@{}l@{\quad}r@{\;}c@{\;}l@{}}
\multicolumn{4}{@{}l@{}}{\textbf{Types}}\\[2pt]
\gcat{elementary} & \tau &\bnfdef&
  \kw{uint}n \bnfalt \kw{int}n \bnfalt \kw{address} \bnfalt \kw{bool}
  \bnfalt \kw{bytes}n
\\
\gcat{ABI} & \alpha &\bnfdef&
  \tau
  \bnfalt \alpha[n] \bnfalt \alpha[\,]
  \bnfalt \kw{bytes} \bnfalt \kw{string}
  \bnfalt (\overline{\alpha})
\\
\gcat{storage types} & T &\bnfdef&
  \tau
  \bnfalt T[n] \bnfalt T[\,]
  \bnfalt \kw{bytes} \bnfalt \kw{string}
  \bnfalt (\overline{T})
  \bnfalt \kw{mapping}(\tau\;\texttt{=>}\;T)
  \bnfalt S
  \bnfalt C
\\[6pt]
\multicolumn{4}{@{}l@{}}{\textbf{Paths}}\\[2pt]
\gcat{path} & \aref &\bnfdef&
  x \bnfalt \aref.f \bnfalt \aref[\anexp]
\\[6pt]
\multicolumn{4}{@{}l@{}}{\textbf{Expressions (effect-free; may revert)}}\\[2pt]
\gcat{expression} & \anexp &\bnfdef&
  n \bnfalt b \bnfalt \texttt{"}\mathit{bs}\texttt{"} \bnfalt \kw{bytes}n(n)
  \bnfalt \aref
  \bnfalt \mathit{env}
  \bnfalt \anexp \oplus \anexp
  \bnfalt \ominus\, \anexp
  \bnfalt \anexp[\anexp]
  \bnfalt \anexp\;\texttt{?}\;\anexp:\anexp
\\ &&\bnfalt&
  \anexp\;\kw{is}\;\tau
  \bnfalt T(\anexp)
  \bnfalt \anexp.i
  \bnfalt [\,\overline{\anexp}\,]
  \bnfalt \kw{tuple}(\overline{\anexp})
  \bnfalt S(\{\overline{f{:}\,\anexp}\})
  \bnfalt \anexp.f
  \bnfalt \anexp[\anexp:\anexp]
  \bnfalt \aref.\kw{length}
\\ &&\bnfalt&
  \kw{new}\;T[\,](\anexp)
  \bnfalt \kw{new}\;\kw{bytes}(\anexp)
  \bnfalt \kw{keccak256}(\anexp)
  \bnfalt \kw{sdiv}(\anexp, \anexp)
  \bnfalt \kw{srem}(\anexp, \anexp)
\\ &&\bnfalt&
  \kw{abi.encodePacked}(\overline{\alpha(\anexp)})
  \bnfalt \kw{abi.encodeWithSelector}(f, \overline{\anexp})
  \bnfalt \kw{abi.decode}(\anexp, (\alpha))
\\ &&\bnfalt&
  \anexp.\kw{code.length}
  \bnfalt \anexp.\kw{codehash}
  \bnfalt \anexp.\kw{balance}
  \bnfalt \kw{blockhash}(\anexp)
  \bnfalt \kw{extCodePrefix}(\anexp, \anexp)
  \bnfalt \kw{addrOf}(\anexp)
\\
\gcat{binary op} & \oplus &\bnfdef&
  \texttt{+} \bnfalt \texttt{-} \bnfalt \texttt{*} \bnfalt \texttt{/} \bnfalt \texttt{\%}
  \bnfalt \texttt{**}
  \bnfalt \texttt{==} \bnfalt \texttt{!=}
  \bnfalt \texttt{<} \bnfalt \texttt{<=} \bnfalt \texttt{>} \bnfalt \texttt{>=}
  \bnfalt \texttt{\&\&} \bnfalt \texttt{||}
\\ &&\bnfalt&
  \texttt{\&} \bnfalt \texttt{|} \bnfalt \texttt{\^{}} \bnfalt \texttt{<<} \bnfalt \texttt{>>}
  \bnfalt \texttt{\&}[\tau] \bnfalt \texttt{|}[\tau] \bnfalt \texttt{\^{}}[\tau]
  \bnfalt \texttt{<<}[\tau] \bnfalt \texttt{>>}[\tau]
\\
\gcat{unary op} & \ominus &\bnfdef&
  \texttt{!} \bnfalt \texttt{-} \bnfalt \texttt{\~{}} \bnfalt \texttt{\~{}}[\tau]
\\
\gcat{environment} & \mathit{env} &\bnfdef&
  \kw{msg.sender} \bnfalt \kw{msg.value} \bnfalt \kw{msg.sig} \bnfalt \kw{msg.data}
  \bnfalt \kw{tx.origin} \bnfalt \kw{tx.gasprice}
\\ &&\bnfalt&
  \kw{this} \bnfalt \kw{address}(\kw{this}).\kw{balance}
  \bnfalt \kw{block.timestamp} \bnfalt \kw{block.number} \bnfalt \kw{block.chainid}
\\ &&\bnfalt&
  \kw{block.coinbase} \bnfalt \kw{block.gaslimit}
  \bnfalt \kw{block.prevrandao} \bnfalt \kw{block.basefee}
\\[6pt]
\multicolumn{4}{@{}l@{}}{\textbf{Statements}}\\[2pt]
\gcat{statement} & \astmt &\bnfdef&
  \alpha\;x = \anexp
  \bnfalt T\;\kw{storage}\;x = \aref
  \bnfalt \aref = \anexp
  \bnfalt \kw{require}(\anexp)
  \bnfalt \kw{uint256}\;x = \kw{gasleft}()
\\ &&\bnfalt&
  \kw{if}\,(\anexp)\,\{\stmts\}\;(\kw{else}\;\{\stmts\})^{?}
  \bnfalt \kw{while}\,(\anexp)\,\{\stmts\}
  \bnfalt \kw{for}\,(\stmts;\, \anexp;\, \stmts)\,\{\stmts\}
  \bnfalt \kw{break}
  \bnfalt \kw{continue}
  \bnfalt \kw{return}\;\overline{\anexp}
\\ &&\bnfalt&
  \aref.\kw{push}(\anexp^{?})
  \bnfalt \aref.\kw{pop}()
  \bnfalt \kw{delete}\;\aref
  \bnfalt \kw{emit}\;E(\overline{\anexp})
\\ &&\bnfalt&
  T\;\kw{storage}^{?}\;x = f(\overline{\anexp})
  \bnfalt \alpha\;x = \anexp.f\{\kw{value}{:}\,\anexp\}^{?}(\overline{\anexp})
  \bnfalt \alpha\;x = \anexp.f\{\kw{view}\}(\overline{\anexp})
\\ &&\bnfalt&
  (\kw{bool}\;x, \kw{bytes}\;y) = \anexp.\kw{call}\{\kw{value}{:}\,\anexp\}^{?}(\anexp)
  \bnfalt (\kw{bool}\;x, \kw{bytes}\;y) = \anexp.\kw{staticcall}(\anexp)
\\ &&\bnfalt&
  (\kw{bool}\;x, \kw{bytes}\;y) = \anexp.\kw{delegatecall}(\anexp)
  \bnfalt \kw{address}\;x = \kw{new}\;C\{\kw{value}{:}\,\anexp,\,\kw{salt}{:}\,\anexp\}(\overline{\anexp})
\\ &&\bnfalt&
  \kw{try}\;\anexp.f\{\kw{value}{:}\,\anexp\}^{?}(\overline{\anexp})\;
    \kw{returns}\,(\alpha\;x)\,\{\stmts\}\;\kw{catch}\,(\kw{bytes}\;y)\,\{\stmts\}
\\[6pt]
\multicolumn{4}{@{}l@{}}{\textbf{Declarations}}\\[2pt]
\gcat{transition} & t &\bnfdef&
  \kw{function}\;f(\overline{\alpha\;x})\;\kw{external}\;\kw{payable}^{?}\;
    \kw{returns}\,(\overline{\alpha})^{?}\;\{\stmts\}
\\
\gcat{function} & \mathit{fn} &\bnfdef&
  \kw{function}\;f(\overline{T\;\kw{storage}^{?}\;x})\;\kw{internal}\;
    \kw{returns}\,(\overline{T\;\kw{storage}^{?}})^{?}\;\{\stmts\}
\\
\gcat{receive} & r &\bnfdef&
  \kw{receive}()\;\kw{external}\;\kw{payable}\;\{\stmts\}
\\
\gcat{fallback} & \mathit{fb} &\bnfdef&
  \kw{fallback}((\kw{bytes}\;x)^{?})\;\kw{external}\;\kw{payable}^{?}\;
    \kw{returns}\,(\kw{bytes})^{?}\;\{\stmts\}
\\
\gcat{contract} & \mathit{ctr} &\bnfdef&
  \kw{contract}\;C\;\{\;
    \overline{\kw{struct}\;S\;\{\overline{T\;f};\}}\;
    \overline{T\;x};\;
    \overline{T\;\kw{transient}\;x};\;
    \overline{\alpha\;\kw{constant}\;x = \anexp};
\\ &&&
  \phantom{\kw{contract}\;C\;\{\;}
    \overline{\tau\;\kw{immutable}\;x};\;
    \kw{constructor}(\overline{\alpha\;x})\;\kw{payable}^{?}\;\{\stmts\}\;
    \overline{t}\;
    \overline{\mathit{fn}}\;
    r^{?}\;
    \mathit{fb}^{?}
  \;\}
\end{array}
\]
\caption{Syntax of \lang. Superscript $^{?}$ marks
optional elements and $\overline{\kw{overline}}$ a possibly empty sequence.}
\label{fig:stele-syntax}
\end{figure}

%% file: figs/stele-judgments.tex
\begin{figure}[t]
\footnotesize
\noindent\makebox[\linewidth][l]{$\begin{array}{@{}l@{\quad}l@{\;}l@{\;}l@{}}
\multicolumn{4}{@{}l@{}}{\textbf{Values}}\\[2pt]
\gcat{value} & \aval &\bnfdef&
  n
  \bnfalt b
  \bnfalt a
  \bnfalt \mathit{bs}_{n}
  \bnfalt \mathit{bs}
  \bnfalt [\,\overline{\aval}\,]
  \bnfalt (\overline{\aval})
  \bnfalt S\{\overline{f = \aval}\}
  \bnfalt \kw{ref}\,(\hat{\aref} : T)
  \bnfalt \kw{unit}
\\[6pt]
\multicolumn{4}{@{}l@{}}{\textbf{Frames}}\\[2pt]
\multicolumn{4}{@{}l@{}}{%
  \gcat{local store}\quad\locals \bnfdef \overline{x \mapsto \aval}
  \qquad
  \gcat{immutables}\quad\imms \bnfdef \overline{x \mapsto \aval}
  \qquad
  \gcat{frame}\quad\frm \bnfdef \langle \mathit{ctr},\, \locals,\, \imms \rangle}
\\[6pt]
\multicolumn{4}{@{}l@{}}{\textbf{Configuration}}\\[2pt]
\gcat{configuration} & \cfg &\bnfdef&
  \langle \backend,\, \backend_{\mathsf{t}},\, \mathsf{abi},\, \mathsf{initcode},\, \mathsf{deploy} \rangle
\\
\multicolumn{4}{@{}l@{}}{%
  \gcat{ABI operations}\quad\mathsf{abi} \bnfdef \langle \mathsf{encode},\, \mathsf{decode} \rangle
  \qquad
  \gcat{storage interface}\quad\backend \bnfdef
  \langle \mathsf{read},\, \mathsf{write},\, \mathsf{clear},\, \mathsf{length},\,
  \mathsf{push},\, \mathsf{pop},\, \layout \rangle}
\\[6pt]
\multicolumn{4}{@{}l@{}}{\textbf{Results}}\\[2pt]
\gcat{expression result} & r &\bnfdef&
  \rok{\aval} \bnfalt \rrev \bnfalt \rerr
\\
\gcat{outcome} & \outc &\bnfdef&
  \oOk{\frm}{\evmst} \bnfalt \oRet{\frm}{\evmst}{\avals}
  \bnfalt \oBrk{\frm}{\evmst} \bnfalt \oCont{\frm}{\evmst}
  \bnfalt \oRev \bnfalt \oStatic \bnfalt \oNoMatch \bnfalt \oNoDecode
\end{array}$}
\par\medskip\noindent\makebox[\linewidth][l]{$\begin{array}{@{}l@{\hspace{3em}}l@{}}
\multicolumn{2}{@{}l@{}}{\textbf{Semantic functions and judgments}\text{ (all indexed by \cfg)}}\\[2pt]
\evalE{\frm}{\evmst}{\anexp} = r
  & \text{expression evaluation} \\
\mathsf{resolve}(\frm, \evmst, \aref) = \rok{(\hat{\aref} : T)} \bnfalt \rrev \bnfalt \rerr
  & \text{path resolution} \\
\readSt{\cfg}{\frm, \evmst}{\aref} = r \qquad
\writeSt{\cfg}{\frm, \evmst}{\aref}{\aval} = \rok{(\frm', \evmst')} \bnfalt \rrev \bnfalt \rerr
  & \text{storage access} \\
\backend.\mathsf{read}(\hat{\aref}, T, \evmst) = r \qquad
\backend.\mathsf{write}(\hat{\aref}, T, \aval, \evmst) = \rok{\evmst'} \bnfalt \rrev \bnfalt \rerr
  & \text{storage interface operations} \\
\execS{\frm}{\evmst}{\astmt}{\outc}
  & \text{statement execution} \\
\execB{\frm}{\evmst}{\stmts}{\outc}
  & \text{block execution} \\
\mathsf{call}_{\cfg}(\evmst, a, \mathit{val}, \mathit{cd}, p) \rightsquigarrow (z, \evmst', o)
  & \text{message call, via } \CallEVM \\
\mathsf{delegatecall}_{\cfg}(\evmst, a, \mathit{cd}) \rightsquigarrow (z, \evmst', o)
  & \text{delegate call, via } \CallEVM \\
\mathsf{new}_{\cfg}(\evmst, C, \mathit{val}, \avals, \mathit{salt}) \rightsquigarrow (a, \evmst', z)
  & \text{contract creation, via } \CreateEVM \\
\mathsf{selector}(\mathit{ctr}, \mathit{cd}) = t \bnfalt \bot \qquad
\mathsf{decode}^{\mathsf{cd}}_{\cfg}(t, \mathit{cd}) = \locals \bnfalt \bot \qquad
\mathsf{fbArgs}(t, \mathit{cd}) = \locals
  & \text{dispatch} \\
\mathsf{init}(\sigma, \sigma_0, g, A, I) = \evmst
  & \text{initial \EVM state of a run} \\
\steleexec{\cfg}{\mathit{ctr}}{\sigma, g, A, I}{\outc}
  & \text{message-level execution} \\
\cfg;\, \mathit{ctr} \vdash (\avals, \sigma, g, A, I) \Downarrow_{\mathsf{ctor}} \outc
  & \text{constructor execution}
\end{array}$}
\caption{Runtime values, frames, results, configuration and judgments of the relational semantics of \lang.}
\label{fig:stele-judgments}
\end{figure}

%% file: figs/stele-extcall.tex
\begin{figure}[t]
\small
\begin{mathpar}
\inferrule*[lab=\rulename{ExtCall}]
  { \evalE{\frm}{\evmst}{\anexp_{r}} = \rok{a} \\
    \evalE{\frm}{\evmst}{\anexp_{v}} = \rok{\mathit{val}} \\
    \evalE{\frm}{\evmst}{\overline{\anexp}} = \rok{\avals} \\\\
    \mathsf{encode}_{\cfg}(f, \avals) = \mathit{cd} \\
    \mathsf{call}_{\cfg}(\evmst, a, \mathit{val}, \mathit{cd}, p)
      \rightsquigarrow (\kw{true},\, \evmst',\, o) \\
    \mathsf{decode}_{\cfg}(f, o) = \avals' }
  { \execS{\frm}{\evmst}
      {\alpha\;y = \anexp_{r}.f\{\kw{value}{:}\anexp_{v}\}(\overline{\anexp})}
      {\oOk{\frm[y \mapsto \avals'\!\downarrow]}{\evmst'}} }

\inferrule*[lab=\rulename{ExtCallFail}]
  { \evalE{\frm}{\evmst}{\anexp_{r}} = \rok{a} \\
    \evalE{\frm}{\evmst}{\anexp_{v}} = \rok{\mathit{val}} \\
    \evalE{\frm}{\evmst}{\overline{\anexp}} = \rok{\avals} \\\\
    \mathsf{encode}_{\cfg}(f, \avals) = \mathit{cd} \\
    \mathsf{call}_{\cfg}(\evmst, a, \mathit{val}, \mathit{cd}, p)
      \rightsquigarrow (\kw{false},\, \evmst',\, o) }
  { \execS{\frm}{\evmst}
      {\alpha\;y = \anexp_{r}.f\{\kw{value}{:}\anexp_{v}\}(\overline{\anexp})}
      {\oRev} }

\end{mathpar}
\caption{Typed external calls of \lang.}
\label{fig:stele-extcall}
\end{figure}

%% file: Refinement.tex
The top-level refinement statement for a contract in \EquiVM reads
$\mathsf{contractRefinement}_{\cfg}\; (\mathit{initcode},\; \mathit{ctr})$.
It connects the creation bytecode $\mathit{initcode}$ and the \lang contract
$\mathit{ctr}$ under a configuration \cfg 
(storage interfaces, ABI operations, and deployment rule, shown in
\cref{sec:stele-semantics}). 
Roughly, it states that the deployment bytecode refines the \lang constructor,
and that the bytecode that is generated refines the \lang API of the contract.
The refinement relations are shown in \cref{fig:equiv-rules}.
\input{figs/refinement}\label{fig:refinement}

Before looking at the relations, let us first consider what the relations
relate.
For the deployment refinement, they should relate the initcode, \ie, the
bytecode that runs during deployment, with the \lang constructor.
For the runtime code, the answer is less obvious.
The runtime bytecode is not, in general, constant; it is computed by the initcode
and may depend on its parameters or the EVM state.
Consequently, the obligations for the runtime code are tied directly to
the initcode execution, making isolated proof construction complicated.
For all major source languages, however, initialization of immutable variables
is the sole direct mechanism for modifying the deployed code.
We leverage this pattern to streamline the refinement relation, by
expressing the runtime bytecode as a function $\mathit{rc}$
from immutables stores assigning values to immutables ($\iota$,
\cref{sec:stele-semantics}) to concrete bytecode.
The runtime code is therefore a template with the immutables' values spliced
in.
The deployment refinement shows that the constructor returns
$\mathit{rc}(\imms)$ for a store $\imms$ containing the immutable values it sets.
The runtime refinement is then stated for $\mathit{rc}(\imms)$, with $\imms$
ranging over every store a deployment can set.

\textbf{Contract refinement.}
The contract-level statement (\cref{fig:equiv-rules}) first requires the
existence of a runtime template $\mathit{rc}$, which is a function from an
immutable store $\iota$ to the bytecode.
It then quantifies over every account map ($\sigma$), checkpoint ($\sigma_{0}$),
gas allowance ($g$), substate ($A$), execution environment ($I$) and argument
list ($\avals$), such that $I$ has installed the initcode
$\mathsf{deploy}_{\cfg}(\mathit{initcode}, \avals)$, 
calldata are empty ($I.\mathsf{calldata} = \varepsilon$, as $\CreateEVM$ runs
the initcode without calldata and the constructor's arguments are part of the
code)
and the write permission is set ($I_{w} = \kw{true}$, which always holds for a
creation run).
Its conclusion requires that deployed initcode refines the \lang
constructor and sets the immutables to values recorded in $\imms^{?}$ (which may
be $\bot$ if the deployment did not succeed),
and that, whenever a store $\imms$ was successfully set, the runtime code
$\mathit{rc}(\imms)$ refines the contract whose immutables take the values
recorded in $\imms$.

The constructor judgment $\mathsf{ctorRefines}_{\cfg}(\mathit{ctr}, \avals,
\mathit{rc};\, \sigma, \sigma_{0}, g, A, I) \Downarrow \imms^{?}$ has two cases.
\rulename{C-Exec} runs the \EVM's code-execution function $\Xexec$ on the
deployed initcode and the \lang constructor on $\avals$, the initcode
parameters, and requires their results to agree under $\approx_{\mathit{rc}}$.
It reports the immutable store when the constructor returns, and $\bot$
otherwise.
\rulename{C-OutOfGas} accepts a deployment that exhausts its gas without
further conditions, and reports $\bot$.

The result agreement relation for constructors, $r \approx_{\mathit{rc}} \outc$,
has three cases. 
In the first case, which indicates successful runs, the relation requires that the
account maps are equal ($\sigma' = \evmst'\!.\sigma$), that the constructor
returned no values ($\avals$ is $\bot$ or the empty list), and that the returned
runtime bytes are $\mathit{rc}(\frm.\imms)$.
In the revert case, the relation requires that both the \EVM and the \lang results
are reverts.
In the third case, an \EVM \code{INVALID} result is related to a \lang
revert.
This is required because \code{solc} before version 0.8 compiled failed
assertions and arithmetic panics, such as division by zero, to the
\code{INVALID} instruction rather than to a revert, while later versions and
\lang express these failures as reverts. Without this case, no contract
compiled by such a version could be refined.

\textbf{Runtime refinement.}
The runtime judgment $\mathsf{runtimeRefinement}_{\cfg}(\mathit{code},
\mathit{ctr}, \imms)$ (\cref{fig:equiv-rules}) quantifies over every account map
($\sigma$), checkpoint ($\sigma_{0}$), gas allowance ($g$), substate ($A$) and
execution environment ($I$), such that $I$ has the runtime bytecode installed
($I.\mathsf{code} = \mathit{code}$) and its calldata is shorter than $2^{256}$
bytes ($\lvert I.\mathsf{calldata}\rvert < 2^{256}$).
The last restriction is required so that in the \EVM model \code{CALLDATASIZE},
as a 256-bit word, reports the length of calldata without wrapping.
In practice no \EVM execution can have calldata of that length, since every
calldata byte costs at least four gas, so this length would cost more than
$10^{77}$ gas, while a whole block's gas limit is below $10^{9}$ gas.
The relation is parametrized by well-formedness preconditions $\mathit{wf}$
and $\gamma$, such that additional restrictions can be imposed on $\sigma$, $I$,
and $g$.
Few contracts require such additional restrictions in their verification, and we
explain this in detail in \cref{sec:casestudies}.
Its conclusion requires that the bytecode refines the contract in this context,
$\mathsf{refines}_{\cfg}(\mathit{ctr}, \imms;\, \sigma, \sigma_{0}, g, A, I)$.

The relation $\mathsf{refines}_{\cfg}$ has two cases.
\rulename{R-Exec} requires that the \EVM's code-execution function $\Xexec$ runs
on the bytecode, \lang's message-level judgment (\cref{fig:stele-judgments}) on
the contract, on the same inputs, and their results to agree under $\approx$.
\rulename{R-OutOfGas} accepts a run that exhausts its gas, without further
conditions.

The relation $r \approx \outc$ has six cases.
The first case relates successful results.
The \EVM result is bytes $o$ with final account map $\sigma'$, and the \lang
result is $\oRet{\frm}{\evmst'}{\avals}$. The relation requires that the account
maps are equal ($\sigma' = \evmst'\!.\sigma$) and that the returned bytes agree
with the returned values ($o \sim_{\overline{\tau}} \avals$). 
The next three cases relate reverting \EVM results with their \lang counterparts.
The \EVM run reverts and the \lang side is either a revert result, or
a result indicating no matching entry point or calldata decoding failure.
We deliberately do not match the revert payload in the last two cases, since it
is not consistent across compilers and versions.
In the fifth case, an \code{INVALID} return agrees with \lang revert, as before.
In the sixth case, an \EVM static-mode violation, raised when the bytecode was
entered through \code{STATICCALL} and attempts a state change, agrees with
a \lang result indicating the same error.

The refinement relation leaves the out-of-gas case unspecified.
An execution that exhausts its gas is accepted whatever the specification does.
\lang does not model gas, and such an execution has no effect on the state.
Because the judgment holds for every gas allowance, a call escapes it only if it
runs out of gas under every allowance.
Revert payloads and the final gas and substate, which records logs, refunds and
access lists, are not compared, since \lang models none of them.

\textbf{What the refinement relation means.}
We prove the following corollary in \Lean (behavioral inclusion). If
$\mathsf{runtimeRefinement}_{\cfg}(\mathit{code}, \mathit{ctr}, \imms)$
holds, then for every $\sigma, \sigma_{0}, g, A, I$ with
$I.\mathsf{code} = \mathit{code}$ and
$\lvert I.\mathsf{calldata}\rvert < 2^{256}$, the result
$r = \Xexec(\sigma, \sigma_{0}, g, A, I)$ satisfies one of the following:
\begin{enumerate*}%
\item the bytecode runs out of gas, or
\item the bytecode run is matched by a \lang execution with the same result.
\end{enumerate*}
Every behavior of terminating bytecode is a behavior of the specification. 
Because \lang treats gas nondeterministically, a \lang spec may admit more
behaviors than the bytecode.
A second corollary shows that refined bytecode never halts with an exception
other than out of gas, \code{INVALID} (matched by \lang revert), or a
static-mode violation (matched by \lang static-mode violation).

To demonstrate the refinement's adequacy we prove a property about the \lang
spec of a contract, and we transfer it to the bytecode using the refinement.
In particular, for the ERC20 contract of \cref{sec:overview} we prove the
supply invariant
$\mathsf{Inv}(\sigma, a)\eqdef\mathit{totalSupply}_{\sigma[a]} = \textstyle\sum_{b} \mathit{balanceOf}_{\sigma[a]}(b)$.
We write $\mathit{ctr}$ for its \lang spec, \cfg for its configuration, and
$\mathit{initcode}$ and $\mathit{code}$ for its initcode and runtime bytecode.
Assuming that the layout maps distinct storage references to distinct
slots, we prove on the spec that the constructor establishes the invariant
and that every call preserves it:
\begin{enumerate}[leftmargin=*,nosep]
\item $\begin{aligned}[t]
  &\forall\, \sigma, \sigma_{0}, g, A, I, n, \frm, \evmst', \avals.\;
    0 \le n < 2^{256} \,\wedge\,
    (\forall b.\; \mathit{balanceOf}_{\sigma[I_{a}]}(b) = 0)\\
  &\qquad\wedge\;
    \cfg;\, \mathit{ctr} \vdash ([n], \sigma, \sigma_{0}, g, A, I)
      \Downarrow_{\mathsf{ctor}} \oRet{\frm}{\evmst'}{\avals}
    \;\Longrightarrow\; \mathsf{Inv}(\evmst'\!.\sigma, I_{a})
  \end{aligned}$
\item $\forall\, \sigma, \sigma_{0}, g, A, I, \frm, \evmst', \avals.\;
    \mathsf{Inv}(\sigma, I_{a}) \,\wedge\,
    \cfg;\, \mathit{ctr} \vdash_{\varnothing} (\sigma, \sigma_{0}, g, A, I)
      \Downarrow_{\mathsf{msg}} \oRet{\frm}{\evmst'}{\avals}
    \;\Longrightarrow\; \mathsf{Inv}(\evmst'\!.\sigma, I_{a})$
\end{enumerate}
Then, under the same assumption, using the refinement we obtain:
\begin{enumerate}[leftmargin=*,nosep]
\item $\begin{aligned}[t]
  &\forall\, \sigma, \sigma_{0}, g, A, I, \avals, \sigma', g', A', o.\;
    I.\mathsf{code} = \mathsf{deploy}_{\cfg}(\mathit{initcode}, \avals) \,\wedge\,
    I.\mathsf{calldata} = \varepsilon \,\wedge\, I_{w} = \kw{true}\\
  &\qquad\wedge\; (\forall b.\; \mathit{balanceOf}_{\sigma[I_{a}]}(b) = 0)\\
  &\qquad\wedge\;
    \Xexec(\sigma, \sigma_{0}, g, A, I) = \mathsf{success}\,(\sigma', g', A')\;o
    \;\Longrightarrow\; \mathsf{Inv}(\sigma', I_{a}) \,\wedge\, o = \mathit{code}
  \end{aligned}$
\item $\begin{aligned}[t]
  &\forall\, \sigma, \sigma_{0}, g, A, I, \sigma', g', A', o.\;
    I.\mathsf{code} = \mathit{code} \,\wedge\,
    \lvert I.\mathsf{calldata}\rvert < 2^{256} \,\wedge\,
    \mathsf{Inv}(\sigma, I_{a})\\
  &\qquad\wedge\;
    \Xexec(\sigma, \sigma_{0}, g, A, I) = \mathsf{success}\,(\sigma', g', A')\;o
    \;\Longrightarrow\; \mathsf{Inv}(\sigma', I_{a})
  \end{aligned}$
\end{enumerate}
The last two statements prove this property directly about the bytecode
without referring to the \lang spec or the refinement relation.

%% file: figs/refinement.tex
\begin{figure}[tp]
\footnotesize

\noindent\makebox[\linewidth][l]{$\begin{array}{@{}l@{\;\;}c@{\;\;}l@{}}
\judgbox{\mathsf{contractRefinement}^{\mathit{wf},\gamma}_{\cfg}(\mathit{initcode}, \mathit{ctr})}
  & \eqdef
  & \begin{aligned}[t]
    &\exists\, \mathit{rc}.\;
      \forall\, \sigma,\, \sigma_{0},\, g,\, A,\, I,\, \avals.\\[1pt]
    &\quad I.\mathsf{code} = \mathsf{deploy}_{\cfg}(\mathit{initcode}, \avals) \,\wedge\,
      I.\mathsf{calldata} = \varepsilon \,\wedge\,
      I_{w} = \kw{true}\\[1pt]
    &\quad\Longrightarrow\;
      \exists\, \imms^{?}.\;
      \mathsf{ctorRefines}_{\cfg}(\mathit{ctr}, \avals, \mathit{rc};\,
        \sigma, \sigma_{0}, g, A, I) \Downarrow \imms^{?}\\[1pt]
    &\quad\;\wedge\;
      \forall\, \imms.\; \imms^{?} = \imms \Longrightarrow
      \mathsf{runtimeRefinement}^{\mathit{wf},\gamma}_{\cfg}(\mathit{rc}(\imms), \mathit{ctr},
        \imms\!\restriction_{\mathit{ctr}})
    \end{aligned}
\end{array}$}

\par\medskip
\noindent\makebox[\linewidth][l]{$\begin{array}{@{}l@{\;\;}c@{\;\;}l@{}}
\judgbox{\mathsf{runtimeRefinement}^{\mathit{wf},\gamma}_{\cfg}(\mathit{code}, \mathit{ctr}, \imms)}
  & \eqdef
  & \begin{aligned}[t]
    &\forall\, \sigma,\, \sigma_{0},\, g,\, A,\, I.\;
      I.\mathsf{code} = \mathit{code} \,\wedge\,
      \lvert I.\mathsf{calldata}\rvert < 2^{256}\\[1pt]
    &\quad\wedge\; \mathit{wf}(\sigma, I) \,\wedge\, \gamma(g)
      \;\Longrightarrow\;
      \mathsf{refines}_{\cfg}(\mathit{ctr}, \imms;\,
        \sigma, \sigma_{0}, g, A, I)
    \end{aligned}
\end{array}$}
\par\medskip

\noindent\judgbox{\mathsf{ctorRefines}_{\cfg}(\mathit{ctr}, \avals, \mathit{rc};\,
  \sigma, \sigma_{0}, g, A, I) \Downarrow \imms^{?}}\hfill\mbox{}
\begin{mathpar}
\inferrule*[lab=\rulename{C-Exec}]
  { \Xexec(\sigma, \sigma_{0}, g, A, I) = r \\
    \cfg;\, \mathit{ctr} \vdash (\avals, \sigma, \sigma_{0}, g, A, I)
      \Downarrow_{\mathsf{ctor}} \outc \\
    r \approx_{\mathit{rc}} \outc }
  { \mathsf{ctorRefines}_{\cfg}(\mathit{ctr}, \avals, \mathit{rc};\,
      \sigma, \sigma_{0}, g, A, I) \Downarrow \mathsf{imms}(\outc) }

\inferrule*[lab=\rulename{C-OutOfGas}]
  { \Xexec(\sigma, \sigma_{0}, g, A, I)
      = \mathsf{exception}\;\mathsf{OutOfGas} }
  { \mathsf{ctorRefines}_{\cfg}(\mathit{ctr}, \avals, \mathit{rc};\,
      \sigma, \sigma_{0}, g, A, I) \Downarrow \bot }
\end{mathpar}

\noindent\judgbox{r \approx_{\mathit{rc}} \outc}\hfill\mbox{}
\begin{mathpar}
\inferrule
  { \sigma' = \evmst'\!.\sigma \\
    \avals \in \{\bot, [\,]\} \\
    o = \mathit{rc}(\frm.\imms) }
  { \mathsf{success}\,(\sigma', g', A')\;o
      \;\approx_{\mathit{rc}}\; \oRet{\frm}{\evmst'}{\avals} }

\inferrule
  { \, }
  { \mathsf{revert}\; g'\,o \;\approx_{\mathit{rc}}\; \oRev }

\inferrule
  { \, }
  { \mathsf{exception}\;\mathsf{Invalid} \;\approx_{\mathit{rc}}\; \oRev }
\end{mathpar}

\noindent\judgbox{\mathsf{refines}_{\cfg}(\mathit{ctr}, \imms;\,
  \sigma, \sigma_{0}, g, A, I)}\hfill\mbox{}
\begin{mathpar}
\inferrule*[lab=\rulename{R-Exec}]
  { \Xexec(\sigma, \sigma_{0}, g, A, I) = r \\
    \cfg;\, \mathit{ctr} \vdash_{\imms} (\sigma, \sigma_{0}, g, A, I)
      \Downarrow_{\mathsf{msg}} \outc \\
    r \approx \outc }
  { \mathsf{refines}_{\cfg}(\mathit{ctr}, \imms;\,
      \sigma, \sigma_{0}, g, A, I) }

\inferrule*[lab=\rulename{R-OutOfGas}]
  { \Xexec(\sigma, \sigma_{0}, g, A, I)
      = \mathsf{exception}\;\mathsf{OutOfGas} }
  { \mathsf{refines}_{\cfg}(\mathit{ctr}, \imms;\,
      \sigma, \sigma_{0}, g, A, I) }
\end{mathpar}

\noindent\judgbox{r \approx \outc}\hfill\mbox{}
\begin{mathpar}
\inferrule
  { \sigma' = \evmst'\!.\sigma \\
    o \sim_{\overline{\tau}} \avals }
  { \mathsf{success}\,(\sigma', g', A')\;o
      \;\approx\; \oRet{\frm}{\evmst'}{\avals} }

\inferrule
  { \, }
  { \mathsf{revert}\; g'\,o \;\approx\; \oRev }

\inferrule
  { \, }
  { \mathsf{revert}\; g'\,o \;\approx\; \oNoMatch }

\inferrule
  { \, }
  { \mathsf{revert}\; g'\,o \;\approx\; \oNoDecode }

\inferrule
  { \, }
  { \mathsf{exception}\;\mathsf{Invalid} \;\approx\; \oRev }

\inferrule
  { \, }
  { \mathsf{exception}\;\mathsf{StaticViolation} \;\approx\; \oStatic }
\end{mathpar}

\noindent\judgbox{o \sim_{\overline{\tau}} \avals}\hfill\mbox{}
\begin{mathpar}
\inferrule
  { \mathsf{enc}_{\overline{\tau}}(\avals) = o }
  { o \sim_{\overline{\tau}} \avals }

\inferrule
  { \mathsf{enc}_{\overline{\tau}}(\mathsf{zero}(\overline{\tau})) = o }
  { o \sim_{\overline{\tau}} \bot }
\end{mathpar}
\caption{The refinement relations.}
\label{fig:equiv-rules}
\end{figure}

%% file: Reasoning.tex
We develop a generic proof library to help refinement proofs stay tractable.
The library is 49k lines and comprises 2.6k theorems.
The goal is to factor out the common parts of reasoning and to provide an
abstraction that allows one to reason about the bytecode execution without
having to manually unfold the execution function all the time.

\input{figs/reach-def}
\input{figs/reach-rules}

\textbf{The reach relation.}
The library's central abstraction is a reach relation of \cref{fig:reach-def}
(\code{RD} in the development). 
A \emph{cursor} exposes the
six components that straight-line code manipulates: the program counter,
the stack, the memory with its active-word count, the return data, and the
account map, where storage lives.
The relation $\reachJ{s_{0}}{k}{C}{c}$ says: either the whole run out of
$s_{0}$ exhausts its gas, or after $k$ steps and $C$ gas the run sits at an
intermediate state matching cursor $c$, with the remainder of the run
equal as a term to the run from that state. 
(In the figure, $\XEVM_{f}(s)$ abbreviates the instruction iterator of
\cref{sec:evm} with fuel $f$ and the code's jump-destination set, $s
\blacktriangleright c$ pins the machine state and account map of $s$ to the
cursor $c$, and the checkpoint $\sigma_{0}$, which no opcode changes, is carried
unchanged.)
This makes the intermediate state existentially hidden.
Everything that a later step of the proof can depend on is in the cursor, so a
proof never needs to touch a concrete machine state.
In addition, out-of-gas is a disjunct inside the relation. Every combinator maps
the disjunct to itself, so the possibility of gas exhaustion threads through an
entire proof without a single case split and is discharged exactly once, at the
top, into the \rulename{R-OutOfGas} case of the refinement relation.
Crucially, the reach relation composes by transitivity. 
This makes lemmas about single opcodes chain into whole execution traces. 
In effect, the proof performs forward symbolic execution of the bytecode, one
opcode at a time.
The library has a reach lemma for each opcode, totalling 201 combinators. 
We show some of them in \cref{fig:reach-rules}. 

The library was built after our first two proofs, which stepped through the
bytecode by hand, with one out-of-gas case split and five bookkeeping facts per
opcode. Rewriting them with the combinators cut the proofs from 2868 to 1203
lines and from 2390 to 493, as each opcode now costs one combinator application
instead of about a dozen lines (Appendix~B of the supplementary material shows
the two styles side by side).

\textbf{The rest of the library.}
The library also provides contract-agnostic lemmas on $256$-bit word
arithmetic, byte-level memory and the compiler's storage layout,
storage and account maps including packed and transient slots, and ABI
encoding and decoding. 
Common \code{solc} patterns, such as the dispatcher, the non-payable guard,
getters, checked arithmetic and reentrancy locks, are proved once, in the
library. 
On the \lang side, lemmas compose statement executions, compute dispatch, and
couple external calls, deployment and loops with their bytecode counterparts.

%% file: figs/reach-def.tex
\begin{figure}[t]
\small
\[
\begin{array}{@{}l@{\quad}r@{\;\;}c@{\;\;}l@{}}
\gcat{cursor}
  & c
  & \Coloneqq
  & \cursor{\mathit{pc}}{\overline{w}}{m}{\mathit{aw}}{r}{\sigma}
\\[6pt]
\gcat{reach invariant}
  & \reachJ{s_{0}}{k}{C}{c}
  & \eqdef
  & \begin{aligned}[t]
    & \XEVM_{g+1}(s_{0}) = \oog
      \;\vee\;
      \exists s.\;
      \XEVM_{g+1}(s_{0}) = \XEVM_{g+1-k}(s)
      \;\wedge\; s \blacktriangleright c \\
    & \quad\wedge\; \mu_{g}(s) = g - C
      \;\wedge\; k \le C \le g
      \;\wedge\; \mathsf{env}(s) = I
      \;\wedge\; s.\sigma_{0} = s_{0}.\sigma_{0}
    \end{aligned}
\\[6pt]
\gcat{terminal forms}
  & \reachRet{s_{0}}{\sigma}{o}
  & \eqdef
  & \XEVM_{g+1}(s_{0}) = \oog \;\vee\;
    \exists s.\; \XEVM_{g+1}(s_{0}) = \mathsf{success}\,s\;o
      \wedge s.\sigma = \sigma
\\[3pt]
  & \reachRev{s_{0}}
  & \eqdef
  & \XEVM_{g+1}(s_{0}) = \oog \;\vee\;
    \exists g'\, o.\; \XEVM_{g+1}(s_{0}) = \mathsf{revert}\;g'\,o
\\[3pt]
  & s_{0} \Downarrow_{\mathsf{static}}
  & \eqdef
  & \XEVM_{g+1}(s_{0}) = \oog \;\vee\;
    \XEVM_{g+1}(s_{0}) = \mathsf{exception}\;\mathsf{StaticViolation}
\\[3pt]
  & s_{0} \Downarrow_{\mathsf{inv}}
  & \eqdef
  & \XEVM_{g+1}(s_{0}) = \oog \;\vee\;
    \XEVM_{g+1}(s_{0}) = \mathsf{exception}\;\mathsf{Invalid}
\end{array}
\]
\caption{Cursors, the reach relation and its terminal forms, over an ambient environment
$I$, gas budget $g$, and start state $s_{0}$ ($\oog$ abbreviates
$\mathsf{exception}\;\mathsf{OutOfGas}$).}
\label{fig:reach-def}
\end{figure}

%% file: figs/reach-rules.tex
\begin{figure}[t]
\small
\begin{mathpar}
\inferrule*[lab=\rulename{JumpiT}]
  { \reachJ{s_{0}}{k}{C}{\cursor{\mathit{pc}}{a \cdot b \cdot \overline{w}}{m}{\mathit{aw}}{r}{\sigma}} \\
    \mathsf{decode}(I_{b}, \mathit{pc}) = \mathtt{JUMPI} \\
    b \ne 0 \\ a \in D_{J}(I_{b}) \\ |\overline{w}| \le 1024 }
  { \reachJ{s_{0}}{k+1}{C+10}
      {\cursor{a}{\overline{w}}{m}{\mathit{aw}}{r}{\sigma}} }

\inferrule*[lab=\rulename{Sstore}]
  { \reachJ{s_{0}}{k}{C}{\cursor{\mathit{pc}}{\mathit{slot} \cdot v \cdot \overline{w}}{m}{\mathit{aw}}{r}{\sigma}} \\
    \mathsf{decode}(I_{b}, \mathit{pc}) = \mathtt{SSTORE} \\ I_{w} = \kw{true} \\
    |\overline{w}| \le 1024 }
  { \exists k'\, C'.\;
    \reachJ{s_{0}}{k'}{C'}
      {\cursor{\mathit{pc}+1}{\overline{w}}{m}{\mathit{aw}}{r}
        {\sigma[I_{a}][\mathit{slot}] \mapsto v}} }

\inferrule*[lab=\rulename{Call}]
  { \reachJ{s_{0}}{k}{C}
      {\cursor{\mathit{pc}}{g_{a} \cdot a \cdot 0 \cdot \mathit{io} \cdot \mathit{is} \cdot \mathit{oo} \cdot \mathit{os} \cdot \overline{w}}{m}{\mathit{aw}}{r}{\sigma}} \\
    \mathsf{decode}(I_{b}, \mathit{pc}) = \mathtt{CALL} \\ I_{e} < 1024 \\
    |\overline{w}| + 1 \le 1024 }
  { \exists\, \sigma'\, z\, o\, A_{\mathit{in}}\, g_{c}\, k'\, C'.\;\;
    (\sigma', z, o) \in \CallEVM(\sigma, s_{0}.\sigma_{0}, A_{\mathit{in}}, I_{a} \!\to\! a,
        m[\mathit{io}\,..\,\mathit{io}{+}\mathit{is}], g_{c}) \\
    \wedge\;
    \reachJ{s_{0}}{k'}{C'}
      {\cursor{\mathit{pc}+1}{z \cdot \overline{w}}{m[\mathit{oo} \mapsto o]}{\mathit{aw}'}{o}{\sigma'}}
    \;\wedge\; |o| < 2^{256} }

\inferrule*[lab=\rulename{Loop}]
  { \forall a.\; \mathit{Inv}\;0\;a \to
      \reachJ{s_{0}}{k}{C}{c_{\mathsf{hd}}(a)} \to
      \exists k'\, C'.\; \reachJ{s_{0}}{k'}{C'}{c_{\mathsf{exit}}} \\\\
    \forall v\, a.\; \mathit{Inv}\;(v{+}1)\;a \to
      \reachJ{s_{0}}{k}{C}{c_{\mathsf{hd}}(a)} \to
      \exists a'\, k'\, C'.\; \mathit{Inv}\;v\;a' \wedge
        \reachJ{s_{0}}{k'}{C'}{c_{\mathsf{hd}}(a')} }
  { \forall v\, a.\; \mathit{Inv}\;v\;a \to
      \reachJ{s_{0}}{k}{C}{c_{\mathsf{hd}}(a)} \to
      \exists k'\, C'.\; \reachJ{s_{0}}{k'}{C'}{c_{\mathsf{exit}}} }
\end{mathpar}
\caption{Selected combinators of the reach algebra, stated over the ambient $I$, $g$, $s_{0}$ of
\cref{fig:reach-def}.}
\label{fig:reach-rules}
\end{figure}

%% file: Evaluation.tex
Our evaluation asks three questions.
\begin{enumerate*}
\item Can agents produce refinement proofs, and at what cost in tokens and time?
\item How much human input do they need?
\item What does it cost to maintain a proof when the source or the compiler
settings change?
\end{enumerate*}
We target real production contracts and we modify neither the source code nor
the bytecode that the compilation pipelines produce.

We have used the framework to prove 35 real contracts, totalling 11,146 lines of
compiled source code. 
We report detailed telemetry on 9 benchmarks (8,229 lines of compiled source
code) which are the most challenging ones and were proved with the current
version of the framework under systematic measurement.
Another 26 (\cref{tab:eval-others}) were proved with earlier versions of the
framework, and were then ported incrementally to newer versions.
They are much smaller and less challenging than the first group.
Some of them were exploratory examples from the design phase of the framework.
Their proofs drove the development of the library, and the agents' proof
work interleaved with library work.

\input{figs/eval-contracts}\label{tab:contracts}
\input{figs/eval-proofs}\label{tab:proofs}
\input{figs/eval-spec-sessions}\label{tab:spec}
\input{figs/eval-repair}\label{tab:repair}

\textbf{Benchmark set.} 
\Cref{tab:eval-contracts} lists the nine contracts.
Safe is the implementation contract of the Safe smart-account wallet, EAS is the
registry of the Ethereum Attestation Service and Attester an example wrapper
around it from the EAS examples repository, Morpho Blue and Comet are the core
contracts of the Morpho and Compound~III lending protocols, UniswapV3Pool is a
Uniswap~v3 exchange pool and PoolManager the Uniswap~v4 contract that holds
every pool.
The Attester is not a production contract but we include it because it contains
two nested loops.
For Morpho Blue, Comet, UniswapV3Pool, PoolManager and LlamaLend Vault, the runtime
bytecode we prove is byte-identical to a mainnet deployment, up to immutables
and, for Comet, the metadata hash.
Attester has no deployment and is compiled as the upstream repositories
configure it.
EAS and MetaMorpho were deployed from \code{solc} 0.8.18 and 0.8.26, which carry
a known compiler bug that we hit during specification
(SOL-2025-1~\cite{SC:Solidity25:sol20251}, explained below). 
We recompiled with 0.8.32, the first fixed version, and 0.8.37, with the
upstream settings otherwise unchanged.
Safe is compiled with 0.8.35, where upstream defaults to 0.7.6.
The configurations span \code{solc} 0.7.6 to 0.8.37, both compilation pipelines (five
of the nine through \texttt{via-IR}), optimizer settings from one run with a
custom Yul step sequence (Comet) to 44\,444\,444 runs (PoolManager).
The compiled code contains 48 loops, six of them nested.
The largest bytecode, PoolManager, is 24\,009 bytes, 567 bytes under the
24\,576-byte limit that the EVM imposes on deployed code (EIP-170).
All prompts for the following are listed in section C of the appendix.
We count prompts that carry human content (goal, free-text steering, replies to
agent questions, and approvals). 
Resumes after a block or a crash are not counted.

In terms of optimizations, the \code{via-IR} pipeline performs SSA-based
common-subexpression elimination, constant folding, dead-code and dead-store
elimination, store-to-load forwarding, loop-invariant code motion, inlining and
function specialization, but no loop unrolling, partial-redundancy elimination
or induction-variable optimization. The legacy pipeline performs only
block-local common-subexpression elimination, constant folding, jump
simplification, block deduplication and small-function inlining, and no loop
transformations.

\textbf{Proof construction.} 
All proof runs used OpenAI's Codex CLI with gpt-6-astra at maximum reasoning
effort, on two ChatGPT Pro accounts, OpenAI's flat-rate subscription at
\$200 per month each. The whole evaluation ran within one billing month.
\Cref{tab:eval-proofs} reports proof telemetry and every line counts a
proof from the beginning to completion.
Morpho Blue, UniswapV3Pool, PoolManager, and LlamaLend Vault were proved from the
initial prompt alone.
Safe's and Attester's additional prompt was the same.
Both had blocked within three hours on a bug in the \EVM model: a memory read of
$2^{64}$ bytes or more returned empty bytes, while the \lang side carried the
full data.
We fixed the model, and the message told the agents to pull the change.
The second prompt of EAS restarted the proof in a new session run after the
first lost connection, eighteen hours of work in. 
The fresh session completed the proof in four hours.
The proof of Comet was the most prompt-heavy proof requiring 6
prompts over almost 26 hours.
Twelve minutes in, it asked permission to change the proof scaffold in order to
restrict two \code{uint8} immutables to values below 256 in intermediate lemmas.
After 21 hours, it reported that the loop of \code{absorb} over its account
array allocates memory without bound, so given more gas than any block can
supply, the compiler's allocation check reverts where \lang succeeds. 
We narrowed the theorem with a gas bound ($g < 8.9 \times 10^{15}$).
After 22 hours, it reported a mismatch between the library's calldata decoder
and \code{solc} 0.8.15, which validates the elements of an \code{address[]} argument
lazily, on access. 
A static call with a malformed element reaches a storage write before the
element is inspected, while the \lang decoder rejected the calldata upfront. 
We asked for an explanation and then authorized a decoder mode for that compiler
version (two prompts).
After 26 hours, it asked to place a size bound on the deployed initcode in
the constructor theorem.
MetaMorpho's five prompts are three goals, one per session, and two approvals.
Seventeen minutes into the first session, the agent asked, as for Comet, to
restrict two \code{uint8} immutables to values below 256 in intermediate lemmas.
Seven hours into the third session, with all 76 runtime theorems proved, it
asked to place a size bound on the initcode in the constructor theorem.
The proof was done in three sessions because of repeated cybersecurity blocks.

Four of the nine theorems carry side conditions and the other five are
unconditional.
Comet and PoolManager both bound the gas a call starts with, below
$8.9 \times 10^{15}$ and $3.2 \times 10^{32}$ respectively.
The bounds exclude executions that no block can fund, in which \code{solc}'s
memory-allocation checks revert before gas runs out, while \lang, which has no
memory, succeeds.
Comet's and MetaMorpho's constructor theorems further cover only deployments
whose initcode, with the encoded constructor arguments, is shorter than
$2^{256}$ bytes.
The condition excludes no deployment that can actually happen.
PoolManager's theorem requires the slot count of \code{extsload(bytes32 s,
uint256 n)} to satisfy $224 + 32\cdot\text{\code{n}} < 2^{256}$.
The function copies \code{n} storage words into memory, which is byte-addressed,
so the bytecode computes the byte length as $32\cdot\text{\code{n}}$ without an
overflow check.
For larger counts this arithmetic wraps, and the function returns a malformed
array that does not follow the ABI encoding and that \lang cannot express.
LlamaLend Vault's theorem assumes that its five address-typed storage variables hold
values below $2^{160}$. Vyper does not mask addresses read from storage, so the
bytecode and the specification agree only when those slots hold canonical
addresses.

\textbf{Specification.}
For seven of these contracts, we also report detailed telemetry for specification
construction in \cref{tab:eval-spec-sessions} (the other two were specified
before we had the specification automation in place and were not systematically
measured).
In a specification session we hand an agent the contract's sources with their
import closure and the compiler settings.
The agent runs the scripts to compile the sources, disassemble the
bytecode, generate a draft \lang specification translated from the compiler's
AST, the proof skeleton and the block summaries.
The agent closes the draft's holes, runs the differential suite against the
bytecode and fixes what disagrees, then audits every block of the bytecode
against the specification.
When a contract has inline assembly, the autogenerated specification has holes
that the agent fills with a high-level account of its behavior. 

The differential tests and the semantic audit caught specification errors before
the proofs started.
Most errors were about order (the auto-translated specification evaluated a
check earlier than the bytecode does).
Another class of errors are checks the compiler inserts and the source does not
contain, such as the memory-allocation guards discussed above, which the
specification must state explicitly.
Three misspecification findings concern ABI decoding, where \code{solc} is more
permissive than the specification's decoder was.
The EAS bytecode accepts calldata in which the pointer to a dynamic field is
negative (\code{solc} checks the pointer as a signed number, so $2^{256}-32$ passes as
$-32$), and the address arithmetic wraps, so the field is read from an earlier
word of the calldata. The library's decoder, and hence the specification,
treated pointers as unsigned and rejected the call.
In the same contract, \code{multiAttest} takes an array of requests, which the
bytecode decodes as it processes them and the specification decoded upfront.
This creates a mismatch in static call context as a valid first request followed
by a malformed second one makes the bytecode halt at the first request's
storage write, before it decodes the second, while the specification rejects
the second request upfront and reverts.
Morpho Blue validates a \code{uint8} argument only after writing the nonce,
while the specification rejected it while decoding.
To handle these, we add a decoder mode, specific to these contracts, that reproduces 
the behavior of the bytecode.
Lastly, the audits of EAS's \code{\_attest} and of MetaMorpho's
\code{setSupplyQueue} hit a known compiler bug~\cite{SC:Solidity25:sol20251}.
When a \code{bytes} value or a dynamic array in storage is replaced, the
bytecode's clearing loop computes its end slot with wrapping arithmetic, so if
the sum wraps past $2^{256}$ the loop exits at once and the old data survives.
\lang clears it properly.
In both cases the agent reproduced the case, which the random tests had not
caught, and we instructed it to move the scaffold to a newer compiler version
that fixes the bug (0.8.32 for EAS, 0.8.37 for MetaMorpho).
EAS's sixteen prompts are two goals (the first session was abandoned after
Codex's cyber-safety filter had cut eight consecutive turns short, and resumed
in a second one with a new goal), two on the compiler bug, seven on the decoder
fixes, and two about the security false alarms.
Beyond the goal prompt, most of the other inputs approve edits outside the
scaffold directory (the differential-test scripts and two fixes to the skeleton
generator) and adding the well-formedness condition to PoolManager.

\textbf{Proof repair.}
To measure the cost of proof repair we use the Safe contract (\cref{tab:eval-repair}). 
For source changes, we step the Safe source back to an earlier commit of its
GitHub history and then forward, commit by commit, repairing the proof at each
step.
For compiler changes, we separately raise the optimizer runs from 200 to
1\,000\,000 and switch to \texttt{via-IR}.
A repair session receives the starting version's proof and the target commit or
compiler setting, may consult no other version, recompiles, updates the
specification until the differential tests pass, and repairs the proof in place.
Each source change was repaired from the goal prompt alone in under two hours,
against eighteen for the original proof.
The thousands of changed proof lines are mostly relocated program counters and
block summary references. New reasoning is confined to the
changed paths.
Raising the optimizer runs trades code size for gas. 
The compiler emits precomputed masks and shifted constants as larger push
literals, so every address moves, and reorders some code.
The repair relocates addresses and summary references and adjusts a few stack
bounds.
Switching to \texttt{via-IR} is largely a new proof.
Function bodies are placed inside their dispatcher arms, shared helpers are
specialized or inlined, and decoding, allocation and return code are generated
from scratch.
The specification changed in six places, all of them compiler-inserted checks
(five memory allocation guards and a calldata check).
Of the fourteen prompts, one is the initial goal, four settle the decoder
difference, seven approve changes in the specification and a new well-formedness
condition, and two are nudges.
The new condition requires the executing account to be present in the account
map.

\input{figs/eval-others}\label{tab:others}

%% file: figs/eval-contracts.tex
\begin{table}[tp]
\caption{Benchmark contracts. Names link to the sources. $^{\mathrm{IR}}$ marks
contracts compiled through the \texttt{via-IR} pipeline. \textbf{Runs} is the
optimizer's runs setting (optimizer on for every contract).
$^{\dagger}$Custom Yul optimizer-step sequence. \textbf{EVM} is the EVM version
(hard fork) the compiler targeted. \textbf{Fn/View} is the ABI surface, as
state-changing/view functions. \textbf{Source} counts non-blank, non-comment
Solidity lines and \textbf{Runtime} is the deployed bytecode in bytes.
\textbf{Deployed} says whether the verified runtime bytecode is byte-identical
to a mainnet deployment, up to the values of immutables ($^{\ddagger}$identical
except for the compiler's 32-byte metadata hash). \textbf{Loops} counts loop
statements, \textbf{for/while} counts their kind, \textbf{Nest} is the deepest
loop nesting, \textbf{Body} is the longest loop in source lines, and
\textbf{Asm} counts inline assembly blocks. Source, Loops and Asm are given as
contract/compiled, for the declarations of the contract itself, and declarations
reachable by its compiled closure (counting only real dependencies). Complete
compiler invocations are listed in Appendix~D of the supplementary material.}
\label{tab:eval-contracts}
\centering
\scriptsize
\setlength{\tabcolsep}{2.5pt}
\begin{tabular}{@{}l!{\color{black!30}\vrule}l!{\color{black!30}\vrule}l!{\color{black!30}\vrule}r!{\color{black!30}\vrule}l!{\color{black!30}\vrule}r!{\color{black!30}\vrule}r!{\color{black!30}\vrule}r!{\color{black!30}\vrule}r!{\color{black!30}\vrule}r!{\color{black!30}\vrule}r!{\color{black!30}\vrule}r!{\color{black!30}\vrule}r@{}}
\hline
\textbf{Contract} & \textbf{solc} & \textbf{Runs}
  & \textbf{EVM} & \textbf{Fn/View} & \textbf{Source} & \textbf{Runtime}
  & \textbf{Deployed} & \textbf{Loops} & \textbf{for/while} & \textbf{Nest} & \textbf{Body} & \textbf{Asm} \\
\hline
\href{https://github.com/safe-global/safe-smart-account/blob/77901a5a1ad835b74ad3b72f73a8412cfe491c57/contracts/Safe.sol}{Safe}
  & 0.8.35 & 200 & shanghai & 15/16 & 251/657 & 11\,874
  & no & 1/5 & 3/2 & 1 & 78 & 6/24 \\
\href{https://github.com/ethereum-attestation-service/eas-contracts/blob/aefff5aea2b61b4d86f5af016bc51b6acf3685fc/contracts/EAS.sol}{EAS}
  & 0.8.32$^{\mathrm{IR}}$ & 1\,000\,000 & paris & 12/10 & 478/715 & 17\,208
  & no & 15/15 & 14/1 & 2 & 61 & 0/0 \\
\href{https://github.com/ethereum-attestation-service/eas-contracts-example/blob/d2864b166a08f9b3f9314f8b302316d67f227462/contracts/Attester.sol}{Attester}
  & 0.8.26 & 1\,000\,000 & paris & 4/0 & 78/112 & 3\,865
  & no & 4/4 & 4/0 & 2 & 22 & 0/0 \\
\href{https://github.com/morpho-org/morpho-blue/blob/55d2d99304fb3fb930c688462ae2ccabb1d533ad/src/Morpho.sol}{Morpho Blue}
  & 0.8.19$^{\mathrm{IR}}$ & 999\,999 & paris & 17/11 & 311/503 & 15\,623
  & yes & 1/1 & 1/0 & 1 & 7 & 1/5 \\
\href{https://github.com/compound-finance/comet/blob/f766f51583c23acc33b2a7824654ef2029a96804/contracts/CometWithExtendedAssetList.sol}{Comet}
  & 0.8.15$^{\mathrm{IR}}$ & 1$^{\dagger}$ & london & 17/51 & 677/891 & 18\,599
  & yes$^{\ddagger}$ & 5/5 & 5/0 & 1 & 19 & 6/6 \\
\href{https://github.com/Uniswap/v3-core/blob/e3589b192d0be27e100cd0daaf6c97204fdb1899/contracts/UniswapV3Pool.sol}{UniswapV3Pool}
  & 0.7.6 & 800 & istanbul & 9/17 & 615/1\,611 & 22\,142
  & yes & 1/4 & 2/2 & 1 & 90 & 0/30 \\
\href{https://github.com/Uniswap/v4-core/blob/46c6834698c48bc4a463a86d8420f4eb1d7f3b75/src/PoolManager.sol}{PoolManager (v4)}
  & 0.8.26$^{\mathrm{IR}}$ & 44\,444\,444 & cancun & 21/12 & 205/2\,038 & 24\,009
  & yes & 0/4 & 3/1 & 1 & 94 & 0/93 \\
\href{https://github.com/curvefi/curve-stablecoin/blob/f19a18abef6e75e3e08b1aa427ede724162bce57/curve_stablecoin/lending/Vault.vy}{LlamaLend Vault}
  & vyper 0.4.3 & codesize & cancun & 17/29 & 321/397 & 8\,635
  & yes & 0/0 & 0/0 & 0 & 0 & 0/0 \\
\href{https://github.com/morpho-org/metamorpho-v1.1/blob/5a0717b2c863060ff452ca324716083fe6d09374/src/MetaMorphoV1_1.sol}{MetaMorpho}
  & 0.8.37$^{\mathrm{IR}}$ & 200 & cancun & 34/42 & 510/1\,305 & 19\,708
  & no & 9/10 & 10/0 & 1 & 41 & 0/10 \\
\hline
\end{tabular}
\end{table}

%% file: figs/eval-proofs.tex
\begin{table}[tp]
\caption{Benchmark proofs. \textbf{Spec} and \textbf{Proof} count non-blank,
non-comment \Lean lines of the specification, in \lang's surface syntax, and of
the proof development. \textbf{Blocks} is the part of Proof that is
script-generated block summaries. \textbf{WF} marks proofs whose refinement
theorem carries side conditions. \textbf{Sessions} counts the proof sessions of
the run. \textbf{Prompts} counts the effective human inputs to the proof
session. \textbf{In}, \textbf{Cached} and \textbf{Out} are fresh input,
cached input and output tokens, in millions. \textbf{Time} is active proof time
(excluding stalled gaps). \textbf{Check} is the time, in seconds, for \Lean to
check the finished proof.}
\label{tab:eval-proofs}
\centering
\footnotesize
\setlength{\tabcolsep}{2pt}
\begin{tabular}{@{}l!{\color{black!30}\vrule}r!{\color{black!30}\vrule}r!{\color{black!30}\vrule}r!{\color{black!30}\vrule}l!{\color{black!30}\vrule}l!{\color{black!30}\vrule}r!{\color{black!30}\vrule}r!{\color{black!30}\vrule}r!{\color{black!30}\vrule}r!{\color{black!30}\vrule}r!{\color{black!30}\vrule}r!{\color{black!30}\vrule}r@{}}
\hline
\textbf{Contract} & \textbf{Spec} & \textbf{Proof} & \textbf{Blocks} & \textbf{WF}
  & \textbf{Model} & \textbf{Sessions} & \textbf{Prompts} & \textbf{In} & \textbf{Cached}
  & \textbf{Out} & \textbf{Time} & \textbf{Check} \\
\hline
Safe             & 546   & 58\,986  & 27\,114 & -- & gpt-6-astra & 1 & 2 & 7.9 & 292.8 & 1.4 & 18h\,14m & 656 \\
EAS              & 854   & 79\,681  & 30\,563 & -- & gpt-6-astra & 2 & 2 & 11.6 & 508.8 & 2.5 & 22h\,10m & 932 \\
Attester         & 74    & 19\,879  & 8\,101 & -- & gpt-6-astra & 1 & 2 & 2.5 & 88.1 & 0.5 & 6h\,09m & 240 \\
Morpho Blue      & 728   & 84\,090  & 48\,676 & -- & gpt-6-astra & 1 & 1 & 7.8 & 293.6 & 1.4 & 14h\,11m & 956 \\
Comet            & 1\,317 & 155\,651 & 91\,539 & yes & gpt-6-astra & 1 & 6 & 13.3 & 436.6 & 2.4 & 25h\,40m & 1\,962 \\
UniswapV3Pool    & 1\,684 & 162\,053 & 89\,965 & -- & gpt-6-astra & 1 & 1 & 13.5 & 659.4 & 3.7 & 30h\,17m & 2\,739 \\
PoolManager (v4) & 1\,884 & 145\,336 & 76\,392 & yes & gpt-6-astra & 1 & 1 & 14.7 & 621.7 & 3.4 & 27h\,09m & 1\,245 \\
LlamaLend Vault  & 615   & 38\,126  & 17\,214 & yes & gpt-6-astra & 1 & 1 & 3.7 & 158.6 & 0.8 & 9h\,16m & 411 \\
MetaMorpho       & 1\,429 & 199\,933 & 106\,117 & yes & gpt-6-astra & 3 & 5 & 19.7 & 698.5 & 3.7 & 34h\,30m & 929 \\
\hline
\end{tabular}
\end{table}

%% file: figs/eval-spec-sessions.tex
\begin{table}[tp]
\caption{Specification sessions. Sessions, Prompts, In, Cached, Out and Time are as
in \cref{tab:eval-proofs}. \textbf{Difftest} counts the randomized differential-test cases run
against the bytecode before proving.}
\label{tab:eval-spec-sessions}
\centering
\footnotesize
\setlength{\tabcolsep}{3pt}
\begin{tabular}{@{}l!{\color{black!30}\vrule}r!{\color{black!30}\vrule}r!{\color{black!30}\vrule}r!{\color{black!30}\vrule}r!{\color{black!30}\vrule}r!{\color{black!30}\vrule}r!{\color{black!30}\vrule}r@{}}
\hline
\textbf{Contract} & \textbf{Sessions} & \textbf{Prompts} & \textbf{In}
  & \textbf{Cached} & \textbf{Out} & \textbf{Time} & \textbf{Difftest} \\
\hline
EAS              & 2 & 16 & 1.5 & 48.4 & 0.3 & 4h\,08m & 1\,325 \\
Morpho Blue      & 1 & 1 & 0.5 & 19.8 & 0.1 & 0h\,59m & 1\,575 \\
Comet            & 1 & 5 & 0.6 & 18.9 & 0.1 & 1h\,24m & 3\,625 \\
UniswapV3Pool    & 1 & 4 & 1.1 & 56.2 & 0.2 & 5h\,12m & 1\,525 \\
PoolManager (v4) & 1 & 2 & 0.9 & 18.3 & 0.1 & 1h\,32m & 1\,875 \\
LlamaLend Vault  & 1 & 1 & 0.5 & 21.4 & 0.1 & 1h\,02m & 2\,525 \\
MetaMorpho       & 1 & 6 & 1.1 & 36.1 & 0.1 & 2h\,35m & 4\,025 \\
\hline
\end{tabular}
\end{table}

%% file: figs/eval-repair.tex
\begin{table}[tp]
\caption{Proof repair on Safe. Each row carries the proof from one version
of the contract to the next, in a single session. \textbf{Source} counts
non-blank, non-comment Solidity lines added/removed among the declarations
compiled into Safe. \textbf{Runtime} is the deployed bytecode of the target
version in bytes. \textbf{Spec} and \textbf{Proof} count non-blank,
non-comment \Lean lines added/removed in the specification, in \lang's
surface syntax, and in the written proof. The remaining columns are as in
\cref{tab:eval-proofs}. $^{\dagger}$The repaired theorem carries a side
condition: the executing account must exist.}
\label{tab:eval-repair}
\centering
\footnotesize
\setlength{\tabcolsep}{2pt}
\begin{tabular}{@{}l!{\color{black!30}\vrule}l!{\color{black!30}\vrule}r!{\color{black!30}\vrule}r!{\color{black!30}\vrule}r!{\color{black!30}\vrule}r!{\color{black!30}\vrule}r!{\color{black!30}\vrule}r!{\color{black!30}\vrule}r!{\color{black!30}\vrule}r!{\color{black!30}\vrule}r@{}}
\hline
\textbf{Step} & \textbf{Versions} & \textbf{Source} & \textbf{Runtime}
  & \textbf{Spec} & \textbf{Proof} & \textbf{Prompts} & \textbf{In}
  & \textbf{Cached} & \textbf{Out} & \textbf{Time} \\
\hline
step back & 77901a5 $\to$ cdb2eb5 & +11/$-$58 & 11\,755 & +5/$-$38 & +3\,216/$-$3\,917
  & 1 & 0.7 & 23.1 & 0.1 & 1h\,41m \\
replay 1  & cdb2eb5 $\to$ 0227664 & +24/$-$10 & 11\,499 & +4/$-$3 & +2\,468/$-$2\,445
  & 1 & 0.4 & 16.8 & 0.1 & 0h\,44m \\
replay 2  & 0227664 $\to$ 0e91d99 & +28/$-$0 & 11\,749 & +22/$-$0 & +2\,821/$-$2\,020
  & 1 & 0.4 & 12.2 & 0.1 & 0h\,39m \\
replay 3  & 0e91d99 $\to$ 77901a5 & +6/$-$1 & 11\,874 & +4/$-$2 & +1\,781/$-$1\,431
  & 1 & 0.4 & 21.3 & 0.1 & 1h\,23m \\
\hline
optimizer & 200 $\to$ 1\,000\,000 runs & 0 & 14\,838 & 0 & +2\,380/$-$2\,221
  & 1 & 0.4 & 11.0 & 0.0 & 0h\,28m \\
via-IR$^{\dagger}$ & legacy $\to$ via-IR & 0 & 11\,003 & +6/$-$3 & +11\,437/$-$7\,501
  & 14 & 4.9 & 161.0 & 0.8 & 11h\,03m \\
\hline
\end{tabular}
\end{table}

%% file: figs/eval-others.tex
\begin{table}[tp]
\caption{Contracts proved with earlier versions of the framework. Columns are
as in \cref{tab:eval-contracts} and \cref{tab:eval-proofs}. The DSS row
covers the 19 MakerDAO contracts and reports their sums; its WF entry is
the number of them whose theorem carries side conditions (Cure and
Clipper).}
\label{tab:eval-others}
\centering
\footnotesize
\setlength{\tabcolsep}{2.5pt}
\begin{tabular}{@{}l!{\color{black!30}\vrule}l!{\color{black!30}\vrule}r!{\color{black!30}\vrule}r!{\color{black!30}\vrule}r!{\color{black!30}\vrule}r!{\color{black!30}\vrule}r!{\color{black!30}\vrule}l!{\color{black!30}\vrule}r!{\color{black!30}\vrule}r!{\color{black!30}\vrule}r@{}}
\hline
\textbf{Contract} & \textbf{solc} & \textbf{Fn/View} & \textbf{Source}
  & \textbf{Runtime} & \textbf{Loops} & \textbf{Asm} & \textbf{WF} & \textbf{Spec}
  & \textbf{Proof} & \textbf{Check} \\
\hline
\href{https://github.com/makerdao/dss}{DSS (19 contracts)}
  & 0.6.12 & 156/171 & 1\,972/2\,120 & 79\,268 & 4/4 & 6/6 & 2 & 3\,233 & 627\,990 & 7\,722 \\
\href{https://github.com/gnosis/canonical-weth/blob/master/contracts/WETH9.sol}{WETH9}
  & 0.5.16 & 5/6 & 47/47 & 1\,763 & 0/0 & 0/0 & yes & 66 & 9\,404 & 154 \\
\href{https://github.com/nounsDAO/nouns-monorepo}{Auction (Nouns)}
  & 0.8.23 & 11/9 & 78/198 & 6\,150 & 0/0 & 0/0 & -- & 221 & 11\,777 & 291 \\
\href{https://github.com/Uniswap/v2-core/blob/v1.0.1/contracts/UniswapV2Pair.sol}{UniswapV2Pair}
  & 0.5.16 & 10/17 & 155/262 & 8\,833 & 0/1 & 0/1 & -- & 342 & 110\,842 & 905 \\
\href{https://github.com/OpenZeppelin/openzeppelin-contracts/blob/master/contracts/access/Ownable2Step.sol}{Ownable2Step}
  & 0.8.35 & 3/2 & 1/54 & 629 & 0/0 & 0/0 & -- & 34 & 2\,327 & 65 \\
\href{https://github.com/OpenZeppelin/openzeppelin-contracts/blob/master/contracts/access/AccessControl.sol}{AccessControl}
  & 0.8.35 & 3/4 & 3/70 & 1\,033 & 0/0 & 0/0 & -- & 48 & 6\,127 & 111 \\
\href{https://github.com/OpenZeppelin/openzeppelin-contracts/blob/master/contracts/utils/Pausable.sol}{Pausable}
  & 0.8.35 & 2/3 & 12/49 & 455 & 0/0 & 0/0 & -- & 31 & 1\,422 & 57 \\
\href{https://github.com/OpenZeppelin/openzeppelin-contracts/blob/master/contracts/token/ERC6909/ERC6909.sol}{ERC6909}
  & 0.8.35 & 4/4 & 0/117 & 2\,102 & 0/0 & 0/0 & -- & 63 & 15\,533 & 472 \\
\hline
\end{tabular}
\end{table}

%% file: related.tex
\textbf{Translation validation.}
Translation validation is closely related to our work.
It checks that the output of a translator correctly implements the input after
each run.
The seminal work of Pnueli et
al.~\cite{TransVal:Pnueli98:translationvalidation} introduced the approach and
demonstrated validation of one program from the synchronous language Signal to C.
The validator is tailored for the shape of the code emitted by the translator in
question.
Necula~\cite{TransVal:Necula00:translationvalidation} applied the approach to
validate the optimization passes, one at a time, of GCC on its RTL intermediate
language, using symbolic evaluation and constraint solving.
Transformations included branch optimization, common subexpression elimination,
loop transformations, register allocation, and instruction scheduling.
The approach does not require compiler cooperation, but the validator assumes
that every target branch corresponds to a source branch, and that at
corresponding program points source and target variables are related only by
equalities.
Huang et al.~\cite{TransVal:Huang06:registerallocation} validate a single pass,
register allocation, by checking that the allocator's output preserves the data
dependences of its input.
Tristan and Leroy~\cite{TransVal:Tristan08:instructionscheduling,
TransVal:Tristan09:lazycodemotion} formally verify validators for CompCert's
instruction scheduling and lazy code motion in Rocq, and integrate them into
CompCert, showing that an unverified optimization paired with a verified
validator can live inside a verified compiler. CompCert comprises more passes in
this style, such as its parser~\cite{Compcert:Jourdan12:validatinglr}.
Alive2~\cite{TransVal:Lopes21:alive2} is a general-purpose validator for LLVM.
It checks refinement between the LLVM IR of a function before and after an
intraprocedural optimization pass, or a batch of passes, by encoding both in
SMT. The validator is bounded and unrolls loops up to a fixed bound.
DESTIVA~\cite{TransVal:Mikek26:destiva} speeds Alive2 up by instrumenting LLVM
to tag the instructions each pass changes, validating only the changed fragment.
The result is sound only if the tags are correct and applies to loop-free
functions.

More closely related to our technique are translation validation approaches that
target whole compilers. 
Rival~\cite{TransVal:Rival04:symbolictransfer} validates \code{gcc}'s
compilation of C to PowerPC assembly directly, -O2 pipeline minus loop
unrolling. 
It relies on the compiler's debugging information that gives the correspondence
between source and assembly program points and variables. For optimized code
this information was partly wrong and had to be repaired by hand.
Matched points are then checked for local equivalence with a fixed set of
hand-written rewrite rules.
On three embedded programs of up to 74k lines of C without dynamic allocation,
compiled at low optimization, it raised no false alarms and took under ten
seconds.
The debugging information, OCaml certifier, both semantic encodings and the
prover are trusted, including the soundness of the hand-written rewrite rules.

Sewell et al.~\cite{TransVal:Sewell13:sel4tv} validate the ARM binary of the
verified seL4 kernel~\cite{TransVal:Klein09:sel4}, 9,500 lines of C compiled by
\code{gcc}.
It validates the generated code at function level. 
The binary is decompiled into logic with a proof-producing decompiler, both
sides are translated to a common graph language, and refinement is proved per
function with SMT solvers, matching each loop at a synchronization point found
by heuristics.
At \code{-O1} all functions validate except three with nested loops, in one hour
of proof time; at \code{-O2} 145 of 225 do.
The kernel source code had to be manually edited: typedefs were changed to
eliminate struct padding, and two functions were rewritten.

Dynamo~\cite{TransVal:Rose24:dynamo} validates the unoptimized LLVM IR of a C
procedure (emitted by \code{clang}) against the optimized x86 assembly produced
by \code{clang}, \code{gcc} or \code{icc}, one procedure at a time and without
compiler hints. On bzip2 it validates 64 of 72 procedures at  \code{-O1} and 54
at \code{-O2}, in about six CPU hours and all procedures are below 190
instructions. 
The tool, its semantic models and the SMT solvers are trusted (one solver
returned an unsound model during the experiments).

The main distinction of \EquiVM is that it delivers a machine-checked
certificate, which is replayed rather than revalidated and composes with
source-level proofs.
The comparison is not entirely fair, since LLM agents change the economics of
proof search.
On our benchmarks no source change was required, nested loops were handled,
and every contract we tried was proved.
\EquiVM certificates are more expensive in terms of hours of agent time and 
cost in money.

\textbf{Verified compilation.}
A verified compiler is proved, once and for all, to preserve the semantics of
every program it compiles. The idea is more than half a century old
~\cite{CompCorr:Mccarthy67:correctnessof,
Compcorr:Milner72:provingcompilercorrectness}, and landmark mechanized results
such as CompCert~\cite{Compcert:Leroy09:formalverification} and
CakeML~\cite{CakeML:Kumar:cakeml} established that it scales to realistic
optimizing compilers.
The guarantee typically establishes that the generated code refines the 
source language, for every program produced by the compiler.
The technique is tied to a single source language and a single toolchain, and,
historically, it has taken years of expert effort to establish and maintain.
LLM-driven proof engineering is lowering the cost of
building verified compilers.
Paraskevopoulou~\cite{LLM:Paraskevopoulou26:mgproofs} demonstrates that LLMs can
be used to accelerate verified compiler development in the context of the
CertiRocq
compiler~\cite{Coq:Anand17:certicoq,Proofs:Paraskevopoulou19:compositionalopt},
showing that an A-normal form transformation can be verified in a matter of days
with human guidance but without human-written code.
We are not aware of a fully verified compiler for an EVM language.
DeepSEA~\cite{SC:Sjoberg24:deepseaevm}, verified in Rocq, retargets the DeepSEA
language for certified system software~\cite{SC:Sjoberg19:deepsea} to \EVM
bytecode.
The language restricts loops to a few terminating patterns, has no mutable
locals, and the backend proofs about pointers are listed as incomplete.
Verity~\cite{LLM:Marchand26:verity} is an EVM contract language embedded in \Lean
with a compiler to Yul IR, which uses the unverified \code{solc} compiler
to produce EVM bytecode. 
The compiler and its verification leverage LLM agents.
To our knowledge, the scope of the theorem is narrow: it does not cover external
calls or loops, and covers only scalar parameters and results.
Recent approaches~\cite{SC:Powdr26:yulcompiler, SC:Robinson26:solidus} are in
progress and partial and sit below \code{solc}'s unverified Solidity-to-Yul
frontend.

In terms of the strength of the compiler theorem, most compiler correctness 
statements do not cover interaction with arbitrary code.
Patterson and Ahmed~\cite{Compcorr:Patterson19:thenext700} order correctness
theorems by what compiled code can be linked with, placing arbitrary target code
at the far end of the spectrum.
Multi-language semantics~\cite{General:Matthews07:operationalsemantics} is the
only technique known to scale to truly arbitrary code. 
Perconti and Ahmed~\cite{Compcorr:Perconti14:verifyinganopencompiler} use it to
verify a two-pass compiler from System~F to a typed lambda calculus with heap references. 
Compiled code links with any well-typed target component.
DimSum~\cite{Compcorr:Sammler23:dimsum} uses multi-language semantics but
replaces syntactic boundaries with semantic wrappers.
Compiled code links with assembly libraries that have no source-level
counterpart.
Both are developed for idealized languages.
\EquiVM relies on multi-language semantics for expressing interaction with
unrestricted EVM bytecode and is the only work we know of to apply it
to a real-world setting.
There is a rich line of work extending CompCert to compositional compiler
statements~\cite{Compcert:Kang16:lightweightverification,
Compcert:Stewart15:compositional,
Compcorr:Ramananandro15:acompositionalsemantics,CompCert:Gu15:deepspecifications,
CompCert:Wang19:anabstractstackbased}, placing different restrictions on the
external components.
The most permissive one, CompCertO~\cite{Compcert:Koenig21:compcerto}, still
requires that external components follow CompCert's memory model and calling
conventions.

\textbf{Certifying compilation.}
Another approach is to build a certifying
compiler~\cite{PCC:Necula98:certifyingcompiler} that along with the generated
code, emits a certificate that it is correct, exploiting the compiler's own
knowledge of the code it generates. 
Blech and Poetzsch-Heffter~\cite{PCC:Blech07:certifyingcodegenphase} demonstrate
the feasibility of a certifying code generator that produces Rocq certificates
for semantic preservation. They show that the approach works for small programs
and identify the certificate checking as the bottleneck.
Blech and Grégoire~\cite{PCC:Blech08:certifyingcodegen} scale this to a
multi-pass compiler for an imperative intermediate language to MIPS assembly and
they prove the independent checker correct in Rocq.

Li et al.~\cite{PCC:Li07:proofproducingcompiler} and Myreen et
al.~\cite{PCC:Myreen09:extensiblecompilation} 
compile pure functions defined in HOL4 to machine code for multiple
architectures, emitting for each run a machine-checked theorem that the code
implements the function. 
This work employs decompilation into
logic~\cite{PCC:Myreen08:machinecodemultiarch,
PCC:Myreen12:decompilationimproved} that extracts from machine code a pure HOL
function together with a certificate theorem relating the two, composing
per-instruction Hoare triples. All further reasoning then happens at the level
of the extracted pure function.

Cogent~\cite{PCC:OConnor16:cogent} is a restricted functional language for
systems code whose certifying compiler emits C code, a shallow embedding in
Isabelle/HOL, and a proof that the C refines the embedding.

Crellvm~\cite{PCC:Kang18:crellvm} instruments LLVM's register promotion and
global value numbering passes to emit with each translation a proof in a
relational Hoare logic, checked by Rocq. The custom inference rules are
unverified. Validation found four miscompilation bugs.

Closest to our domain, Krijnen et al.~\cite{TransVal:Krijnen24:smartcontracts}
add translation certification to the Plutus Tx compiler for Cardano. Each
compiler run emits a certificate (a Rocq proof object relating the source to the
compiled code through a chain of translation relations), so validation produces
a machine-checkable artifact. Their certifier is built around the compiler and
also exposes its intermediate representations.

In all certifying approaches above the certificate is produced by the compiler
itself, so each applies to a single toolchain.

\textbf{Certifying synthesis.}
A complementary way to obtain low-level code with a certificate is to
synthesize both together. SuSLik~\cite{Synth:Polikarpova19:suslik} synthesizes
heap-manipulating programs from separation-logic specifications, and its
derivations can be translated into Rocq
certificates~\cite{Synth:Watanabe21:certifiedsuslik}.
Refinement-based frameworks generate verified LLVM from
Isabelle/HOL~\cite{Synth:Lammich19:isabellellvm}.
Vericert extends CompCert to certified high-level synthesis of
hardware~\cite{Synth:Herklotz21:vericert}. 

\textbf{Smart-contract verification.}
There are several frameworks that bypass compiler correctness and work on 
verifying the generated bytecode directly.
Notably, the Certora Prover~\cite{SC:Grossman24:certora, SC:Certora25:prover}
is a commercial framework that decompiles bytecode to an intermediate 
representation and  checks specifications written in its CVL language by 
using SMT solvers. 
Loops are only unrolled up to a bound, and the decompiler, the VC gen and the
SMT solver are trusted.
Frameworks like Halmos~\cite{a16z:halmos} and hevm~\cite{SC:Dxo24:hevm} utilize
symbolic execution engines to reason about all possible contract states directly
on the emitted EVM bytecode. Their handling of loops is inherently incomplete.
Similarly, Kontrol~\cite{rv:kontrol} leverages
KEVM~\cite{SC:Hildenbrandt18:kevm} (an executable formal semantics of the EVM in
the K framework) to lift standard Foundry tests into reachability logic proofs,
utilizing symbolic execution and SMT. 
The same semantics underlies the verification of the Ethereum 2.0 deposit
contract by Park et al.~\cite{SC:Park20:deposit}, who prove the incremental
Merkle tree algorithm in K, hand-write a bytecode-level specification that
refines each transition of the model to concrete EVM state, and verify the
Vyper-compiled bytecode against it with the KEVM verifier.
The refinement is trusted. The effort found several Vyper compiler bugs.

Beyond the \EVM, ConCert~\cite{SC:Annenkov21:concert} verifies contracts in Rocq
and extracts them to Liquidity, targeting the Tezos blockchain, and Midlang,
targeting Concordium, through MetaRocq's verified
erasure~\cite{Coq:Sozeau19:themetacoq} and a verified optimization pass.
Scilla~\cite{SC:Sergey19:scilla}, the contract language of Zilliqa, avoids the
gap between source and deployed code altogether: motivated by compiler bugs and
the difficulty of auditing \EVM bytecode, it deploys contracts as source and
executes them with an (unverified) interpreter.
The Move Prover~\cite{SC:Zhong20:moveprover} checks programs in Move, the
language used by the Sui blockchain, against formal specifications
automatically, by translation to Boogie.

\textbf{LLMs and mechanized proof.}
Until recently, LLM-assisted proving has mostly been evaluated by reproving
theorems from existing developments~\cite{baldur-2023, proof-automation-llms,
Thompson:2025:Rango, fscq-case-study}.
Recent reports focus on using LLMs and agentic development to synthesize novel
proofs with minimum human involvement. 
Ioannidis et al.~\cite{LLM:Ioannidis26:proofspromptly} use AI agents to verify
thousands of lines of F$^\star$ and Pulse code,
including extending the standard library with verified data structures and
concurrency primitives, re-implementing and verifying MSQuic's receive buffer,
and porting a 30K-line verified OCaml garbage collector from Low$^\star$ to
Pulse. 
The work took three experts a month and they estimate that manual proof
engineering would have taken six months.
Sergey~\cite{sergey2026movelean} reports that the 39\,000-line mechanization of
type soundness for the Move borrow checker in \Lean using LLM agents took about
27 working days.
LeetProof~\cite{Synth:Feng26:multimodal} is an agentic pipeline over
Velvet~\cite{Proofs:Gladshtein26:velvet}, a multi-modal verifier in \Lean. 
It validates specifications by property-based testing before synthesis,
decomposes the task along verification conditions, and hands the remaining
obligations to AI provers.

%% file: Appendix.tex
\section{Selected Rules of the \lang Semantics}
\label{app:rules}
\Cref{fig:stele-stmt-rules} shows selected statement rules of \lang,
\cref{fig:stele-call-rules} the external-call boundary, and
\cref{fig:stele-exec-rules} the message-level execution of a \lang contract.
Every judgment is indexed by the configuration \cfg and the immutables store
\imms, both elided.
$\frm[x \mapsto \aval]$ updates a local binding.
$\avals\!\downarrow$ collapses a returned value list. An absent or empty list
becomes $\kw{unit}$, a singleton its value, and any other list a tuple.
$\mathbb{W}$ is the set of \EVM words.
$\mathsf{perm}(\evmst)$ is the write permission of the execution environment.
It is cleared when the contract was entered through \code{STATICCALL}.
$\mathsf{local}(\aref)$ holds when the root of $\aref$ is a local variable,
which is decided statically from its declaration. Writes to persistent or
transient storage and calls that transfer value require the permission.
Local assignments and zero-value calls do not.
$\mathsf{selector}$ matches the first four calldata bytes against the
selectors of the contract's transitions.
$\mathsf{decode}^{\mathsf{cd}}_{\cfg}$ decodes the calldata into the
transition's parameter store.
$\mathsf{fbArgs}$ binds the fallback's parameter, either none or the raw
calldata.
$\mathsf{init}(\sigma, \sigma_0, g, A, I)$ is the \EVM state with account map
$\sigma$, checkpoint $\sigma_0$, gas $g$, substate $A$ and execution
environment $I$.
In \cref{fig:stele-call-rules}, $p$ is the opcode's permission bit, which is
$\kw{false}$ for \code{staticcall}.
$\evmst\!\upharpoonright_{\sigma', A'}$ is $\evmst$ with its account map and
substate replaced.
$\mathsf{access}(\evmst, a)$ is the substate of $\evmst$ with $a$ added to the
accessed accounts.
\input{figs/stele-stmt-rules}
\input{figs/stele-call-rules}
\input{figs/stele-exec-rules}

\section{Before and After \code{RD}}
\label{app:proofs}

\Cref{fig:reach-run} shows a segment lemma from the ERC20 development. The
statement is a reach fact at the decoder's entry cursor. The proof obtains
the incoming reach fact and threads it through one combinator per opcode
with \code{evm_run}. 
The elaborator inserts the decode obligation (\code{by native_decide}) and the
stack bound (\code{by evm_ov}) at every step. The facts that carry information
are passed explicitly. Here, these are the calldata-length check (\code{hslt})
and the jump-target facts (\code{erc20_jd}). The lemma's statement is the next
cursor, and segment lemmas chain by ordinary function application.

\Cref{fig:reach-hand} shows the same straight-line reasoning as it was written
before the abstraction. Every opcode is a nested out-of-gas case split carrying
five bookkeeping facts by hand, about twelve lines that the combinator chain of
\cref{fig:reach-run} replaces with one.
The library was developed soon after our first two small proofs, and
significantly dropped the proof size (down to 1203 from  2868 and to 493 from
2390).
Normalized by the number of opcode steps each proof reasons about, the
whole-proof cost falls from roughly thirteen to seventeen lines per opcode to
three to five. 
The straight-line stepping that the abstraction replaces (\cref{fig:reach-hand})
shrinks by about a factor of twelve, from twelve lines per opcode to one.
 
\input{figs/reach-hand}

\input{figs/reach-run}

\input{appendix-telemetry}

%% file: figs/stele-stmt-rules.tex
\begin{figure}[t]
\small
\noindent\makebox[\linewidth][l]{\fbox{$\execS{\frm}{\evmst}{\astmt}{\outc}$}}

\vspace{-0.5ex}
\begin{mathpar}
\inferrule*[lab=\rulename{Let}]
  { \evalE{\frm}{\evmst}{\anexp} = \rok{\aval} }
  { \execS{\frm}{\evmst}{\alpha\;x = \anexp}{\oOk{\frm[x \mapsto \aval]}{\evmst}} }

\inferrule*[lab=\rulename{LetRev}]
  { \evalE{\frm}{\evmst}{\anexp} = \rrev }
  { \execS{\frm}{\evmst}{\alpha\;x = \anexp}{\oRev} }

\inferrule*[lab=\rulename{Gas}]
  { w \in \mathbb{W} }
  { \execS{\frm}{\evmst}{\kw{uint256}\;x = \kw{gasleft}()}{\oOk{\frm[x \mapsto w]}{\evmst}} }

\inferrule*[lab=\rulename{Req}]
  { \evalE{\frm}{\evmst}{\anexp} = \rok{\kw{true}} }
  { \execS{\frm}{\evmst}{\kw{require}(\anexp)}{\oOk{\frm}{\evmst}} }

\inferrule*[lab=\rulename{ReqFail}]
  { \evalE{\frm}{\evmst}{\anexp} = \rok{\kw{false}} }
  { \execS{\frm}{\evmst}{\kw{require}(\anexp)}{\oRev} }

\inferrule*[lab=\rulename{Assign}]
  { \evalE{\frm}{\evmst}{\anexp} = \rok{\aval} \\
    \writeSt{\cfg}{\frm, \evmst}{\aref}{\aval} = \rok{(\frm', \evmst')} \\
    \mathsf{local}(\aref) \lor \mathsf{perm}(\evmst) = \kw{true} }
  { \execS{\frm}{\evmst}{\aref = \anexp}{\oOk{\frm'}{\evmst'}} }

\inferrule*[lab=\rulename{AssignStatic}]
  { \evalE{\frm}{\evmst}{\anexp} = \rok{\aval} \\
    \writeSt{\cfg}{\frm, \evmst}{\aref}{\aval} = \rok{(\frm', \evmst')} \\
    \neg\mathsf{local}(\aref) \\
    \mathsf{perm}(\evmst) = \kw{false} }
  { \execS{\frm}{\evmst}{\aref = \anexp}{\oStatic} }

\inferrule*[lab=\rulename{WhileF}]
  { \evalE{\frm}{\evmst}{\anexp} = \rok{\kw{false}} }
  { \execS{\frm}{\evmst}{\kw{while}\,(\anexp)\,\{\stmts\}}{\oOk{\frm}{\evmst}} }

\inferrule*[lab=\rulename{WhileT}]
  { \evalE{\frm}{\evmst}{\anexp} = \rok{\kw{true}} \\
    \execB{\frm}{\evmst}{\stmts}{\oOk{\frm'}{\evmst'}} \\
    \execS{\frm'}{\evmst'}{\kw{while}\,(\anexp)\,\{\stmts\}}{\outc} }
  { \execS{\frm}{\evmst}{\kw{while}\,(\anexp)\,\{\stmts\}}{\outc} }

\inferrule*[lab=\rulename{Ret}]
  { \evalE{\frm}{\evmst}{\overline{\anexp}} = \rok{\avals} }
  { \execS{\frm}{\evmst}{\kw{return}\;\overline{\anexp}}{\oRet{\frm}{\evmst}{\avals}} }

\inferrule*[lab=\rulename{CallInt}]
  { \evalE{\frm}{\evmst}{\overline{\anexp}} = \rok{\avals} \\
    \frm.\mathsf{contract}(f) = (\overline{x}, \stmts_{f}) \\
    \execF{\frm\langle\overline{x} \mapsto \avals\rangle}{\evmst}{\stmts_{f}}
          {\oRet{\frm''}{\evmst'}{\avals'}} }
  { \execS{\frm}{\evmst}{T\;\kw{storage}^{?}\;y = f(\overline{\anexp})}
          {\oOk{\frm[y \mapsto \avals'\!\downarrow]}{\evmst'}} }
\end{mathpar}
\caption{Selected statement rules of \lang.}
\label{fig:stele-stmt-rules}
\end{figure}

%% file: figs/stele-call-rules.tex
\begin{figure}[t]
\small
\noindent\makebox[\linewidth][l]{\fbox{$\mathsf{call}_{\cfg}(\evmst, a, \mathit{val}, \mathit{cd}, p)
  \rightsquigarrow (z,\, \evmst',\, o)$}}

\vspace{-0.5ex}
\begin{mathpar}
\inferrule*[lab=\rulename{$\CallEVM$-Call}]
  { w = \mathsf{word}(\mathit{val}) \\
    w \le \mathsf{balance}(\evmst, \kw{this}) \\
    \mathsf{depth}(\evmst) \ne 1024 \\\\
    \exists\, g\; A_{\mathit{in}}.\;\;
      (\sigma', A', z, o) =
        \CallEVM(\evmst, A_{\mathit{in}},\;
          \kw{this} \xrightarrow{\,w\,} a,\;
          \mathit{cd},\; g,\; \mathsf{depth}(\evmst)+1,\; p \wedge \mathsf{perm}(\evmst)) }
  { \mathsf{call}_{\cfg}(\evmst, a, \mathit{val}, \mathit{cd}, p)
      \rightsquigarrow (z,\, \evmst\!\upharpoonright_{\sigma', A'},\, o) }

\inferrule*[lab=\rulename{$\CallEVM$-NoCall}]
  { \neg\bigl(\mathsf{word}(\mathit{val}) \le \mathsf{balance}(\evmst, \kw{this})
      \;\wedge\; \mathsf{depth}(\evmst) \ne 1024\bigr) \\
    A' = \mathsf{access}(\evmst, a) }
  { \mathsf{call}_{\cfg}(\evmst, a, \mathit{val}, \mathit{cd}, p)
      \rightsquigarrow (\kw{false},\, \evmst\!\upharpoonright_{A'},\, \varepsilon) }

\inferrule*[lab=\rulename{ExtCall}]
  { \evalE{\frm}{\evmst}{\anexp_{r}} = \rok{a} \\
    \evalE{\frm}{\evmst}{\anexp_{v}} = \rok{\mathit{val}} \\
    \evalE{\frm}{\evmst}{\overline{\anexp}} = \rok{\avals} \\
    \mathsf{word}(\mathit{val}) = 0 \lor \mathsf{perm}(\evmst) = \kw{true} \\\\
    \mathsf{encode}_{\cfg}(f, \avals) = \mathit{cd} \\
    \mathsf{call}_{\cfg}(\evmst, a, \mathit{val}, \mathit{cd}, p)
      \rightsquigarrow (\kw{true},\, \evmst',\, o) \\
    \mathsf{decode}_{\cfg}(f, o) = \avals' }
  { \execS{\frm}{\evmst}
      {\alpha\;y = \anexp_{r}.f\{\kw{value}{:}\anexp_{v}\}(\overline{\anexp})}
      {\oOk{\frm[y \mapsto \avals'\!\downarrow]}{\evmst'}} }

\inferrule*[lab=\rulename{ExtCallFail}]
  { \evalE{\frm}{\evmst}{\anexp_{r}} = \rok{a} \\
    \evalE{\frm}{\evmst}{\anexp_{v}} = \rok{\mathit{val}} \\
    \evalE{\frm}{\evmst}{\overline{\anexp}} = \rok{\avals} \\
    \mathsf{word}(\mathit{val}) = 0 \lor \mathsf{perm}(\evmst) = \kw{true} \\\\
    \mathsf{encode}_{\cfg}(f, \avals) = \mathit{cd} \\
    \mathsf{call}_{\cfg}(\evmst, a, \mathit{val}, \mathit{cd}, p)
      \rightsquigarrow (\kw{false},\, \evmst',\, o) }
  { \execS{\frm}{\evmst}
      {\alpha\;y = \anexp_{r}.f\{\kw{value}{:}\anexp_{v}\}(\overline{\anexp})}
      {\oRev} }

\inferrule*[lab=\rulename{ExtCallDecodeRev}]
  { \evalE{\frm}{\evmst}{\anexp_{r}} = \rok{a} \\
    \evalE{\frm}{\evmst}{\anexp_{v}} = \rok{\mathit{val}} \\
    \evalE{\frm}{\evmst}{\overline{\anexp}} = \rok{\avals} \\
    \mathsf{word}(\mathit{val}) = 0 \lor \mathsf{perm}(\evmst) = \kw{true} \\\\
    \mathsf{encode}_{\cfg}(f, \avals) = \mathit{cd} \\
    \mathsf{call}_{\cfg}(\evmst, a, \mathit{val}, \mathit{cd}, p)
      \rightsquigarrow (\kw{true},\, \evmst',\, o) \\
    \mathsf{decode}_{\cfg}(f, o) = \bot }
  { \execS{\frm}{\evmst}
      {\alpha\;y = \anexp_{r}.f\{\kw{value}{:}\anexp_{v}\}(\overline{\anexp})}
      {\oRev} }
\end{mathpar}
\caption{Selected external-call rules of \lang.}
\label{fig:stele-call-rules}
\end{figure}

%% file: figs/stele-exec-rules.tex
\begin{figure}[t]
\small
\noindent\makebox[\linewidth][l]{\fbox{$\steleexec{\cfg}{\mathit{ctr}}{\sigma, g, A, I}{\outc}$}}

\vspace{-0.5ex}
\begin{mathpar}
\inferrule*[lab=\rulename{Ex-Sel}]
  { \mathsf{selector}(\mathit{ctr}, I.\mathsf{calldata}) = t \\
    \mathsf{decode}^{\mathsf{cd}}_{\cfg}(t, I.\mathsf{calldata}) = \locals \\\\
    \evmst = \mathsf{init}(\sigma, \sigma_0, g, A, I) \\
    \execF{\langle \mathit{ctr}, \locals, \imms\rangle}{\evmst}{t.\mathsf{body}}{\outc} }
  { \steleexec{\cfg}{\mathit{ctr}}{\sigma, g, A, I}{\outc} }

\inferrule*[lab=\rulename{Ex-NoDecode}]
  { \mathsf{selector}(\mathit{ctr}, I.\mathsf{calldata}) = t \\
    \mathsf{decode}^{\mathsf{cd}}_{\cfg}(t, I.\mathsf{calldata}) = \bot }
  { \steleexec{\cfg}{\mathit{ctr}}{\sigma, g, A, I}{\oNoDecode} }

\inferrule*[lab=\rulename{Ex-Recv}]
  { I.\mathsf{calldata} = \varepsilon \\
    \mathit{ctr}.\kw{receive} = t \\\\
    \evmst = \mathsf{init}(\sigma, \sigma_0, g, A, I) \\
    \execF{\langle \mathit{ctr}, \varnothing, \imms\rangle}{\evmst}{t.\mathsf{body}}{\outc} }
  { \steleexec{\cfg}{\mathit{ctr}}{\sigma, g, A, I}{\outc} }

\inferrule*[lab=\rulename{Ex-Fb}]
  { \mathsf{selector}(\mathit{ctr}, I.\mathsf{calldata}) = \bot \\
    I.\mathsf{calldata} \neq \varepsilon \lor \mathit{ctr}.\kw{receive} = \bot \\
    \mathit{ctr}.\kw{fallback} = t \\\\
    \mathsf{fbArgs}(t, I.\mathsf{calldata}) = \locals \\\\
    \evmst = \mathsf{init}(\sigma, \sigma_0, g, A, I) \\
    \execF{\langle \mathit{ctr}, \locals, \imms\rangle}{\evmst}{t.\mathsf{body}}{\outc} }
  { \steleexec{\cfg}{\mathit{ctr}}{\sigma, g, A, I}{\outc} }

\inferrule*[lab=\rulename{Ex-NoMatch}]
  { \mathsf{selector}(\mathit{ctr}, I.\mathsf{calldata}) = \bot \\
    I.\mathsf{calldata} \neq \varepsilon \lor \mathit{ctr}.\kw{receive} = \bot \\
    \mathit{ctr}.\kw{fallback} = \bot }
  { \steleexec{\cfg}{\mathit{ctr}}{\sigma, g, A, I}{\oNoMatch} }
\end{mathpar}
\caption{Message-level execution of a \lang contract.}
\label{fig:stele-exec-rules}
\end{figure}

%% file: figs/reach-hand.tex
\begin{figure}[t]
\begin{lstlisting}[style=lean]
  -- 0: JUMPDEST
  have st0 := jumpdest_xstep hcode hpc (by decide)
                (by rw [hstk]; simp only [List.length_cons]; omega)
  by_cases g0 : g.toNat < C + 1
  . exact Or.inl (hX.trans (stepOOG hgas st0 hk hC (by omega)))
  . set s1 := stJumpdest s with hs1
    have hX1 := hX.trans (stepContinue (k := k) (C := C) hgas st0 hk (by omega))
    have hc1 : s1.executionEnv.code = powBytecode := by
      rw [hs1]; simp only [stJumpdest]; exact hcode
    have hp1 : s1.machineState.pc = ⟨157⟩ := by
      rw [hs1]; simp only [stJumpdest]; rw [hpc]; rfl
    have hg1 : s1.machineState.gasAvailable.toNat = g.toNat - (C + 1) := by
      rw [hs1]; simp only [stJumpdest]; rw [toNat_sub_ofNat (by omega)]; omega
    have hk1 : s1.machineState.stack = v :: ret :: R := by
      rw [hs1]; simp only [stJumpdest]; exact hstk
    -- 1: PUSH0
    have st1 := push0_xstep hc1 hp1 (by decide) hk1 (by simp only [List.length_cons]; omega)
    by_cases g1 : g.toNat < C + 3
    . exact Or.inl (hX1.trans (stepOOG (k := k+1) (C := C+1) hg1 st1 (by omega) ...))
    . set s2 := stPush0 s1 with hs2
      -- ... four more `have`s for s2, then one nested block per remaining opcode
\end{lstlisting}
\caption{The same straight-line reasoning before the \code{RD} abstraction:
hand-written per-opcode stepping (Pow proof; two of nine opcodes shown).
Each opcode contributes a nested out-of-gas split and five bookkeeping facts
(code, pc, gas, stack, and the threaded execution equation) --- about twelve
lines, which the combinator chain of \cref{fig:reach-run} replaces with one.}
\label{fig:reach-hand}
\end{figure}

%% file: figs/reach-run.tex
\begin{figure}[t]
\begin{lstlisting}[style=lean]
/-- The one-address tuple decoder accepts calldata with at least one
    static word and jumps to the address element decoder at pc 1874. -/
theorem erc20BalanceOfX_dec1874 (hsz36 : 36 <= I.calldata.size) ... :
    ...
    exists k C, RD erc20Bytecode I g (initState ...) (pc := ⟨1874⟩)
        [⟨4⟩ + ⟨0⟩, I.calldata.size, ⟨2212⟩, ⟨0⟩, ⟨0⟩, ⟨4⟩,
          I.calldata.size, ⟨247⟩, ⟨252⟩, sel]
        solcFreePtrMem ⟨3⟩ ByteArray.empty (cA, sigma) k C := by
  have hslt : UInt256.slt (I.calldata.size - ⟨4⟩) ⟨32⟩ = ⟨0⟩ :=
    solcDecodeLenCheckOk_4_32 hsz36 hszhi hsize
  obtain (k, C, rd) := erc20BalanceOfX_toDecoder ... hreach
  exact (_, _, evm_run rd with [
    jumpdest, push0, push1 ⟨32⟩, dup3, dup5, sub, slt, iszero, push2 ⟨2199⟩,
    jumpiT (by rw [hslt]; decide) erc20_jd,
    jumpdest, push0, push2 ⟨2212⟩, dup5, dup3, dup6, add, push2 ⟨1874⟩,
    jump erc20_jd ])
\end{lstlisting}
\caption{A segment lemma from the ERC20 proof (lightly reformatted).}
\label{fig:reach-run}
\end{figure}

%% file: appendix-telemetry.tex
\input{appendix-prompts}

\section{Compiler Invocations}
\label{app:compile}

\Cref{tab:compile} lists the compiler invocation that produced the bytecode of
each contract of the paper's Tables~1 and~5, as recorded in the checked-in
artifacts. Safe, the DSS contracts, WETH9, Auction and UniswapV2Pair were
compiled on the \code{solc} command line and LlamaLend Vault on the \code{vyper}
command line; the other contracts were compiled through \code{solc}'s
standard-JSON interface, and their settings are shown as the equivalent flags.
Flags selecting output files, base paths and the import remappings that resolve
each contract's vendored dependencies are omitted. Where no
\code{--evm-version} is given, the compiler's default target applies.
$^{\dagger}$Comet additionally runs the Yul optimizer with the custom step
sequence of its upstream Hardhat configuration,
\texttt{dhfoDgvulfnTUtnIf [xa[r]scLM cCTUtTOntnfDIul Lcul Vcul [j] Tpeul
xa[rul] xa[r]cL gvif CTUca[r]LsTOtfDnca[r]Iulc] jmul[jul] VcTOcul jmul}.
EAS and MetaMorpho were rebuilt with a newer compiler than their deployments;
the only source change is the version pragma. The repair rows of Table~4
recompile Safe with \code{--optimize-runs 1000000} and with \code{--via-ir}
added, respectively.

\begin{table}[h]
\caption{Compiler invocations, per contract group.}
\label{tab:compile}
\centering
\footnotesize
\setlength{\tabcolsep}{4pt}
\begin{tabular}{@{}l l l@{}}
\toprule
\textbf{Contracts} & \textbf{Compiler} & \textbf{Options} \\
\midrule
Safe & solc 0.8.35 & \code{--optimize --optimize-runs 200} \\
 & & \quad \code{--evm-version shanghai --metadata-hash none} \\
EAS & solc 0.8.32 & \code{--optimize --optimize-runs 1000000 --via-ir} \\
 & & \quad \code{--evm-version paris --metadata-hash none} \\
Attester & solc 0.8.26 & \code{--optimize --optimize-runs 1000000} \\
 & & \quad \code{--evm-version paris --metadata-hash none} \\
Morpho Blue & solc 0.8.19 & \code{--optimize --optimize-runs 999999 --via-ir} \\
 & & \quad \code{--evm-version paris} \\
Comet$^{\dagger}$ & solc 0.8.15 & \code{--optimize --optimize-runs 1 --via-ir} \\
UniswapV3Pool & solc 0.7.6 & \code{--optimize --optimize-runs 800 --metadata-hash none} \\
PoolManager (v4) & solc 0.8.26 & \code{--optimize --optimize-runs 44444444 --via-ir} \\
 & & \quad \code{--evm-version cancun --metadata-hash none} \\
LlamaLend Vault & vyper 0.4.3 & \code{--evm-version cancun} \\
 & & \quad (\code{pragma optimize codesize} in the source) \\
MetaMorpho & solc 0.8.37 & \code{--optimize --optimize-runs 200 --via-ir} \\
 & & \quad \code{--evm-version cancun} \\
\midrule
DSS contracts & solc 0.6.12 & \code{--optimize --optimize-runs 200 --metadata-hash none} \\
WETH9 & solc 0.5.16 & \code{--optimize --optimize-runs 200} \\
Auction & solc 0.8.23 & \code{--optimize --evm-version shanghai --metadata-hash none} \\
UniswapV2Pair & solc 0.5.16 & \code{--optimize --optimize-runs 200} \\
Ownable2Step, Pausable, & solc 0.8.35 & \code{--optimize --evm-version shanghai} \\
\quad AccessControl, ERC6909 & & \quad (optimizer runs 200) \\
\bottomrule
\end{tabular}
\end{table}

%% file: appendix-prompts.tex
\lstdefinestyle{prompt}{basicstyle=\ttfamily\scriptsize,breaklines=true,breakatwhitespace=false,columns=fullflexible,keepspaces=true,showstringspaces=false,frame=leftline,framerule=0.4pt,rulecolor=\color{black!40},xleftmargin=6pt,aboveskip=3pt,belowskip=6pt,keywordstyle=,literate={—}{{---}}1 {“}{{``}}1 {”}{{''}}1 {’}{{'}}1 {…}{{\dots}}1 {−}{{-}}1}

\section{Human Inputs}
\label{app:prompts}

This appendix lists every human input to the sessions behind the benchmark
proofs, specification sessions and proof repair tables of the main paper,
verbatim and in the order given, from the goal to the input that completed it.
Each input carries its timestamp, its class and whether it is counted in the
Prompts column. Counted inputs are the goal, free-text steering and replies to
the agent's questions. Not counted are resumes (\code{/goal resume}), bare
continues and feedback (\eg, \code{/feedback false cybersecurity alarms}). 
These resume the blocked initial goal after successful resolution of an issue
(with an already counted human prompt), security block (common in some
contracts), or disconnected session.%
\footnote{We hit a few of these in one
session: \code{stream disconnected before completion: An error occurred while
processing your request. You can retry your request, or contact us through our
help center at help. openai. com if the error persists. Please include the
request ID 674b16a0-57b1-4a8b-a05e-80414c25b88a in your message.} We had to
start a new session for the proof to resume.}
Inputs before the first goal, after the completion call, and inputs that never
reached the agent are omitted.

\subsection{Proof sessions}

\subsubsection{Safe}
Counted inputs: 2 over 1 session

\paragraph{Session \texttt{01-proof-01a117dd}} {\sloppy Safe refinement proof (Benchmarks/Safe); goal complete 2026-10-08 22:30Z. Paused once (blocked) for a fix on main, resumed after the user's message. Goal set 2026-10-07T19:37:46Z; completed 2026-10-08T22:30:32Z. Counted 2 of 3 inputs between the goal and its completion; goal 1, resume 1, steer 1. 3 input(s) after completion omitted.\par}
\begin{enumerate}[leftmargin=*,itemsep=2pt]
\item \textbf{goal}, counted (2026-10-07T19:37:46Z):
\begin{lstlisting}[style=prompt]
Your goal is to complete the refinement proof of Benchmarks/Safe. Follow the directions in Misc/prompt.md. The proof should be completed and-to-end without sorries.
\end{lstlisting}
\item \textbf{steer}, counted (2026-10-08T07:21:26Z):
\begin{lstlisting}[style=prompt]
A fix has been pushed to main. Please pull this. You only need to rebuild Solm/, and Reasoning/. **Do not rebuild examples or the rest of the benchmarks**. Now readWithPadding returns everything. Every outOfGas argument has to be reasoned explicitly, and the corresponding constructor of the refinement should be used. For everything else, you will need to add an explicit well-formedness condition in the WF refinement relation. Before adding a WF condition, make sure that the case that needs it cannot be handled with a direct proof argument.
\end{lstlisting}
\item \textbf{resume}, not counted (2026-10-08T07:21:32Z):
\begin{lstlisting}[style=prompt]
/goal resume
\end{lstlisting}
\end{enumerate}

\subsubsection{EAS}
Counted inputs: 2 over 2 sessions

\paragraph{Session \texttt{01-proof-first-run-blocked-01a11f52}} {\sloppy EAS main-contract proof, first session; every turn ended with a stream disconnect, goal left blocked 2026-10-10 12:54Z. Goal set 2026-10-09T06:21:20Z; not completed in this session. Counted 1 of 3 inputs between the goal and its completion; goal 1, resume 2.\par}
\begin{enumerate}[leftmargin=*,itemsep=2pt]
\item \textbf{goal}, counted (2026-10-09T06:21:20Z):
\begin{lstlisting}[style=prompt]
You are working on the proof of Benchmarks/EAS/EAS/. You goal is to finish the proof without any sorries or admits. Your instructions are in Misc/prompt.md.
\end{lstlisting}
\item \textbf{resume}, not counted (2026-10-10T05:41:31Z):
\begin{lstlisting}[style=prompt]
/goal resume
\end{lstlisting}
\item \textbf{resume}, not counted (2026-10-10T12:52:35Z):
\begin{lstlisting}[style=prompt]
/goal resume
\end{lstlisting}
\end{enumerate}

\paragraph{Session \texttt{02-proof-resumed-01a125e8-1c84}} {\sloppy EAS main-contract proof, resumed in a new session with a new goal; goal complete 2026-10-10 17:12Z. Goal set 2026-10-10T13:02:55Z; completed 2026-10-10T17:12:13Z. Counted 1 of 1 inputs between the goal and its completion; goal 1.\par}
\begin{enumerate}[leftmargin=*,itemsep=2pt]
\item \textbf{goal}, counted (2026-10-10T13:02:55Z):
\begin{lstlisting}[style=prompt]
/goal You are resuming the proof of Benchmarks/EAS/EAS/ from a previous agent.  You goal is to finish the proof without any sorries or admits or extra assumptions. You are not allowed to make changes outside of your working directory. Do not rerun the differential testing or redo the initial audit. Initial instructions are in Misc/prompt.md.
\end{lstlisting}
\end{enumerate}

\subsubsection{Attester}
Counted inputs: 2 over 1 session

\paragraph{Session \texttt{01-proof-01a117fb}} {\sloppy EAS Attester (wrapper) proof; goal complete 2026-10-08 11:11Z. Same fix-on-main message as Safe. Goal set 2026-10-07T20:09:41Z; completed 2026-10-08T11:11:56Z. Counted 2 of 3 inputs between the goal and its completion; goal 1, resume 1, steer 1.\par}
\begin{enumerate}[leftmargin=*,itemsep=2pt]
\item \textbf{goal}, counted (2026-10-07T20:09:41Z):
\begin{lstlisting}[style=prompt]
Your job is to complete the proof of Benchmarks/EAS. Your directions are in Misc/prompt.md. The proof should be complete without sorries.
\end{lstlisting}
\item \textbf{steer}, counted (2026-10-08T07:22:00Z):
\begin{lstlisting}[style=prompt]
A fix has been pushed to main. Please pull this. You only need to rebuild Solm/, and Reasoning/. **Do not rebuild examples or the rest of the benchmarks**. Now readWithPadding returns everything. Every outOfGas argument has to be reasoned explicitly, and the corresponding constructor of the refinement should be used. For everything else, you will need to add an explicit well-formedness condition in the WF refinement relation. Before adding a WF condition, make sure that the case that needs it cannot be handled with a direct proof argument.
\end{lstlisting}
\item \textbf{resume}, not counted (2026-10-08T07:22:05Z):
\begin{lstlisting}[style=prompt]
/goal resume
\end{lstlisting}
\end{enumerate}

\subsubsection{Morpho Blue}
Counted inputs: 1 over 1 session

\paragraph{Session \texttt{01-proof-01a11c03}} {\sloppy Morpho Blue proof; goal complete 2026-10-09 05:11Z. First turn killed by a Codex restart, resumed by the goal loop. Goal set 2026-10-08T14:58:30Z; completed 2026-10-09T05:11:49Z. Counted 1 of 1 inputs between the goal and its completion; goal 1. 1 input(s) after completion omitted.\par}
\begin{enumerate}[leftmargin=*,itemsep=2pt]
\item \textbf{goal}, counted (2026-10-08T14:58:30Z):
\begin{lstlisting}[style=prompt]
you are working on the proof of Benchmarks/Morpho/MorphoBlue/. Your goal is to finish the proof without sorries or axioms. Follow instructions on Misc/prompt.md
\end{lstlisting}
\end{enumerate}

\subsubsection{Comet}
Counted inputs: 6 over 1 session

\paragraph{Session \texttt{01-proof-01a11c06}} {\sloppy Comet proof; goal complete 2026-10-09 20:02Z. Blocked four times (approval, fix on main, ABI decoder mode, constructor bound); post-completion discussion narrowed the decoder change. Goal set 2026-10-08T14:59:59Z; completed 2026-10-09T20:02:00Z. Counted 6 of 10 inputs between the goal and its completion; goal 1, resume 4, steer 5. 8 input(s) after completion omitted.\par}
\begin{enumerate}[leftmargin=*,itemsep=2pt]
\item \textbf{goal}, counted (2026-10-08T14:59:59Z):
\begin{lstlisting}[style=prompt]
You are working on the proof of Benchmarks/CompoundIII/Comet/. Your goal is to finish the proof without sorries or axioms. Follow the instructions on Misc/prompt.md
\end{lstlisting}
\item \textbf{steer}, counted (2026-10-08T15:12:36Z):
\begin{lstlisting}[style=prompt]
I approve
\end{lstlisting}
\item \textbf{resume}, not counted (2026-10-08T15:12:45Z):
\begin{lstlisting}[style=prompt]
/goal resume
\end{lstlisting}
\item \textbf{steer}, counted (2026-10-09T11:52:44Z):
\begin{lstlisting}[style=prompt]
Please pull the latest main to narrow the theorem with a resource bound on gas.
\end{lstlisting}
\item \textbf{resume}, not counted (2026-10-09T11:52:52Z):
\begin{lstlisting}[style=prompt]
/goal resume
\end{lstlisting}
\item \textbf{steer}, counted (2026-10-09T13:17:15Z):
\begin{lstlisting}[style=prompt]
Please explain clearly and briefly what exactly is the ABI decoding fix that is required.
\end{lstlisting}
\item \textbf{steer}, counted (2026-10-09T13:22:05Z):
\begin{lstlisting}[style=prompt]
Ok, please edit the decoder to add a new mode and only change code inside this mode to fix the missmatch. Name the mode after the solc compiler version used. Only change code inside this mode and do not chanage the decoder otherwise inluding other modes or general purpuse code. Recompile Solm and Reasoning but not Examples and other benchmarks. Then keep working on the proof.
\end{lstlisting}
\item \textbf{resume}, not counted (2026-10-09T13:22:15Z):
\begin{lstlisting}[style=prompt]
/goal resume
\end{lstlisting}
\item \textbf{steer}, counted (2026-10-09T16:55:19Z):
\begin{lstlisting}[style=prompt]
instead of putting the limit there, can you instead create a copy of the ConstructorRefinement relation that enforces this bound? Thank you.
\end{lstlisting}
\item \textbf{resume}, not counted (2026-10-09T17:05:14Z):
\begin{lstlisting}[style=prompt]
/goal resume
\end{lstlisting}
\end{enumerate}

\subsubsection{UniswapV3Pool}
Counted inputs: 1 over 1 session

\paragraph{Session \texttt{01-proof-01a11f5a}} {\sloppy UniswapV3 Pool proof; goal complete 2026-10-10 12:46Z, one uninterrupted turn. Goal set 2026-10-09T06:30:21Z; completed 2026-10-10T12:46:51Z. Counted 1 of 1 inputs between the goal and its completion; goal 1. 1 input(s) after completion omitted.\par}
\begin{enumerate}[leftmargin=*,itemsep=2pt]
\item \textbf{goal}, counted (2026-10-09T06:30:21Z):
\begin{lstlisting}[style=prompt]
You are working on the proof of Benchmarks/UniswapV3/Pool/. You goal is to finish the proof without any sorries or admits. You are not allowed to rewrite the differential testing suite and do not ever attempt to do it. Your instructions are in Misc/prompt.md.
\end{lstlisting}
\end{enumerate}

\subsubsection{PoolManager (v4)}
Counted inputs: 1 over 1 session

\paragraph{Session \texttt{01-proof-01a120bd}} {\sloppy UniswapV4 PoolManager proof (Lefteris' account on proof-machine); goal complete 2026-10-10 16:07Z. First turn killed by a Codex restart, resumed by the goal loop. Goal set 2026-10-09T12:58:11Z; completed 2026-10-10T16:07:04Z. Counted 1 of 1 inputs between the goal and its completion; goal 1.\par}
\begin{enumerate}[leftmargin=*,itemsep=2pt]
\item \textbf{goal}, counted (2026-10-09T12:58:11Z):
\begin{lstlisting}[style=prompt]
Implement a refienent proof for Benchmarks/Uniswap4PoolManager/, per the Misc/prompt.md guide.
\end{lstlisting}
\end{enumerate}

\subsubsection{LlamaLend Vault}
Counted inputs: 1 over 1 session

\paragraph{Session \texttt{01-proof-01a13454}} {\sloppy LlamaLend Vault (Vyper 0.4.3, curve-stablecoin lending/Vault.vy; the first Vyper benchmark) proof, Lefteris' account on proof-machine; goal complete 2026-10-13 17:58Z. First turn ended at 14:33Z with a server\_overloaded error (model at capacity), resumed by the user with /goal resume at 15:00Z. Goal set 2026-10-13T08:15:54Z; completed 2026-10-13T17:58:08Z. Counted 1 of 2 inputs between the goal and its completion; goal 1, resume 1.\par}
\begin{enumerate}[leftmargin=*,itemsep=2pt]
\item \textbf{goal}, counted (2026-10-13T08:15:54Z):
\begin{lstlisting}[style=prompt]
Implement a refinement proof for Benchmarks/CurveVault/, per the Misc/prompt.md guide.
\end{lstlisting}
\item \textbf{resume}, not counted (2026-10-13T15:00:19Z):
\begin{lstlisting}[style=prompt]
/goal resume
\end{lstlisting}
\end{enumerate}

\subsubsection{MetaMorpho}
Counted inputs: 5 over 3 sessions

\paragraph{Session \texttt{01-proof-01a11c98}} {\sloppy MetaMorpho V1.1 proof, session 1 (proof-machine-3, Misc/prompt.md): blocked 17:47Z pending approval of a scaffold fix, approved ('I approve.') and resumed; last turn failed on a false cybersecurity flag; goal left blocked 2026-10-08 21:35Z. Goal set 2026-10-08T17:39:42Z; not completed in this session. Counted 2 of 3 inputs between the goal and its completion; goal 1, resume 1, steer 1.\par}
\begin{enumerate}[leftmargin=*,itemsep=2pt]
\item \textbf{goal}, counted (2026-10-08T17:39:42Z):
\begin{lstlisting}[style=prompt]
You are working on the proof of Benchmarks/Morpho/MetaMorphoV1_1/. Your goal is to finish the proof without any sorries or axioms. Your instructions are in Misc/prompt.md.
\end{lstlisting}
\item \textbf{steer}, counted (2026-10-08T17:56:46Z):
\begin{lstlisting}[style=prompt]
I approve.
\end{lstlisting}
\item \textbf{resume}, not counted (2026-10-08T17:56:51Z):
\begin{lstlisting}[style=prompt]
/goal resume
\end{lstlisting}
\end{enumerate}

\paragraph{Session \texttt{02-proof-resumed-01a11f61}} {\sloppy MetaMorpho V1.1 proof, session 2 (proof-machine-3, resumed in a new session with a new goal): second turn cut by a Codex restart, last turn failed on a false cybersecurity flag; goal left blocked 2026-10-10 02:18Z. Goal set 2026-10-09T06:40:10Z; not completed in this session. Counted 1 of 1 inputs between the goal and its completion; goal 1. 1 undelivered input(s) omitted.\par}
\begin{enumerate}[leftmargin=*,itemsep=2pt]
\item \textbf{goal}, counted (2026-10-09T06:40:10Z):
\begin{lstlisting}[style=prompt]
You are resuming the proof of Benchmarks/Morpho/MetaMorphoV1_1/ from a previous agent.  You goal is to finish the proof without any sorries or admits or extra assumptions. You are not allowed to make changes outside of your working directory. Do not rerun the differential testing suite or redo the audit. Initial instructions are in Misc/prompt.md.
\end{lstlisting}
\end{enumerate}

\paragraph{Session \texttt{03-proof-completing-01a13938}} {\sloppy MetaMorpho V1.1 proof, completing session (Lefteris' account, new goal from the merged main with the new ABI decoding API): all 76 runtime proofs by 13:50Z, blocked once for the constructor code-size bound, approved by the user ('Apply the proposed patch'); goal complete 2026-10-14 18:39Z. Goal set 2026-10-14T07:03:32Z; completed 2026-10-14T18:39:12Z. Counted 2 of 3 inputs between the goal and its completion; goal 1, resume 1, steer 1.\par}
\begin{enumerate}[leftmargin=*,itemsep=2pt]
\item \textbf{goal}, counted (2026-10-14T07:03:32Z):
\begin{lstlisting}[style=prompt]
You are resuming the proof of Benchmarks/Morpho/MetaMorphoV1_1/ from a previous agent.  You goal is to finish the proof without any sorries or admits or extra assumptions. You are not allowed to make changes outside of your working directory. You can used the latest changes to the ABI decoding API to fix any blocker present in the current proof. Do not rerun the differential testing suite or redo the audit. Initial instructions are in Misc/prompt.md
\end{lstlisting}
\item \textbf{steer}, counted (2026-10-14T14:25:15Z):
\begin{lstlisting}[style=prompt]
Apply the proposed patch
\end{lstlisting}
\item \textbf{resume}, not counted (2026-10-14T14:25:18Z):
\begin{lstlisting}[style=prompt]
/goal resume
\end{lstlisting}
\end{enumerate}

\subsection{Specification sessions}

\subsubsection{EAS}
Counted inputs: 16 over 2 sessions

\paragraph{Session \texttt{01-scaffold-first-run-blocked-01a11c80}} {\sloppy EAS scaffold, first session; found a solc ABI-decoding mismatch, then eight turns failed on false cybersecurity alarms; goal blocked. Goal set 2026-10-08T17:14:42Z; not completed in this session. Counted 7 of 13 inputs between the goal and its completion; feedback 1, goal 1, resume 5, steer 6.\par}
\begin{enumerate}[leftmargin=*,itemsep=2pt]
\item \textbf{goal}, counted (2026-10-08T17:14:42Z):
\begin{lstlisting}[style=prompt]
pasted text file: /root/.codex-account2/attachments/d0073dea-db04-49e4-a0b8-308d653bcb6e/pasted-text-1.txt. Read this file before continuing.
\end{lstlisting}
Pasted attachment, recovered from the agent's read of the file:
\begin{lstlisting}[style=prompt]
Scaffold EAS (`EAS.sol`) as a proof-ready benchmark, following `Misc/scaffold-prompt.md` to the
end.

Sources: https://github.com/ethereum-attestation-service/eas-contracts at the commit of the
mainnet deployment 0xA1207F3BBa224E2c9c3c6D5aF63D0eb1582Ce587 (verified exact match:
https://etherscan.io/address/0xA1207F3BBa224E2c9c3c6D5aF63D0eb1582Ce587#code). Main unit
`contracts/EAS.sol`, contract `EAS`; the closure is `contracts/` plus the OpenZeppelin files the
verification lists. Use the Etherscan standard-json input as the authority for settings,
remappings and metadata.

Layout: `Benchmarks/EAS/` is the family root (it already holds `Attester/`). Sources go to
`Benchmarks/EAS/eas-contracts/`, the scaffold to `Benchmarks/EAS/EAS/`, module
`Benchmarks.EAS.EAS`.

Settings, exactly the deployment's, metadata included: solc 0.8.18 (v0.8.18+commit.87f61d96),
optimizer with 1,000,000 runs, no via-IR, default EVM version, metadata as in the verification
input:
  scripts/scaffold.py all --dir Benchmarks/EAS/EAS --solc <solc-0.8.18> \
      --sources-root Benchmarks/EAS/eas-contracts --main contracts/EAS.sol --contract EAS \
      --runs 1000000 --metadata-hash <as verified> \
      --remapping @openzeppelin/contracts/=<path in the sources root> \
      --module Benchmarks.EAS.EAS
`runtime.hex` must be byte-exact with the deployed code, metadata tail included. Record commit,
solc commit hash and the comparison in HANDOFF.md.

`ecrecover` (inside OpenZeppelin's SignatureChecker) is a call to precompile address 1; it has
no differential coverage on this machine, which is not a failure: note it in HANDOFF.md.

`DiffTarget.lean`: callees for a schema registry returning a record with a resolver and one
without, a resolver whose attest/revoke return true, an ERC-1271 signer, and a recipient that
accepts ETH, so that `lake exe solm-difftest --only EAS --count 50` reaches attest, multiAttest,
revoke, multiRevoke, timestamp, multiTimestamp and revokeOffchain.
\end{lstlisting}
\item \textbf{steer}, counted (2026-10-08T18:21:35Z):
\begin{lstlisting}[style=prompt]
What do you want to fix? Can you check if this is a compiler bug?
\end{lstlisting}
\item \textbf{steer}, counted (2026-10-08T18:25:43Z):
\begin{lstlisting}[style=prompt]
We should use a different compiler version to avoid the bug, for example 0.8.32.?
\end{lstlisting}
\item \textbf{resume}, not counted (2026-10-08T18:25:49Z):
\begin{lstlisting}[style=prompt]
/goal resume
\end{lstlisting}
\item \textbf{resume}, not counted (2026-10-08T18:37:34Z):
\begin{lstlisting}[style=prompt]
/goal resume
\end{lstlisting}
\item \textbf{resume}, not counted (2026-10-08T18:48:24Z):
\begin{lstlisting}[style=prompt]
/goal resume
\end{lstlisting}
\item \textbf{resume}, not counted (2026-10-08T18:54:27Z):
\begin{lstlisting}[style=prompt]
/goal resume
\end{lstlisting}
\item \textbf{steer}, counted (2026-10-08T19:07:07Z):
\begin{lstlisting}[style=prompt]
why is the goal keep getting security blocks?
\end{lstlisting}
\item \textbf{steer}, counted (2026-10-08T19:12:06Z):
\begin{lstlisting}[style=prompt]
Still stalled because of this false alarm. What can I do? It is blocking progress.
\end{lstlisting}
\item \textbf{resume}, not counted (2026-10-08T19:13:15Z):
\begin{lstlisting}[style=prompt]
/goal resume
\end{lstlisting}
\item \textbf{feedback}, not counted (2026-10-08T19:16:16Z):
\begin{lstlisting}[style=prompt]
/feedback false cybersecurity alarms
\end{lstlisting}
\item \textbf{steer}, counted (2026-10-08T19:23:09Z):
\begin{lstlisting}[style=prompt]
what would be the fix? What is different thann the current decoding? Explain briefly and accuretly.
\end{lstlisting}
\item \textbf{steer}, counted (2026-10-08T19:24:32Z):
\begin{lstlisting}[style=prompt]
Could this be a compiler bug or misspecification? Can you spec the ABI spec of Solidity?
\end{lstlisting}
\end{enumerate}

\paragraph{Session \texttt{02-scaffold-audit-01a11cfd-b936}} {\sloppy EAS scaffold continuation: decoder mode for the solc mismatch (goal 1), then the semantic audit (goal 2); goal complete 2026-10-08 21:46Z. Goal set 2026-10-08T19:30:51Z; completed 2026-10-08T21:46:42Z. Counted 9 of 9 inputs between the goal and its completion; goal 2, steer 7. 1 undelivered input(s) omitted.\par}
\begin{enumerate}[leftmargin=*,itemsep=2pt]
\item \textbf{goal}, counted (2026-10-08T19:30:51Z):
\begin{lstlisting}[style=prompt]
pasted text file: /root/.codex-account2/attachments/a73ec352-c361-4dee-883d-caa22996e2ce/pasted-text-1.txt. Read this file before continuing. We are working on this. You resume the work of a previous agent. The agennt foundn and ABI mismatch. You you check if the missmatch is a compiler bug, according to the official ABI spec? What would be the expected behavior and what is it doing? If tghis is not a big can we handle it with an extra case for this compiler version in the decoder?
\end{lstlisting}
Pasted attachment, recovered from the agent's read of the file:
\begin{lstlisting}[style=prompt]
Scaffold EAS (`EAS.sol`) as a proof-ready benchmark, following `Misc/scaffold-prompt.md` to the
end.

Sources: https://github.com/ethereum-attestation-service/eas-contracts at the commit of the
mainnet deployment 0xA1207F3BBa224E2c9c3c6D5aF63D0eb1582Ce587 (verified exact match:
https://etherscan.io/address/0xA1207F3BBa224E2c9c3c6D5aF63D0eb1582Ce587#code). Main unit
`contracts/EAS.sol`, contract `EAS`; the closure is `contracts/` plus the OpenZeppelin files the
verification lists. Use the Etherscan standard-json input as the authority for settings,
remappings and metadata.

Layout: `Benchmarks/EAS/` is the family root (it already holds `Attester/`). Sources go to
`Benchmarks/EAS/eas-contracts/`, the scaffold to `Benchmarks/EAS/EAS/`, module
`Benchmarks.EAS.EAS`.

Settings, exactly the deployment's, metadata included: solc 0.8.18 (v0.8.18+commit.87f61d96),
optimizer with 1,000,000 runs, no via-IR, default EVM version, metadata as in the verification
input:
  scripts/scaffold.py all --dir Benchmarks/EAS/EAS --solc <solc-0.8.18> \
      --sources-root Benchmarks/EAS/eas-contracts --main contracts/EAS.sol --contract EAS \
      --runs 1000000 --metadata-hash <as verified> \
      --remapping @openzeppelin/contracts/=<path in the sources root> \
      --module Benchmarks.EAS.EAS
`runtime.hex` must be byte-exact with the deployed code, metadata tail included. Record commit,
solc commit hash and the comparison in HANDOFF.md.

`ecrecover` (inside OpenZeppelin's SignatureChecker) is a call to precompile address 1; it has
no differential coverage on this machine, which is not a failure: note it in HANDOFF.md.

`DiffTarget.lean`: callees for a schema registry returning a record with a resolver and one
without, a resolver whose attest/revoke return true, an ERC-1271 signer, and a recipient that
accepts ETH, so that `lake exe solm-difftest --only EAS --count 50` reaches attest, multiAttest,
revoke, multiRevoke, timestamp, multiTimestamp and revokeOffchain.
\end{lstlisting}
\item \textbf{steer}, counted (2026-10-08T19:40:39Z):
\begin{lstlisting}[style=prompt]
You need to add a new decoder mode that captures exactly this.
\end{lstlisting}
\item \textbf{steer}, counted (2026-10-08T19:50:23Z):
\begin{lstlisting}[style=prompt]
can you explain why all of this code was needed? I though it was a single missmatch, and everything else could be the same.
\end{lstlisting}
\item \textbf{steer}, counted (2026-10-08T19:52:16Z):
\begin{lstlisting}[style=prompt]
Just add only what differs and reused as much of the existing decoder as possible.
\end{lstlisting}
\item \textbf{steer}, counted (2026-10-08T20:07:31Z):
\begin{lstlisting}[style=prompt]
why did you change the ABI decoder again and without asking?
\end{lstlisting}
\item \textbf{steer}, counted (2026-10-08T20:07:31Z):
\begin{lstlisting}[style=prompt]
you are only allowd to change the decoder only inside the new mode. Everything else has other dependencies.
\end{lstlisting}
\item \textbf{steer}, counted (2026-10-08T20:30:59Z):
\begin{lstlisting}[style=prompt]
Is it ready for proof?
\end{lstlisting}
\item \textbf{steer}, counted (2026-10-08T20:32:20Z):
\begin{lstlisting}[style=prompt]
why did you mark the goal as achieved then?
\end{lstlisting}
\item \textbf{goal}, counted (2026-10-08T20:35:29Z):
\begin{lstlisting}[style=prompt]
Please complete the semantic audit of EAS so that is ready for proof, as per Misc/scaffold-prompt.md. The scaffold is ready. You need to do the semanttic audit.
\end{lstlisting}
\end{enumerate}

\subsubsection{Morpho Blue}
Counted inputs: 1 over 1 session

\paragraph{Session \texttt{01-scaffold-01a11bbd}} {\sloppy Morpho Blue scaffold; goal complete 2026-10-08 14:38Z, one turn, no human input after the goal. Goal set 2026-10-08T13:39:34Z; completed 2026-10-08T14:38:07Z. Counted 1 of 1 inputs between the goal and its completion; goal 1.\par}
\begin{enumerate}[leftmargin=*,itemsep=2pt]
\item \textbf{goal}, counted (2026-10-08T13:39:34Z):
\begin{lstlisting}[style=prompt]
pasted text file: /root/.codex/attachments/bc955904-b96a-4d72-bd78-2ad134315dd0/pasted-text-1.txt. Read this file before continuing.
\end{lstlisting}
Pasted attachment, recovered from the agent's read of the file:
\begin{lstlisting}[style=prompt]
Scaffold Morpho Blue (`Morpho`) as a proof-ready benchmark, from scratch, following
`Misc/scaffold-prompt.md` to the end (hand-off included). Do not reuse anything from existing
scaffolds or other branches.

Sources: https://github.com/morpho-org/morpho-blue, at the commit of the mainnet deployment
0xBBBBBbbBBb9cC5e90e3b3Af64bdAF62C37EEFFCb (verified on Etherscan as an exact match:
https://etherscan.io/address/0xBBBBBbbBBb9cC5e90e3b3Af64bdAF62C37EEFFCb#code). Main unit
`src/Morpho.sol`, contract `Morpho`; the import closure is `src/interfaces/` and `src/libraries/`.
Confirm that `runtime.hex` is byte-exact with the deployed code and record the commit, the solc
commit hash and the comparison in HANDOFF.md.

Layout: first contract of a repository family, so create
  Benchmarks/Morpho/contracts/     the import closure, as in upstream
  Benchmarks/Morpho/MorphoBlue/    this scaffold, module `Benchmarks.Morpho.MorphoBlue`
Later contracts of the family (Vault V2, …) get sibling directories.

Compiler settings, exactly the deployment's: solc 0.8.19, optimizer enabled with 999,999 runs,
via-IR, EVM version paris:
  scripts/scaffold.py all --dir Benchmarks/Morpho/MorphoBlue --solc <solc-0.8.19> \
      --sources-root Benchmarks/Morpho/contracts --main Morpho.sol --contract Morpho \
      --runs 999999 --via-ir --evm-version paris --metadata-hash '' \
      --module Benchmarks.Morpho.MorphoBlue

What the draft will leave to you, and how it is modelled:
- `abi.encode(...)` under `keccak256` (constructor's DOMAIN_SEPARATOR, the authorization hash
  struct) and `bytes.concat("\x19\x01", ...)`: `abiEncodePacked` of 32-byte words, addresses
  and bools cast to `uint256`.
- `ecrecover`: a call to precompile address 1 with the packed `(digest, v, r, s)`, 32-byte
  return. This machine cannot execute the precompile, so `setAuthorizationWithSig` has no
  differential coverage here; say so in HANDOFF.md rather than treating it as a failure.
- `SafeTransferLib`: `extcodesize` check, low-level call with `abi.encodeCall`, then
  `returndata.length == 0 || abi.decode(returndata, (bool))`. Solm's `||` is strict: put the
  decode under an `if` on the length.
- `extSloads`: `new bytes32[](n)` and a loop of raw `sload`s into it; model the raw slot reads
  as the existing specs do.
- `Id` is `bytes32`; `Position` and `Market` are packed `uint128` structs; `createMarket`
  writes the whole `MarketParams` struct.
- External calls: `IIrm.borrowRate(MarketParams, Market)` (struct arguments), `IOracle.price()`,
  the five `msg.sender` callbacks with `bytes` data, and the ERC20 transfers above; all go into
  the spec's external-call table with the callees' ABIs.

`DiffTarget.lean`: add callees for an IRM returning a rate, an oracle returning a price, an
ERC20 that returns true and one that returns nothing, and a callback receiver; put WAD (1e18),
ORACLE_PRICE_SCALE (1e36), typical LLTVs (0.8e18, 0.945e18), MAX_FEE (0.25e18) and some
timestamps in `words`, so that `lake exe solm-difftest --only Morpho --count 50` reaches
createMarket, supply, borrow, repay, withdraw, liquidate, flashLoan and accrueInterest.

The solc 0.8.19 binary is not on this machine's PATH; give the agent its path or let it install one. The deployment commit I have not pinned myself; the byte-exact check will confirm whichever tag the agent picks.
\end{lstlisting}
\end{enumerate}

\subsubsection{Comet}
Counted inputs: 5 over 1 session

\paragraph{Session \texttt{01-scaffold-01a11bb0}} {\sloppy Comet scaffold; goal complete 2026-10-08 14:49Z, one turn, four question replies. Goal set 2026-10-08T13:25:45Z; completed 2026-10-08T14:49:39Z. Counted 5 of 5 inputs between the goal and its completion; goal 1, reply 4.\par}
\begin{enumerate}[leftmargin=*,itemsep=2pt]
\item \textbf{goal}, counted (2026-10-08T13:25:45Z):
\begin{lstlisting}[style=prompt]
pasted text file: /root/.codex-account2/attachments/bec9aa46-d25c-4000-85d8-02b29bc1e417/pasted-text-1.txt. Read this file before continuing.
\end{lstlisting}
Pasted attachment, recovered from the agent's read of the file:
\begin{lstlisting}[style=prompt]
Scaffold Compound III's `CometWithExtendedAssetList` as a proof-ready benchmark, from scratch,
following `Misc/scaffold-prompt.md` to the end (hand-off included).

Inputs are upstream and the automation only. Do not read or reuse anything from the old
scaffolds (`Benchmarks/Scaffolds/Comet`, `Benchmarks/Scaffolds/CometRewards`,
`Benchmarks/Scaffolds/CompoundIII`) or from other branches: they are stale.

Sources: fetch https://github.com/compound-finance/comet at commit
f766f51583c23acc33b2a7824654ef2029a96804 (the current head of `main`); main unit
`contracts/CometWithExtendedAssetList.sol`, contract `CometWithExtendedAssetList`.

Layout: this is the first contract of a repository family, so create the root
`Benchmarks/CompoundIII/` with the sources shared by every contract of the repository and one
directory per contract:
  Benchmarks/CompoundIII/contracts/   the import closure of the main unit, as in upstream
  Benchmarks/CompoundIII/Comet/       this scaffold, module `Benchmarks.CompoundIII.Comet`
Later contracts of the same repository (CometRewards, …) get sibling directories and reuse
`contracts/`.

Compiler settings, exactly upstream's `hardhat.config.ts`: solc 0.8.15, optimizer enabled with
1 run, via-IR, no EVM version set, metadata at solc's default:
  scripts/scaffold.py all --dir Benchmarks/CompoundIII/Comet --solc <solc-0.8.15> \
      --sources-root Benchmarks/CompoundIII/contracts --main CometWithExtendedAssetList.sol \
      --contract CometWithExtendedAssetList --runs 1 --via-ir --metadata-hash '' \
      --module Benchmarks.CompoundIII.Comet
Record the solc commit hash and the artifact hashes in HANDOFF.md.

What to expect in this contract:
- via-IR: the dispatcher has no global callvalue guard; calldata reads happen in Yul.
- Many immutables (governor, pause guardian, base token and price feed, extension delegate,
  rates, kink, factors, the asset-list contract): the immutable modules and the patched runtime
  are generated; the valuation for the differential target comes from a constructor run.
- A `fallback` that delegatecalls `extensionDelegate` (CometExt): model it with Solm's
  `delegatecall`; the differential target needs a callee for it.
- External calls to price feeds (`latestRoundData`), to the asset list, and ERC20 transfers via
  `doTransferIn`/`doTransferOut` with return-data checks in assembly: modelled by their effect,
  with the ABI entries in the spec's external-call table.
- Loops over the asset list in `getAssetInfoByAddress`, `isBorrowCollateralized`,
  `isLiquidatable`, `absorbInternal`; `absorb` loops over accounts and calls `absorbInternal`.
- Runtime about 18.6 KB: expect many block summaries and a long audit; nothing is skipped.

Give `DiffTarget.lean` the callees the success paths need (a price feed returning a round, an
asset list, an ERC20, the extension delegate) and the amounts in `words`, so that
`lake exe solm-difftest --only CometWithExtendedAssetList --count 50` reaches supply, withdraw,
transfer, absorb and buyCollateral.
\end{lstlisting}
\item \textbf{reply}, counted (2026-10-08T13:27:57Z), to the agent's question ``The pinned hardhat.config.ts also sets a custom Yul optimizer sequence, which the supplied scaffold command omits. Which compiler configuration should define the benchmark?'':
\begin{lstlisting}[style=prompt]
Match hardhat.config.ts exactly, including its Yul sequence (recommended)
\end{lstlisting}
\item \textbf{reply}, counted (2026-10-08T13:59:26Z), to the agent's question ``May I apply /tmp/compound-remove-dangling-target.patch? It removes two missing CometRewards imports, its unused target definition, and its list entry from Tests/DiffTest/Targets.lean so the required solm-difftest command can build.'':
\begin{lstlisting}[style=prompt]
Apply the prepared registry cleanup
\end{lstlisting}
\item \textbf{reply}, counted (2026-10-08T14:07:48Z), to the agent's question ``May I apply /tmp/compound-immutable-accessors.patch to scripts/proof\_skeleton.py and regenerate the skeleton? It fixes the immutable accessor build timeouts; all 25 accessors and their word lemmas pass with the replacement template.'':
\begin{lstlisting}[style=prompt]
Apply the verified generator fix
\end{lstlisting}
\item \textbf{reply}, counted (2026-10-08T14:24:09Z), to the agent's question ``May I apply /tmp/compound-fallback-generator.patch? It stops generating false no-dispatch/revert claims for fallback contracts and leaves the unmatched-calldata refinement as an explicit proof stub.'':
\begin{lstlisting}[style=prompt]
Apply the fallback generator correction
\end{lstlisting}
\end{enumerate}

\subsubsection{UniswapV3Pool}
Counted inputs: 4 over 1 session

\paragraph{Session \texttt{01-scaffold-01a11c95-3b6d}} {\sloppy UniswapV3 Pool scaffold; goal complete 2026-10-08 23:07Z. Goal set 2026-10-08T17:56:09Z; completed 2026-10-08T23:07:44Z. Counted 4 of 4 inputs between the goal and its completion; goal 1, steer 3.\par}
\begin{enumerate}[leftmargin=*,itemsep=2pt]
\item \textbf{goal}, counted (2026-10-08T17:56:09Z):
\begin{lstlisting}[style=prompt]
pasted text file: /root/.codex-account2/attachments/ec0f79eb-136c-4d67-b97c-a495a10cfad3/pasted-text-1.txt. Read this file before continuing.
\end{lstlisting}
Pasted attachment, recovered from the agent's read of the file:
\begin{lstlisting}[style=prompt]
Scaffold Uniswap V3 `UniswapV3Pool` as a proof-ready benchmark, following
`Misc/scaffold-prompt.md` to the end.

Sources: https://github.com/Uniswap/v3-core at the commit of the mainnet deployment; main unit
`contracts/UniswapV3Pool.sol`, contract `UniswapV3Pool`; the closure is `contracts/` (libraries,
interfaces, `NoDelegateCall`). Pools are created by the factory
0x1F98431c8aD98523631AE4a59f267346ea31F984 through CREATE2 with no constructor arguments: the
constructor reads `parameters()` from its deployer and the six immutables are patched into the
runtime. Use the factory's Etherscan verification (its input embeds the pool) as the authority
for settings; for byte-exactness compare the template, patched with a deployed pool's immutable
values (for example 0x88e6A0c2dDD26FEEb64F039a2c41296FcB3f5640), with that pool's code.

Layout: family root `Benchmarks/UniswapV3/` (V4 gets its own root later); sources in
`Benchmarks/UniswapV3/v3-core/`, scaffold in `Benchmarks/UniswapV3/Pool/`, module
`Benchmarks.UniswapV3.Pool`.

Settings, exactly upstream's: solc 0.7.6 (v0.7.6+commit.7338295f), optimizer with 800 runs, no
via-IR, default EVM version, metadata hash none:
  scripts/scaffold.py all --dir Benchmarks/UniswapV3/Pool --solc <solc-0.7.6> \
      --sources-root Benchmarks/UniswapV3/v3-core --main contracts/UniswapV3Pool.sol \
      --contract UniswapV3Pool --runs 800 --metadata-hash none \
      --module Benchmarks.UniswapV3.Pool
Record commit, solc commit hash and the comparison in HANDOFF.md.

The library assembly (`FullMath`, `TickMath`, `UnsafeMath`) is pure arithmetic: model it as the
exact mathematical operations.

`DiffTarget.lean`: callees for the deployer (`parameters()` returning the six values), the
factory (`owner()`), two tokens whose `balanceOf` the callback receiver can raise, and callback
receivers for mint, swap and flash; the tick-spacing and typical sqrt-price words in `words`, so
that `lake exe solm-difftest --only UniswapV3Pool --count 50` reaches initialize, mint, burn,
collect, swap, flash, increaseObservationCardinalityNext and observe.
\end{lstlisting}
\item \textbf{steer}, counted (2026-10-08T18:36:17Z):
\begin{lstlisting}[style=prompt]
yes
\end{lstlisting}
\item \textbf{steer}, counted (2026-10-08T19:13:08Z):
\begin{lstlisting}[style=prompt]
you should do the semantic audit in parallel to waiting.
\end{lstlisting}
\item \textbf{steer}, counted (2026-10-08T20:37:51Z):
\begin{lstlisting}[style=prompt]
Can you explain why is this tetst suite taking so long to run?
\end{lstlisting}
\end{enumerate}

\subsubsection{PoolManager (v4)}
Counted inputs: 2 over 1 session

\paragraph{Session \texttt{01-scaffold-01a12043}} {\sloppy UniswapV4 PoolManager scaffold (Lefteris' account); goal complete 2026-10-09 12:20Z; four turns failed on false cybersecurity alarms. Goal set 2026-10-09T10:44:19Z; completed 2026-10-09T12:20:38Z. Counted 2 of 8 inputs between the goal and its completion; continue 3, goal 1, resume 3, steer 1.\par}
\begin{enumerate}[leftmargin=*,itemsep=2pt]
\item \textbf{goal}, counted (2026-10-09T10:44:19Z):
\begin{lstlisting}[style=prompt]
Apply the process of Misc/scaffold-prompt.md for the UniswapV4 Pool manager, as in the repository https://github.com/Uniswap/v4-core/blob/main/src/PoolManager.sol, with the settings as described in the repo's foundry.toml file, in directory Benchmarks/Scaffolds/UniswapV4PoolManager and module Benchmarks.Scaffolds.UniswapV4PoolManager. Do not declare the goal complete until the prompt's targets are done.
\end{lstlisting}
\item \textbf{steer}, counted (2026-10-09T10:53:59Z):
\begin{lstlisting}[style=prompt]
If a well-formedness condition can fix this, as in the style of Benchmarks/Dss/Core/Common, use that
\end{lstlisting}
\item \textbf{continue}, not counted (2026-10-09T11:39:25Z):
\begin{lstlisting}[style=prompt]
continue
\end{lstlisting}
\item \textbf{resume}, not counted (2026-10-09T11:42:23Z):
\begin{lstlisting}[style=prompt]
/goal resume
\end{lstlisting}
\item \textbf{resume}, not counted (2026-10-09T11:59:25Z):
\begin{lstlisting}[style=prompt]
/goal resume
\end{lstlisting}
\item \textbf{continue}, not counted (2026-10-09T12:00:29Z):
\begin{lstlisting}[style=prompt]
continue
\end{lstlisting}
\item \textbf{continue}, not counted (2026-10-09T12:05:16Z):
\begin{lstlisting}[style=prompt]
continue
\end{lstlisting}
\item \textbf{resume}, not counted (2026-10-09T12:19:13Z):
\begin{lstlisting}[style=prompt]
/goal resume
\end{lstlisting}
\end{enumerate}

\subsubsection{LlamaLend Vault}
Counted inputs: 1 over 1 session

\paragraph{Session \texttt{01-scaffold-01a13409}} {\sloppy LlamaLend Vault (Vyper) scaffold, Lefteris' account; goal complete 2026-10-13 08:03Z, one turn, no human input after the goal. Goal set 2026-10-13T07:01:04Z; completed 2026-10-13T08:03:04Z. Counted 1 of 1 inputs between the goal and its completion; goal 1.\par}
\begin{enumerate}[leftmargin=*,itemsep=2pt]
\item \textbf{goal}, counted (2026-10-13T07:01:04Z):
\begin{lstlisting}[style=prompt]
Apply the process of Misc/scaffold-prompt.md for the curve-stablecoing lending Vault, of the repository https://github.com/curvefi/curve-stablecoin/blob/f19a18abef6e75e3e08b1aa427ede724162bce57/curve_stablecoin/lending/Vault.vy, with the settings as described in the repo's foundry.toml file, in directory Benchmarks/CurveVault and module Benchmarks.CurveVault. Do not declare the goal complete until the prompt's targets are done.
\end{lstlisting}
\end{enumerate}

\subsubsection{MetaMorpho}
Counted inputs: 6 over 1 session

\paragraph{Session \texttt{01-scaffold-01a11be9}} {\sloppy MetaMorpho scaffold (proof-machine-3, pasted goal); goal complete 2026-10-08 17:37Z, including resumed work. Goal set 2026-10-08T14:27:56Z; completed 2026-10-08T17:36:43Z. Counted 6 of 8 inputs between the goal and its completion; goal 1, reply 1, resume 2, steer 4.\par}
\begin{enumerate}[leftmargin=*,itemsep=2pt]
\item \textbf{goal}, counted (2026-10-08T14:27:56Z):
\begin{lstlisting}[style=prompt]
pasted text file: /root/.codex/attachments/930c59af-27c0-4441-816c-1c7a52da3ee5/pasted-text-1.txt. Read this file before continuing.
\end{lstlisting}
Pasted attachment, recovered from the agent's read of the file:
\begin{lstlisting}[style=prompt]
Scaffold MetaMorpho V1.1 (`MetaMorphoV1_1`) as a proof-ready benchmark, from scratch, following
`Misc/scaffold-prompt.md` to the end (hand-off included). Do not reuse anything from existing
scaffolds or other branches.

Sources: https://github.com/morpho-org/metamorpho-v1.1 at the commit of the mainnet deployment,
main unit `src/MetaMorphoV1_1.sol`, contract `MetaMorphoV1_1`. Its imports are relative
(`../lib/openzeppelin-contracts/...`, `../lib/morpho-blue/...`), so keep the repository's `src/`
and `lib/` layout under the sources root with the submodules at their pinned commits.

Deployment: vaults are immutable instances created by `MetaMorphoV1_1Factory` (Ethereum and
Base, metadata hash included). Take the factory address from Morpho's addresses page, pick one
mainnet vault it created that is verified on Etherscan, and use that verification's
standard-json input as the authority for every setting.

Compiler settings, exactly upstream's, metadata included: solc 0.8.26, optimizer enabled with
200 runs, via-IR, EVM version cancun, metadata hash ipfs with the CBOR appended (Foundry's
defaults; `foundry.toml` sets only via-ir, runs and evm_version):
  scripts/scaffold.py all --dir Benchmarks/Morpho/MetaMorphoV1_1 --solc <solc-0.8.26> \
      --sources-root Benchmarks/Morpho/metamorpho-v1.1 --main src/MetaMorphoV1_1.sol \
      --contract MetaMorphoV1_1 --runs 200 --via-ir --evm-version cancun --metadata-hash ipfs \
      --module Benchmarks.Morpho.MetaMorphoV1_1
The deployed vault's code must equal the template with the two immutables (`MORPHO`,
`DECIMALS_OFFSET`) patched to that vault's values, metadata tail included; if the tail differs,
find out why (source paths, remappings, settings) and fix the input rather than accepting it.
Record commit, solc commit hash, vault and factory addresses and the comparison in HANDOFF.md.

Layout: sibling of the Morpho Blue scaffold under `Benchmarks/Morpho/`; the sources of this
repository live in their own directory, `Benchmarks/Morpho/metamorpho-v1.1/`, because they are a
different repository with its own dependencies.

What the draft will leave to you, and how it is modelled:
- The OpenZeppelin bases are inlined by the translator: `ERC4626` (`SafeERC20` with its assembly
  call and return-data check, by effect; `mulDiv` as exact arithmetic), `ERC20Permit`
  (`ecrecover`: a call to precompile address 1; no differential coverage on this machine, say so
  in HANDOFF.md), `Ownable2Step`, `Multicall` (self-`delegatecall` per entry, revert bubbling by
  effect).
- `marketParams.id()`: `keccak256(abi.encode(...))` of five static fields, written as
  `abiEncodePacked` of 32-byte words with addresses cast to `uint256`.
- Calls to Morpho Blue: `supply`, `withdraw` (two return values), `position`, `market`,
  `idToMarketParams` (struct returns), `accrueInterest`. The generated external-call table does
  not fill struct or multi-value returns; write those entries by hand from the Blue ABI.
- `try MORPHO.supply(...) {} catch {}` and `try MORPHO.withdraw(...) {} catch {}` are
  `checkedCall`s; the loops over `supplyQueue`, `withdrawQueue` and `allocations`, and
  `new bool[](n)` / `new Id[](n)` in `updateWithdrawQueue`, are ordinary.
- Keys are `address` and `Id` (`bytes32`); the pending-value structs and the queues are plain
  storage.

`DiffTarget.lean`: put the Morpho Blue scaffold's bytecode (`Benchmarks/Morpho/MorphoBlue`) at
the vault's `MORPHO` address in `callees`, plus an ERC20 that returns true and one that returns
nothing; add WAD (1e18), a few `Id` words from created markets and typical caps to `words`, so
that `lake exe solm-difftest --only MetaMorphoV1_1 --count 50` reaches deposit, withdraw,
submitCap/acceptCap, setSupplyQueue, updateWithdrawQueue and reallocate. Where the real Blue
state makes success unreachable, add a second callee that answers the Blue interface with fixed
values and say which paths it covers.
\end{lstlisting}
\item \textbf{reply}, counted (2026-10-08T14:43:56Z), to the agent's question ``This branch has no Benchmarks/Morpho/MorphoBlue scaffold. For the required differential-test callee, should I generate a fresh Blue bytecode module at that path from pinned upstream sources, or use a Blue scaffold at another local location?'':
\begin{lstlisting}[style=prompt]
Generate the Blue bytecode module fresh
\end{lstlisting}
\item \textbf{steer}, counted (2026-10-08T16:25:24Z):
\begin{lstlisting}[style=prompt]
Are you sure this is needed and these two aren't equivalent. If if is needed explain clearly and briefly and report a short cex.
\end{lstlisting}
\item \textbf{steer}, counted (2026-10-08T16:33:43Z):
\begin{lstlisting}[style=prompt]
Can you verify if this is how storage arrays work for new Solidity versions? Which solc version is this one for? Or is this particular for this one contract?
\end{lstlisting}
\item \textbf{steer}, counted (2026-10-08T16:50:22Z):
\begin{lstlisting}[style=prompt]
Alright. I am thinking that we should compile with a version that doesn't have this bug then.
\end{lstlisting}
\item \textbf{resume}, not counted (2026-10-08T17:00:36Z):
\begin{lstlisting}[style=prompt]
/goal resume
\end{lstlisting}
\item \textbf{steer}, counted (2026-10-08T17:16:18Z):
\begin{lstlisting}[style=prompt]
Is the spec blocked or fixed? I do not understand.
\end{lstlisting}
\item \textbf{resume}, not counted (2026-10-08T17:16:53Z):
\begin{lstlisting}[style=prompt]
/goal resume
\end{lstlisting}
\end{enumerate}

\subsection{Repair sessions}

\subsubsection{step back}
Counted inputs: 1 over 1 session

\paragraph{Session \texttt{01-step-back-77901a5-to-cdb2eb5-01a12f4f}} {\sloppy Backward repair Main (77901a5) $\to$ V0\_cdb2eb5; goal complete 2026-10-12 10:34Z. Goal set 2026-10-12T08:54:16Z; completed 2026-10-12T10:34:59Z. Counted 1 of 1 inputs between the goal and its completion; goal 1.\par}
\begin{enumerate}[leftmargin=*,itemsep=2pt]
\item \textbf{goal}, counted (2026-10-12T08:54:16Z):
\begin{lstlisting}[style=prompt]
Please port the proof of Benchmarks/Safe to a different commit using the instructions in EquiVM/Benchmarks/Safe/repair-prompt.md. The inputs are FROM_COMMIT=77901a5a1ad835b74ad3b72f73a8412cfe491c57 TO_COMMIT=cdb2eb578dbdba4c3f10a47f9a2dd9580773e63a START=Main TARGET=V0_cdb2eb5. Thank you.
\end{lstlisting}
\end{enumerate}

\subsubsection{replay 1}
Counted inputs: 1 over 1 session

\paragraph{Session \texttt{02-replay-1-cdb2eb5-to-0227664-01a12fb4}} {\sloppy Forward repair V0\_cdb2eb5 $\to$ V1\_0227664; goal complete 2026-10-12 11:28Z. Goal set 2026-10-12T10:44:59Z; completed 2026-10-12T11:28:47Z. Counted 1 of 1 inputs between the goal and its completion; goal 1.\par}
\begin{enumerate}[leftmargin=*,itemsep=2pt]
\item \textbf{goal}, counted (2026-10-12T10:44:59Z):
\begin{lstlisting}[style=prompt]
Please port the proof of Benchmarks/Safe to a different commit using the instructions in EquiVM/Benchmarks/Safe/repair-prompt.md. FROM_COMMIT=cdb2eb578dbdba4c3f10a47f9a2dd9580773e63a TO_COMMIT=02276647d5dd3353edc52d8c3e915f7f18dab95a START=V0_cdb2eb5 TARGET=V1_0227664. Thank you.
\end{lstlisting}
\end{enumerate}

\subsubsection{replay 2}
Counted inputs: 1 over 1 session

\paragraph{Session \texttt{03-replay-2-0227664-to-0e91d99-01a12fee-5037}} {\sloppy Forward repair V1\_0227664 $\to$ V2\_0e91d99; goal complete 2026-10-12 12:25Z. Goal set 2026-10-12T11:46:25Z; completed 2026-10-12T12:25:09Z. Counted 1 of 1 inputs between the goal and its completion; goal 1.\par}
\begin{enumerate}[leftmargin=*,itemsep=2pt]
\item \textbf{goal}, counted (2026-10-12T11:46:25Z):
\begin{lstlisting}[style=prompt]
Please port the proof of Benchmarks/Safe to a different commit using the instructions in EquiVM/Benchmarks/Safe/repair-prompt.md. FROM_COMMIT=02276647d5dd3353edc52d8c3e915f7f18dab95a TO_COMMIT=0e91d99ea35f2ec650dd7cb3795f78eeacec7985 START=V1_0227664 TARGET=V2_0e91d99. Thank you.
\end{lstlisting}
\end{enumerate}

\subsubsection{replay 3}
Counted inputs: 1 over 1 session

\paragraph{Session \texttt{04-replay-3-0e91d99-to-77901a5-01a1303c}} {\sloppy Forward repair V2\_0e91d99 $\to$ V3\_77901a5 in the isolated checkout /root/EquiVM-repair; goal complete 2026-10-12 14:34Z. Goal set 2026-10-12T13:11:59Z; completed 2026-10-12T14:34:37Z. Counted 1 of 1 inputs between the goal and its completion; goal 1.\par}
\begin{enumerate}[leftmargin=*,itemsep=2pt]
\item \textbf{goal}, counted (2026-10-12T13:11:59Z):
\begin{lstlisting}[style=prompt]
Please port the proof of Benchmarks/Safe to a different commit using the instructions in EquiVM-repair/Benchmarks/Safe/repair-prompt.md. FROM_COMMIT=0e91d99ea35f2ec650dd7cb3795f78eeacec7985 TO_COMMIT=77901a5a1ad835b74ad3b72f73a8412cfe491c57 START=V2_0e91d99 TARGET=V3_77901a5. Do not search the internet or search other directories other than this Thank you.
\end{lstlisting}
\end{enumerate}

\subsubsection{optimizer}
Counted inputs: 1 over 1 session

\paragraph{Session \texttt{05-optimizer-runs-200-to-1000000-01a131ab-fc45}} {\sloppy Compiler-configuration repair Main (--optimize-runs 200) $\to$ Runs1e6 (--optimize-runs 1000000); goal complete 2026-10-12 20:19Z. Goal set 2026-10-12T19:52:00Z; completed 2026-10-12T20:19:26Z. Counted 1 of 1 inputs between the goal and its completion; goal 1.\par}
\begin{enumerate}[leftmargin=*,itemsep=2pt]
\item \textbf{goal}, counted (2026-10-12T19:52:00Z):
\begin{lstlisting}[style=prompt]
Please port the proof of Benchmarks/Safe to a different compiler configuration using the instructions in EquiVM/Benchmarks/Safe/compiler-prompt.md. Thank you.
\end{lstlisting}
\end{enumerate}

\subsubsection{via-IR}
Counted inputs: 14 over 1 session

\paragraph{Session \texttt{06-via-ir-legacy-to-via-ir-01a1341c-d600}} {\sloppy Compiler-pipeline repair Main (legacy pipeline) $\to$ ViaIR (--via-ir, same sources, compiler and optimizer runs); goal complete 2026-10-13 19:16Z after four blocked stops (a calldata-decoder counterexample and spec question, allocation-guard patch approvals) resumed by the user. Goal set 2026-10-13T07:14:30Z; completed 2026-10-13T19:16:09Z. Counted 14 of 19 inputs between the goal and its completion; goal 1, reply 1, resume 5, steer 12.\par}
\begin{enumerate}[leftmargin=*,itemsep=2pt]
\item \textbf{goal}, counted (2026-10-13T07:14:30Z):
\begin{lstlisting}[style=prompt]
Please port the proof of Benchmarks/Safe to a different compiler configuration using the instructions in EquiVM/Benchmarks/Safe/compiler-prompt-ir.md. Thank you.
\end{lstlisting}
\item \textbf{steer}, counted (2026-10-13T07:39:00Z):
\begin{lstlisting}[style=prompt]
yes apply the patch
\end{lstlisting}
\item \textbf{steer}, counted (2026-10-13T08:31:16Z):
\begin{lstlisting}[style=prompt]
could this be a compiler bug?
\end{lstlisting}
\item \textbf{steer}, counted (2026-10-13T08:39:16Z):
\begin{lstlisting}[style=prompt]
Can this be fixed with a change in specification? If yes, what would this be?
\end{lstlisting}
\item \textbf{steer}, counted (2026-10-13T08:39:40Z):
\begin{lstlisting}[style=prompt]
also condider the additional decoding modes we have available
\end{lstlisting}
\item \textbf{steer}, counted (2026-10-13T08:46:11Z):
\begin{lstlisting}[style=prompt]
modern is not via-ir specific so this is not a good change. Do the smallest required change for this particular example please.
\end{lstlisting}
\item \textbf{steer}, counted (2026-10-13T09:07:56Z):
\begin{lstlisting}[style=prompt]
Please continue with the proof as per initial instructions.
\end{lstlisting}
\item \textbf{resume}, not counted (2026-10-13T09:07:59Z):
\begin{lstlisting}[style=prompt]
/goal resume
\end{lstlisting}
\item \textbf{steer}, counted (2026-10-13T10:41:51Z):
\begin{lstlisting}[style=prompt]
yes you can
\end{lstlisting}
\item \textbf{resume}, not counted (2026-10-13T10:41:55Z):
\begin{lstlisting}[style=prompt]
/goal resume
\end{lstlisting}
\item \textbf{steer}, counted (2026-10-13T11:13:13Z):
\begin{lstlisting}[style=prompt]
apply the local guard fix
\end{lstlisting}
\item \textbf{steer}, counted (2026-10-13T11:52:18Z):
\begin{lstlisting}[style=prompt]
I approve this
\end{lstlisting}
\item \textbf{resume}, not counted (2026-10-13T12:06:06Z):
\begin{lstlisting}[style=prompt]
/goal resume
\end{lstlisting}
\item \textbf{steer}, counted (2026-10-13T12:19:07Z):
\begin{lstlisting}[style=prompt]
I approve
\end{lstlisting}
\item \textbf{resume}, not counted (2026-10-13T12:33:18Z):
\begin{lstlisting}[style=prompt]
/goal resume
\end{lstlisting}
\item \textbf{steer}, counted (2026-10-13T15:48:18Z):
\begin{lstlisting}[style=prompt]
are you working?
\end{lstlisting}
\item \textbf{steer}, counted (2026-10-13T17:15:37Z):
\begin{lstlisting}[style=prompt]
yes
\end{lstlisting}
\item \textbf{resume}, not counted (2026-10-13T17:45:28Z):
\begin{lstlisting}[style=prompt]
/goal resume
\end{lstlisting}
\item \textbf{reply}, counted (2026-10-13T18:42:13Z), to the agent's question ``Approve the additional local setup guard refinement in /tmp/safe-ir-token-payment-guard.patch? It reserves the token encoder's extra 128 bytes by using the bound 2\textasciicircum{}64 $-$ 352 only when payment > 0 and paymentToken != address(0); other calls retain 2\textasciicircum{}64 $-$ 224. This extends the previously approved specification change.'':
\begin{lstlisting}[style=prompt]
Approve the conditional guard
\end{lstlisting}
\end{enumerate}